\documentclass[lettersize,journal]{IEEEtran}
\usepackage{amsmath,amsfonts}
\usepackage{algorithmic}
\usepackage{algorithm}
\usepackage{array}
\ifCLASSOPTIONcompsoc
\usepackage[caption=false, font=normalsize, labelfont=sf, textfont=sf]{subfig}
\else
\usepackage[caption=false, font=footnotesize]{subfig}
\fi
\usepackage{textcomp}
\usepackage{stfloats}
\usepackage{url}
\usepackage{verbatim}
\usepackage{graphicx}
\usepackage{epstopdf}
\usepackage{cite}
\usepackage{color}
\usepackage{amssymb}
\usepackage{multirow}
\usepackage{bbding}
\usepackage{makecell}
\usepackage[T1]{fontenc}

\newcommand{\tablesize}{\fontsize{7.5pt}{12pt}\selectfont}

\def\C{\mathbf{C}}
\def\D{\mathbf{D}}
\def\E{\mathbf{E}}

\def\G{\mathbf{G}}
\def\H{\mathbf{H}}
\def\K{\mathbf{K}}

\def\N{\mathbf{N}}

\def\P{\mathbf{P}}

\def\S{\mathbf{S}}

\def\U{\mathbf{U}}

\def\W{\mathbf{W}}
\def\X{\mathbf{X}}
\def\Y{\mathbf{Y}}

\def\d{\mathbf{d}}

\def\h{\mathbf{h}}
\def\k{\mathbf{k}}

\def\z{\mathbf{z}}
\def\cited#1{{\iffalse #1 \fi}}
\def\1#1{\textcolor{red}{\textbf{#1}}}
\def\2#1{\textcolor{blue}{\textbf{#1}}}
\def\3#1{\textcolor{green}{\textbf{#1}}}

\begin{document}

\title{Deep Equilibrium Convolutional Sparse Coding for Hyperspectral Image Denoising}

\author{Jin Ye, Jingran Wang, Fengchao Xiong,~\IEEEmembership{Member,~IEEE}, Jingzhou Chen,~\IEEEmembership{Member,~IEEE} and Yuntao Qian,~\IEEEmembership{Senior Member,~IEEE}
\thanks {This work was supported in part by  the National Natural Science Foundation of China under Grant 62371237 and the Fundamental Research Funds for the Central Universities under Grant 30923010213. (Corresponding author: Fengchao Xiong.)}
\thanks{Jin Ye, Jingran Wang, Fengchao Xiong and Jingzhou Chen  are with the School of Computer Science and Engineering, Nanjing University of Science and Technology, Nanjing 210094, China. }
\thanks{Yuntao Qian  is with the College of Computer Science, Zhejiang University, Hangzhou 310027, China.}
}
\markboth{Journal of \LaTeX\ Class Files,~Vol.~14, No.~8, February~2024}%
{Shell \MakeLowercase{\textit{et al.}}: A Sample Article Using IEEEtran.cls for IEEE Journals}

% \IEEEpubid{0000--0000/00\$00.00~\copyright~2024 IEEE}

\maketitle

\begin{abstract}
Hyperspectral images (HSIs) play a crucial role in remote sensing but are often degraded by complex noise patterns. Ensuring the physical property of the denoised HSIs is vital for robust HSI denoising, giving  the rise of deep unfolding-based  methods. However, these methods map the optimization of a physical model to a learnable network with a predefined depth, which lacks convergence guarantees. In contrast, Deep Equilibrium (DEQ) models treat the hidden layers of deep networks as the solution to a fixed-point problem and models them as infinite-depth networks, naturally consistent with the optimization. Under the framework of DEQ, we propose a Deep Equilibrium Convolutional Sparse Coding (DECSC) framework that unifies local spatial-spectral correlations, nonlocal spatial self-similarities, and global spatial consistency for robust HSI denoising. Within the convolutional sparse coding (CSC) framework, we enforce shared 2D convolutional sparse representation to ensure global spatial consistency across bands, while unshared 3D convolutional sparse representation captures local spatial-spectral details. To further exploit nonlocal self-similarities, a transformer block is embedded after the 2D CSC. Additionally, a detail enhancement module is integrated with the 3D CSC to promote image detail preservation. We formulate the proximal gradient descent of the CSC model as a fixed-point problem and transform the iterative updates into a learnable network architecture within the framework of DEQ. Experimental results demonstrate that our DECSC method achieves superior denoising performance compared to state-of-the-art methods.

%Hyperspectral images (HSIs) play a crucial role in remote sensing but are often degraded by complex noise patterns. While convolutional sparse coding (CSC) methods have demonstrated success in other low-level image processing, they have not been fully explored for HSI denoising. In this paper, we propose a Deep Equilibrium Convolutional Sparse Coding (DECSC) framework that unifies local spatial-spectral correlations, nonlocal spatial self-similarities, and global spatial consistency for robust HSI denoising. Within the CSC framework, we enforce shared 2D convolutional sparse coding to ensure global spatial consistency across bands, while unshared 3D convolutional sparse coding captures local spatial-spectral details. To further exploit nonlocal self-similarities, a transformer block is embedded after the 2D CSC . Additionally, a detail enhancement module is integrated with the 3D  CSC to promote image detail preservation. We formulate the optimization as a fixed-point problem and transform the iterative updates into a learnable network architecture. Experimental results on benchmark datasets (ICVL and Houston 2018) demonstrate that our DECSC model outperforms state-of-the-art model-driven, data-driven, and hybrid methods, achieving superior denoising performance.
\end{abstract}

\begin{IEEEkeywords}
Hyperspectral image denoising, convolutional sparse coding, deep equilibrium model.
\end{IEEEkeywords}

\section{Introduction}
Hyperspectral image (HSI), with its rich spectral information, is widely used in remote sensing~\cite{cloutis1996review}, medical diagnostics~\cite{fei2019hyperspectral}, agriculture~\cite{kersting2012pre}, and image recognition~\cite{fauvel2012advances}.  Increasing spectral resolution comes at the cost of reducing the number of photons received in each channel. Combined with atmospheric interference, this inevitably introduces noise during the sensing process, which can degrade subsequent applications. Therefore, HSI denoising is a crucial preprocessing step to ensure image quality and enable reliable downstream applications.

The inherent spatial and spectral redundancy in HSIs allows clean HSIs to be represented as the  combinations of few elements from a dictionary, which has driven the success of sparse coding in HSI denoising.    Due to the high dimentionarity of reshaping the whole image into a vector, traditional sparse coding model typically requires partitioning the whole HSIs into multiple overlapping patches, thereby ignoring the shift-invariant properties of the data. As an alternative, convolutional sparse coding (CSC) employs a set of convolutional atoms to represent the image, naturally preserving the spatial relationships between pixels. Owing to this advantage, CSC has been widely applied to various inverse problems~\cite{gu2015convolutional,bao2019convolutional,hu2017convolutional}. In the context of HSI denoising, Xiong~\emph{et al.}~\cite{xiong2022spatial} and Yin~\emph{et al.}~\cite{yin2024deep} extended 2D CSC models to 3D ones by employing 3D convolutions to jointly  model local spatial-spectral correlations but fail to capture the global spectral correlations among bands. More recently, Tu~\emph{et al.}~\cite{Tu2024} enforced shared convolutional sparse coefficients across bands to capture inter-band spatial structural consistency. Nevertheless, this approach overlooks the local spatial-spectral correlations.

%Moreover, all the aforementioned models are based on hand-crafted sparse priors and lack the ability to learn data-driven priors, thereby limiting their representational capacity.

%In addition to prior-based methods, some approaches design network architectures based on iterative optimization algorithms such as ISTA~\cite{zhang2018ista}, half quadratic splitting (HQS)~\cite{aggarwal2018modl}, and alternating direction method of multipliers (ADMM)~\cite{yang2018admm}.

Additionally, the CSC model is based on the hand-crafted sparsity prior and cannot enjoy the data-driven learning from data.  To address this limitation, the deep unfolding technique have been used in~\cite{xiong2022spatial,Tu2024} to transform the iterative optimization process into a deep neural network with a fixed number of weight-tied layers, where each layer mimics one step of the original optimization. One major limitation of the  deep unfolding models is the hardware memory constraints: during training, all intermediate activations must be stored for backpropagation, restricting the model depth and complicating the training.  As a result, such unfolding networks are typically constrained to a small number of layers. However, the fixed number of iterations may not guarantee convergence, and running additional iterations at test time can lead to significant performance degradation~\cite{gilton2021deep,Monga2021}.

% mplicitly-defined infinite-depth networks for
% solving linear inverse problems in imaging. These empirical benefits complement convergence guarantees
% that are unavailable to widely-used deep unrolling methods.

%where the number of iterations is effectively
%infinite – a paradigm that has been beyond the reach of all previous deep unrolling architectures for solving
%inverse problems in imaging

% First, at deployment,
% these systems are optimized to compute image estimates quickly – a desirable property we wish to retain in
% developing new methods. Second, it is challenging to train deep unrolled networks for many iterations due to
% memory limitations of GPUs because the memory required to calculate the backpropagation updates scales
% linearly with the number of unrolled iterations.

In contrast, deep equilibrium models (DEQs)~\cite{bai2019deep} offer a paradigm shift by treating the hidden layers of deep networks as the solution to a fixed-point problem. Instead of unfolding the optimization process into predefined depth, DEQs model infinite-depth networks and directly solve for the equilibrium (fixed) point during the forward pass. This fixed-point formulation makes DEQs particularly suitable for tasks involving iterative refinement, such as those frequently encountered in convex and non-convex optimization for model-driven methods. Additionally, DEQs leverage existing numerical solvers and implicit differentiation for forward evaluation and backward propagation, significantly reducing memory usage and improving training stability. In a nutshell, DEQ offers a manner to transform the traditional model-driven methods into a learnable architecture    with convergence guarantees and preferable interpretability to bypass the limitation of hand-crafted priors.  In the literature,  Gilton~\emph{et al.}~\cite{gilton2021deep} demonstrated the improved effectiveness of DEQs in a variety of  image reconstruction tasks over deep unfolding networks. Zhao~\emph{et al.}\cite{zhao2023deep} extended the DEQ framework to recurrent neural networks (RNNs) and and Plug-and-Play (PnP) algorithms for snapshot compressive imaging. More recently, Geng~\emph{et al.}\cite{geng2023one} applied DEQ to accelerate sampling in diffusion models.

Building upon the strengths of CSC and DEQ, we propose a novel DECSC framework for HSI denoising. To fully exploit both local spatial-spectral and global spectral correlations inherent in HSIs, we decompose the hyperspectral data into two components: (1) global inter-band common (GIC) structures and (2) local spatial-spectral unique (LSU) structures. For modeling the GIC structure, we enforce shared convolutional sparse coefficients across bands, thereby ensuring spatial consistency throughout the spectral dimension. For local structures, we employ 3D convolution to jointly capture fine-grained spatial-spectral dependencies, preserving spatial-spectral details within nearby bands. To overcome the limitations of hand-crafted sparsity priors, we integrate a transformer module to capture the nonlocal spatial self-similarity within the GIC component. Additionally, we introduce a detail enhancement module to promote detail preservation within the LSU component. Within the DEQ framework, we reformulate the estimation of convolutional sparse coefficients as a fixed-point problem, resulting in an implicit infinite-depth learnable network that not only integrates the physical interpretability of the model but also offers guaranteed convergence properties. Extensive experiments on both synthetic and real-world datasets validate the superior denoising performance of the proposed DECSC method.

The remainder of this paper is organized as follows. Section~\ref{related_work} reviews recent advances in HSI denoising and DEQ. Section~\ref{method} details the proposed DECSC, including its network architecture and optimization under the DEQ framework. Section~\ref{experiment} presents extensive experimental results validating our approach. Finally, Section~\ref{conc} concludes the paper.
% In summary, the contributions of this paper are as follows:

% \begin{itemize}
%         \item We introduce a novel approach by modeling HSI as a combination of inter-band common structures and spatial-spectral correlations, and design a network layer based on ISTA for the iterative updating of these components.
%         \item We propose implementations of the model under both the deep unfolding and DEQ frameworks.
%         \item Extensive experiments on synthetic and real noise removal tasks demonstrate that the proposed model outperforms state-of-the-art methods in both objective metrics and subjective visual quality.
% \end{itemize}

\section{Related Work}\label{related_work}
This section presents a concise review of recent advancements in HSI denoising  and DEQ.

\subsection{HSI Denoising}

In recent years, the paradigm of HSI denoising has shifted from model-driven to data-driven methods, with hybrid-driven approaches emerging as a promising direction that integrates the advantages of both.
Defining an effective prior is crucial in traditional model-based methods. The strong spatial and spectral correlations in HSIs enable their representation in a low-dimensional subspace, underpinning the success of low-rankness and sparsity priors. Methods based on low-rank matrix or tensor decomposition~\cite{zheng2020double, He2022NGMeet} and on rank minimization~\cite{Zheng20203DlogTNN, xie2021adaptive, chen2022hyperspectral} denoise HSIs by extracting these low-rank structures. Meanwhile, the intrinsic redundancy of HSIs supports sparse representations over a few  atoms from a dictionary. Peng~\emph{et al.}~\cite{peng2014decomposable} reconstructed clean HSIs using either predefined or learned dictionaries, and Zhao~\emph{et al.}~\cite{Zhao2015Hyperspectral} employed sparse coding to capture both global spatial and local spectral redundancies. Furthermore, Zhuang~\emph{et al.}~\cite{zhuang2020hyperspectral} preserved "rare pixels" containing critical information through collaborative sparsity.  Total variation(TV) regularization--another form of transformed sparsity--has also proven effective to promote piecewise smoothness~\cite{wang2017hyperspectral,Peng2020E-3DTV}. Despite their interpretability and effectiveness, these model-driven approaches often demand extensive parameter tuning and entail high computational costs.

Data-driven methods utilize various deep nerual network (DNN) architectures to model the intrinsic structure of HSI from data. Convolutional neural networks (CNNs)~\cite{yuan2018hyperspectral,chang2018hsi,zhang2019hybrid}  can only model the local spatial-spectral correlation with  limited receptive fields, restricting their ability to capture long-range dependencies that always important in image processing. Transformers treat non-overlapping image patches as sequences and leverage attention mechanisms to model long-range dependencies~\cite{dosovitskiy2020image}. Compared to RGB images, HSIs possess significantly richer spectral resolution and exhibit unique spatial-spectral characteristics. Consequently, transformer-based methods for HSI denoising often emphasize the exploration of both spatial and spectral self-similarity. For instance, Peng~\emph{et al.}~\cite{peng2014decomposable} employed a dual-branch network combining CNNs and transformers to separately capture spectral correlations and  nonlocal spatial self-similarity. Li~\emph{et al.}~\cite{li2022spatialspectral} adopted non-local spatial self-attention and global spectral attention to fully exploit the inherent similarities along spatial and spectral dimensions. More recently, some approaches have combined self-attention with carefully designed low-rank modules to simultaneously capture spatial self-similarity and spectral low-rank structure~\cite{li2023spectral, li2024latent, tan2024low}. However, the quadratic complexity of transformers in handling long sequences leads to substantial computational overhead, limiting their scalability and practicality in HSI applications. To address this, recent works have explored  Mamba~\cite{Gu2023}, a state space model architecture, for HSI denoising~\cite{fu2024ssumamba}. Data-driven nerual networks  often require extensive empirical tuning and trial-and-error architecture design and  can not share the well-established interpretability of model-based methods.

Hybrid-driven methods have emerged as a promising research direction by combining the interpretability of model-driven approaches with the strong representation capabilities of data-driven techniques. To incorporate low-rank priors in HSIs, several studies~\cite{Xiong2022MAC-Net,zhang2023combined,Chen2024,zhuang2023eigenimage2eigenimage} have projected HSIs into low-dimensional spectral subspaces to facilitate efficient restoration. Similarly, sparsity priors have inspired the design of network architectures based on deep unfolding~\cite{xiong2022spatial,Zeng2025,bodrito2021T3SC,xiong2022smds,li2023pixel}, where the network structure is derived by unfolding the optimization process of a model. For instance, Xiong~\emph{et al.}~\cite{Xiong2023} converted the iterative soft-shrinkage algorithm of a multitask sparse representation model into an explainable network for enhanced HSI denoising. In~\cite{Li2023}, a total variation-based denoising model was unfolded into a learnable network. By further integrating a statistical feature injection module and a multiscale degradation guidance module, the method effectively recovers real structural details. Despite these advantages, deep unfolding networks often suffer from high memory consumption,  instability and umerical issues arising in backpropagation, especially as the number of iterations increases, negatively impacting the  reconstruction performance.

\subsection{Deep Equilibrium Model}

Traditional neural networks typically enhance their representational capacity by increasing the number of explicit layers. However, this strategy often leads to increased memory consumption and computational overhead. Recent research has shown that comparable or even superior performance can be achieved by sharing weights across all layers~\cite{bai2018trellis,dehghani2018universal}. Motivated by this insight, Bai~\emph{et al.}~\cite{bai2019deep} introduced the DEQ, which abandons the conventional notion of a finite stack of layers. Instead, DEQ defines an implicit infinite-depth network by formulating a set of analytical conditions whose solution corresponds to the network's output. Rather than unrolling layers, DEQ directly solves for the equilibrium point where the hidden representation becomes stable. This  point can be found using efficient black-box root-finding methods such as Broyden's method or Anderson acceleration. Crucially, DEQ leverages implicit differentiation to compute gradients during backpropagation without storing intermediate activations, enabling memory usage to remain constant at $\mathcal{O}(1)$. To enhance training stability, several follow-up studies have introduced techniques such as convergence-enforcing layers~\cite{winston2020monotone}, Jacobian regularization~\cite{bai2021stabilizing}, and phantom gradients~\cite{geng2021training}. Notably, the iterative optimization process in model-based methods naturally aligns with the equilibrium-seeking principle of DEQ, making it well-suited for tasks in imaging~\cite{gilton2021deep,zhao2023deep} and denoising~\cite{gkillas2023connections}. Beyond optimization, the multiscale fusion can also be framed within an equilibrium formulation. For instance, Bai~\emph{et al.}~\cite{bai2020multiscale} proposed the multiscale deep equilibrium model (MDEQ), where inputs are injected at the highest resolution and propagated implicitly across scales to satisfy a joint equilibrium condition. In summary, the DEQ model provides an attractive manner to build the connection between the fixed-point optimization and learning-based techniques to enjoy both benefits. For this end, we adopt the DEQ model to transform the optimization of the CSC model for HSI denoising.

\section{Method}\label{method}

In this section, we first define the problem and outline the motivation behind our approach. We then detail our DECSC, including its network layer design, forward and backward strategy. The main notations used in this section are listed in Table~\ref{tab:notations}.

\begin{table}
    \caption{List of main notations used in this paper.}\label{tab:notations}
    % \vspace{-0.3cm}
    \centering
    \begin{tabular}{c|c}
    \Xhline{1.2pt}  		
    Notations &  Description\\
    \Xhline{1.2pt}  		
    $\X$ &  the clean HSI\\
    \hline\
    $\Y$ &  the noisy HSI\\
    \hline\
    $\N$ & the additive noise\\
    \hline\
    $\C$ & the GIC tructures\\
    \hline\
    $\U$ &  the LSU tructures\\
    \hline\
    $\K_b$ & the 2D convolutional dictionary for the b-th band of the HSI \\
    \hline\
    $\S$ &  the corresponding sparse coefficients of $\C$ \\
    \hline\
    $\D$ & the 3D convolutional dictionary of $\U$ \\
    \hline\
    $\H$ &  the corresponding sparse coefficients of $\U$ \\
    \hline\
    $\alpha^{t}$ &  the result of the t-th iteration of the DEQ \\
    \hline\
    $\alpha^{*}$ &  the equilibrium point of DEQ \\
    \hline\
    $\z$ &  the noisy image injected into each layer of DEQ \\
    \Xhline{1.2pt}      
\end{tabular}
            %  \vspace{-0.1cm}
\end{table}

\subsection{Convolutional Sparse Coding Model for HSI Denoising}

Let $\Y \in \mathbb{R}^{H \times W \times B}$ be an HSI with  $H \times W$ pixels and $B$  bands. Typically, the observed noisy HSI $\Y$ is modeled as:

\begin{equation} \Y = \X + \N \label{eq:noise} \end{equation} where $\X \in \mathbb{R}^{H \times W \times B}$ is the clean HSI to be recovered, and $\N \in \mathbb{R}^{H \times W \times B}$ represents the additive noise.

Unlike conventional RGB images, HSIs capture the spectral reflectance of each pixel across contiguous wavelengths. Accurately modeling these spatial–spectral structures is critical for effective HSI denoising. While spectral reflectance varies with wavelength, the spatial distribution of scene objects remains largely consistent across bands, resulting in globally shared spatial structures. To preserve such consistency, our model learns different filters (dictionaries) for each band but enforces them to share the same sparse representation coefficients, encouraging the extraction of coherent global patterns.
In addition to these commonalities, HSIs also exhibit local spatial–spectral variations unique to individual bands, which encode fine details. To capture these, we employ 3D filters with unshared weights, allowing the model to extract band-specific sparse representations that retain subtle inter-band differences.
Motivated by these observations, we decompose the clean HSI into two complementary components---global inter-band common (GIC) structures $\C$ and local  spatial-spectral unique (LSU) structures $\U$:

\begin{equation}
\X = \C + \U.
\label{eq:decomposing}
\end{equation}

We leverage the success of sparse representations to capture $\C$ and $\U$. Specifically, each band of $\C$ is modeled as a sparse combination of convolutional atoms from a band-specific dictionary, while enforcing shared sparse coefficients across all bands to reflect common spatial structures:
\begin{equation}
\begin{split}
\C_1 = \K_1 \star \S =& \sum_{m=1}^{M} \k_{1,m} \star \S_m,\\
&\vdots\\
\C_b = \K_b \star \S =& \sum_{m=1}^{M} \k_{b,m} \star \S_m,\\
&\vdots\\
\C_B = \K_B \star \S =& \sum_{m=1}^{M} \k_{B,m} \star \S_m,\\
\end{split}
\label{eq:common_csc_v1}
\end{equation}
where $\star$ denotes the 2D convolution operator, $\K_b = \{\k_{m}\}_{m=1}^M$ is the 2D dictionary for band $b$, and $\S = \{\S_m\}_{m=1}^M$ represents the shared sparse coding.

For the local unique spatial-spectral  component $\U$, we employ a 3D convolutional sparse coding model:

\begin{equation}
\U = \D \star \H = \sum_{j=1}^{J} \d_j \star \h_j,
\label{eq:ss_csc}
\end{equation} where $\D = \{\d_j\}_{j=1}^J$ is the 3D convolutional dictionary and $\H = \{\h_j\}_{j=1}^J$ denotes the corresponding sparse coefficients.

Given Eqs.~(3) and~(4) , our goal is to jointly estimate $\C$ and $\U$ by exploiting their respective sparsity priors. The overall optimization problem is formulated as:

\begin{equation}
\min_{\S, \H}  \frac{1}{2} \left\| \Y - \K \otimes \S - \D \star \H \right\|_F^2 + \lambda_1 \|\S\|_1 + \lambda_2 \|\H\|_1,
\label{eq:opt_problem}
\end{equation}
where $\K = \text{concat}(\K_1, \dots, \K_B)$, $\otimes$ denotes depth-wise convolution across spectral bands, and $\lambda_1$, $\lambda_2$ are regularization parameters controlling the sparsity level.

In addition to global and local structures, HSIs exhibit nonlocal spatial self-similarity, arising from repetitive spatial patterns across the image. To effectively leverage this characteristic, we incorporate a data-driven regularization term $\mathcal{R}(\S)$. Furthermore, the local spatial-spectral correlation primarily captures fine-grained image details, which are crucial for accurate reconstruction. To better exploit these details, we introduce another data-driven regularization term $\mathcal{R}(\H)$.  The resulting objective function is therefore formulated as:

\begin{equation}
\begin{split}
	\min_{\S, \H} &\frac{1}{2} \left\| \Y - \K \otimes \S - \D \star \H \right\|_F^2 + \lambda_1 \|\S\|_1 \\
&+ \lambda_2 \|\H\|_1 + \mu_1 \mathcal{R}(\S) + \mu_2 \mathcal{R}(\H),
\end{split}
\label{eq:opt_problem2}
\end{equation}
where $\mu_1$ and $\mu_2$ balance the contributions of the data-driven regularizations.

To solve Eq.~(\ref{eq:opt_problem2}), we adopt an alternating optimization strategy that updates $\S$ and $\H$ iteratively:
\begin{align}
\min_{\S}  & \frac{1}{2} \left\| \Y - \K \otimes \S - \D \star \H \right\|_F^2 + \lambda_1 \|\S\|_1 + \mu_1 \mathcal{R}(\S),\\
\min_{\H}  & \frac{1}{2} \left\| \Y - \K \otimes \S - \D \star \H \right\|_F^2 + \lambda_2 \|\H\|_1 + \mu_2 \mathcal{R}(\H).
\end{align}
Each subproblem is a constrained sparse coding task and is solved via proximal gradient descent. The iterative updates at the $(t+1)$-th step are given by:

\begin{equation}
\begin{split}
\!\!\S^{(t+1)}\!\!&=\!\!\text{Net}_1\!\!\left(\text{Soft}_{\theta_1}\!\left(\S^{(t)}\!\!+\!\!\K^{T} \otimes \left(\Y\!-\!\!\K\otimes\S^{(t)}\!-\!\!\D \star \H^{(t)} \right) \right) \right), \\
\!\!\H^{(t+1)}\!\!&=\!\!\text{Net}_2\!\!\left(\text{Soft}_{\theta_2} \!\left(\H^{(t)}\!+\!\D^{T} \star \left(\Y\!-\!\!\K\otimes \S^{(t+1)} - \!\!\D\star\H^{(t)}\right)\right)\right)\label{eq:updatesh}
\end{split}
\end{equation} where $(\cdot)^T$ denotes the transposed convolution, and $\text{Soft}_{\theta}(\cdot)$ is the soft-thresholding function:

\begin{equation}
\text{Soft}_{\theta}(x)=\text{sign}(x)\cdot\max(|x|-\theta, 0)	
\end{equation}
The modules $\text{Net}_1$ and $\text{Net}_2$ serve as learned regularizers, corresponding to the constraints $\mathcal{R}(\S)$ and $\mathcal{R}(\H)$, considering that the neural network can serve as a proximal operator.
Once $\S$ and $\H$ are inferred, the clean HSI is reconstructed as:
\begin{equation}
\widehat{\X} = \K \otimes \S + \D \star \H.
\label{eq:estimate}
\end{equation}

%Inspired by the success of deep unfolding, we unroll the iterative optimization process into a learnable neural network. Each stage of the network corresponds to one iteration of a proximal gradient descent step for both $\S$ and $\H$. Specifically, for the $t$-th iteration, the updates are formulated as:
%
%\begin{align} \S^{(t+1)} &= \text{SoftThresh}\left(\S^{(t)} - \eta_1 \nabla_{\S} \mathcal{L}, \theta_1 \right) \\
%     \H^{(t+1)} &= \text{SoftThresh}\left(\H^{(t)} - \eta_2 \nabla_{\H} \mathcal{L}, \theta_2 \right), \end{align}
%
%where $\eta_1$, $\eta_2$ are learnable step sizes, $\theta_1$, $\theta_2$ are learnable thresholds, and $\mathcal{L}$ is the reconstruction loss. The dictionaries $\K_b$ and $\D$ can also be learned end-to-end.

\subsection{Deep Equilibrium Convolutional Sparse Coding Layer}
\subsubsection{Weight-tied Convolutional Sparse Coding Layers}

%By introducing a set of learnable parameters $\Theta = \{\theta_1, \theta_2, \W_{K}, \W_H, \D, \K\}$, where $\W_K = \K^{T}$ and $\W_H = \D^{T}$ represent the transposed convolutional dictionaries, the iterative updates in Eq.~(\ref{eq:updatesh}) can be reformulated into two alternating CSC layers as follows:
%
%\begin{equation}
%    \begin{split}
%    \!\!\S^{(t+1)}\!\!&= \text{Net}_1\left( \text{Soft}_{\theta_1} \left( \S^{(t)} \!\!+\!\W_K \otimes \left( \Y\!\! - \K \otimes \S^{(t)}\!\!- \D \star \H^{(t)} \right) \right) \right), \\
%    \!\!\H^{(t+1)}\!\!&= \text{Net}_2\left( \text{Soft}_{\theta_2} \left( \H^{(t)}\!\! +\!\W_H \star \left(\Y\!\! - \K \otimes \S^{(t+1)}\!\!- \D \star \H^{(t)} \right) \right) \right),
%    \end{split}
%\end{equation}

In fact, both update steps share a common structural form that seeks the fixed-point solution of $\alpha$:
\begin{equation}
\alpha^{(t+1)} = \text{Net}\left(\text{Soft}_{\theta} \left( \alpha^{(t)} + \E^{T} \star \left( \z - \E \star \alpha^{(t)} \right) \right)\right)
\label{eq:deq}
\end{equation}
where z represents the noisy image injected into each layer. By introducing a set of parameters $\Theta=\{\theta, \W_E, \E\}$ with $\W_E=\E^{T}$, Eq.~(\ref{eq:deq}) can be transformed into a learnable network architecture $\alpha^{(t+1)} = f_{\Theta}(\alpha^{(t)}, \z)$ defined as:
   \begin{equation}
    f_{\Theta}(\alpha^{(t)}, \z)= \text{Net}\left(\text{Soft}_{\theta}\left( \alpha^{(t)} + \W_E \star \left( \z - \E \star \alpha^{(t)} \right) \right)\right) \label{eq:deq2}
\end{equation}
Eq.~(\ref{eq:deq2}) is a weight-tied architecture with residual  connections from the input to each layer until convergence.  In other words, each layer incrementally refines the output by building upon the updates from the previous iteration. As the depth increases, the magnitude of these updates gradually diminishes, leading to a regime of diminishing returns, where additional layers contribute progressively less until the network reaches a stable equilibrium.  As pointed in~\cite{bai2019deep},  such a design offers several advantages. First, weight sharing serves as a form of implicit regularization that stabilizes training and promotes generalization. Second, it significantly reduces the number of trainable parameters, resulting in more compact models. Third, unlike deep unfolding models that require a pre-defined, fixed number of layers, this iterative structure can theoretically be unrolled to arbitrary depth, aligning naturally with the principles of fixed-point optimization. This architecture seamlessly integrates with numerical solvers to reach a stable equilibrium and therefore has  convergence guarantee.

\begin{figure}[!h]
    \centering
    \includegraphics[width=1\linewidth]{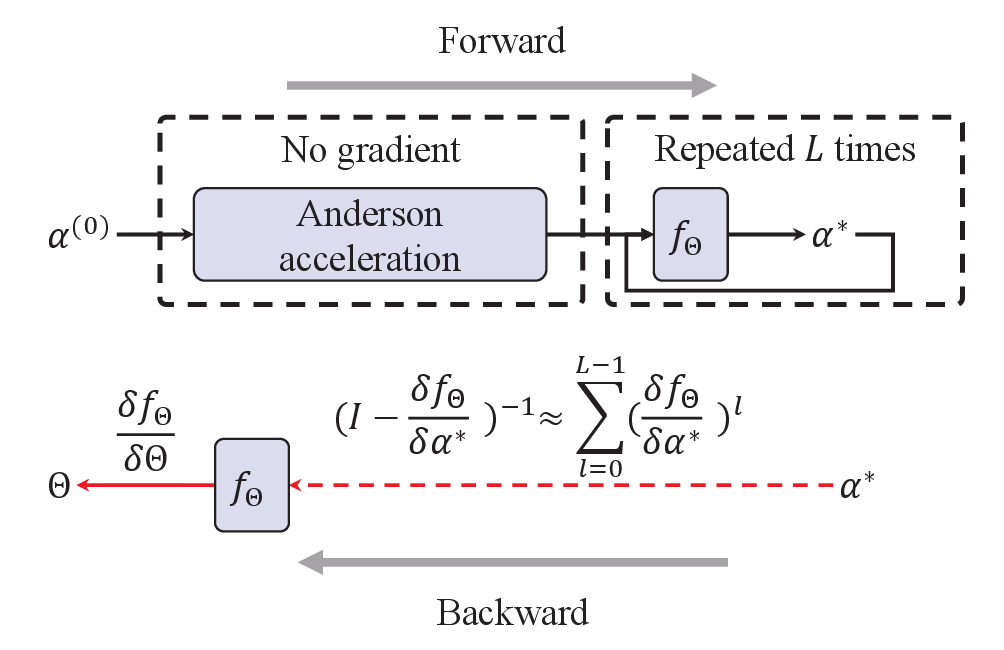}
    \caption{DEQ-based inference of the proposed model. Forward pass uses Anderson acceleration; backward pass uses an unrolling-based phantom gradient.}\label{fig:gradient}
\end{figure}

\begin{figure*}[!ht]
    \centering
    \includegraphics[width=1\linewidth]{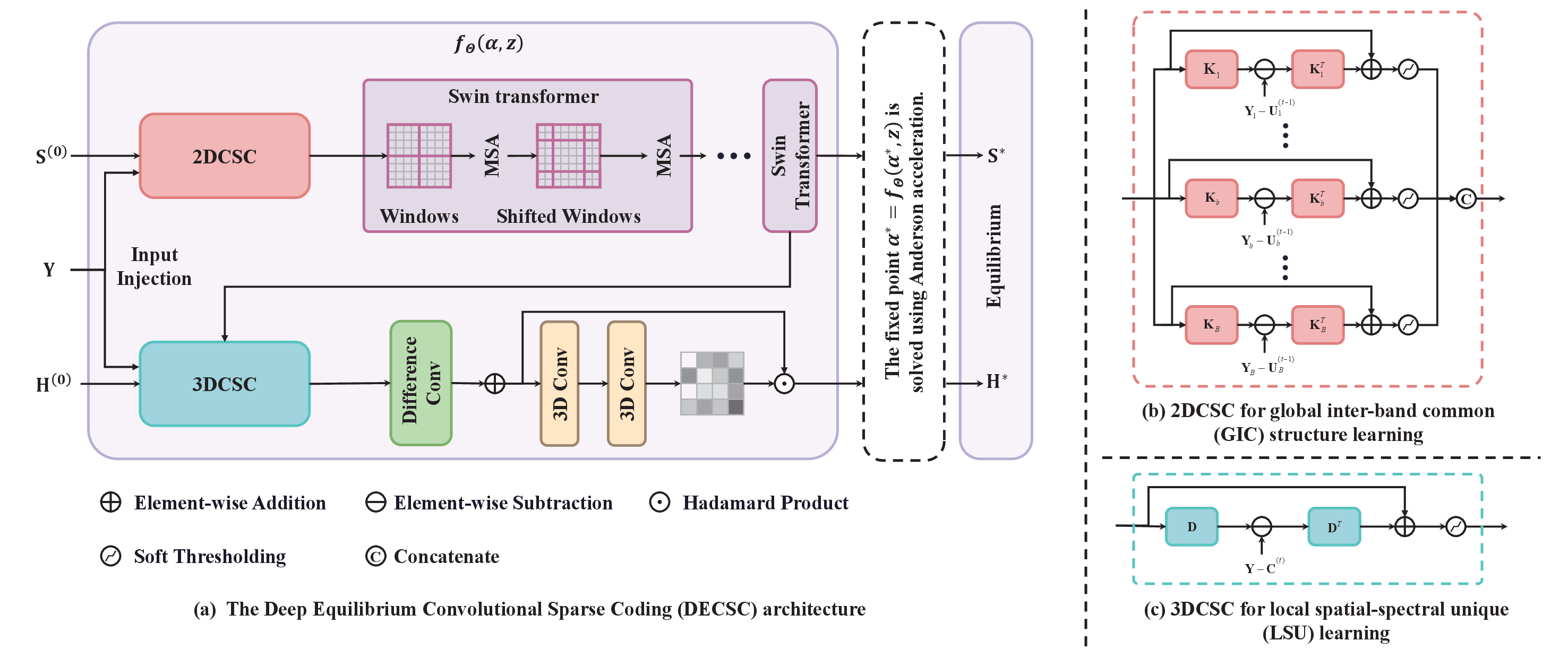} % Reduce the figure size so that it is slightly narrower than the column. Don't use precise values for figure width.This setup will avoid overfull boxes.
    \caption{Illustration of the DECSC architecture. The network layer $f_{\Theta}$ integrates the ISTA backbone, a Swin Transformer for capturing non-local dependencies, and a difference convolution module with an attention mechanism for enhancing fine details. The fixed point is directly solved using Anderson acceleration}
    \label{DECSC}
\end{figure*}

% \subsubsection{Alternating Update Algorithm}

% Given a known dictionary, the clean HSI can be estimated by solving the following regularized optimization problem, which jointly models inter-band common structures and spatial-spectral correlations:
% \begin{equation}
%     \mathop{\arg\min}\limits_{\{ r,h \}} \frac{1}{2} \Vert \Y- \C -\S \Vert_2^2 \\ +\lambda_1 \sum_{i=1}^{I}{{\Vert r_{i}\Vert}_1}+\lambda_2 \sum_{j=1}^{J}{{\Vert h_{j}\Vert}_1}, \label{eq:opt}
% \end{equation} with
% \begin{equation}
%     \begin{alignedat}{2}
%         &\C = \left [k_{1}*r \; , \; \cdots \; ,\; k_{B}*r \right ], \\
%         &\S =d*h, \\
%     \end{alignedat}
% \end{equation} where $\lambda_1$ and $\lambda_2$ are regularization parameters that control the sparsity constraints, and the square brackets $\left [ \cdot \right ]$ denotes the concatenation operation.

\subsubsection{Forward and Backward Pass}

Fig.~\ref{fig:gradient} illustrates the processing flow of the model under the DEQ framework. As aforementioned, the output of a DEQ is the equilibrium point $\alpha^{*}$ where the predefined condition is met:
\begin{equation}
     \alpha^{*}= f_{\Theta}\left ( \alpha^*, \z \right ) \label{eq:dcsc_deq}
\end{equation}
A naive way is to obtain $\alpha^{*}$ by iteratively running Eq.~(\ref{eq:deq}) till convergence, which is time-consuming.  Instead, the equilibrium point can be be faster obtained with any  black-box root-finding algorithm. In our implementation, we use the Anderson acceleration procedure which uses the past updates to identify promising directions to move during the current update to find the  equilibrium point:
\begin{equation}
    \alpha^{(t+1)}=  (1-\beta)\sum \limits_{i=0}^{m} \gamma_i^{(t)} \alpha^{(t-i)}+\beta \sum \limits_{i=0}^{m}\gamma_i^{(t)} f_{\Theta}(\alpha^{(t-i)}, \z)
\end{equation} for the mixing parameter $\beta>0$. Defining  the residual $g_{\Theta}(\alpha^{(t)}, \z)=f_{\Theta}(\alpha^{(t)}, \z)-\alpha^{(t)}$,  $\gamma$ is calculated as
\begin{equation}
    \arg \min_{\gamma}\|\G\gamma\|_2^2, \quad \text{s.t.} \: \sum  \limits_{i=0}^{m} \gamma_i=1
\end{equation}
where $\G=[g_{\Theta}(\alpha^{(t)}, \z), \cdots, g_{\Theta}(\alpha^{(t-m+1)}, \z)] $ is a matrix containing the $m$ past residuals.

Since the forward pass of DEQ does not rely on explicit iterations, gradient computation cannot be automatically performed with Pytorch or Tensorflow, during the backward propagation process.  According to the derivation by Bai~\emph{et al.}~\cite{bai2019deep}, the gradient with respect to  the network parameters  be learned with constant memory using implicit differentiation and is calculated as
\begin{equation}
     \frac{\partial \alpha^*}{\partial \Theta } = \left ( \mathbf{I} - \frac{\partial f_{\Theta}(\alpha^*, \z)}{\partial \alpha^*} \right )^{-1} \frac{\partial f_{\Theta}(\alpha^*, \z)}{\partial \Theta}. \label{eq:deq_dif}
\end{equation}
% Given the loss $\mathcal{L}_{\theta}$, the gradient with respect to the parameters $\theta$ is:
% \begin{equation}
%     \frac{\delta \mathcal{L}_{\theta}}{\delta \theta} = \frac{\delta \mathcal{L}_{\theta}}{\delta A^*}\frac{\delta A^*}{\delta \theta} = \frac{\delta \mathcal{L}_{\theta}}{\delta A^*} \left ( I - \frac{\delta f_{\theta}}{\delta A^*} \right )^{-1} \frac{\delta f_{\theta}}{\delta \theta}. \label{eq:deq_dif}
% \end{equation}
Since the exact computation of the Jacobian-inverse $\left ( \mathbf{I} - \frac{\partial f_{\Theta}(\alpha^*, \z)}{\partial \alpha^*} \right )^{-1}$ requires high computational costs, we approximates it using phantom gradient~\cite{geng2021training}. Specifically, consider the Neumann series expansion of the Jacobian-inverse:
% Considering the computational time and the instability that arises during training of the DEQ, we do not aim to precisely solve its gradient. Instead, we approximate the exact gradient using phantom gradient~\cite{geng2021training} to reduce computational complexity and improve model efficiency.
\begin{equation}
    \mathbf{I}  + \frac{\partial f_{\Theta}}{\partial \alpha^*} + \left ( \frac{\partial f_{\Theta}}{\partial \alpha^*} \right )^{2} + \left ( \frac{\partial f_{\Theta}}{\partial \alpha^*} \right )^{3} \cdots \label{eq:neumann_series}
\end{equation}
The gradient approximated using a L-term Neumann series is equivalent to the differentiation of unrolling $\alpha^* = f_{\Theta}\left ( \alpha^*, \z \right )$ for $L$ steps:
\begin{equation}
    \frac{\partial \alpha^*}{\partial \Theta} = \sum_{l=0}^{L-1} \left ( \frac{\partial f_{\Theta}}{\partial \alpha^*} \right )^l \frac{\partial f_{\Theta}}{\partial \Theta} \approx \left ( \mathbf{I}  - \frac{\partial f_{\Theta}}{\partial \alpha^*} \right )^{-1} \frac{\partial f_{\Theta}}{\partial \Theta} . \label{eq:neumann_series_dif}
\end{equation}
As shown in Fig.~\ref{fig:gradient}, we therby adopt the unrolling-based phantom gradient to approximate the gradient  to obtain an exact solution.

\subsection{Network  Implementation}

\subsubsection{The Overall Architecture}

Fig.~\ref{DECSC} illustrates the architecture and implementation details of the network layer, showing how the components of $\S^{(t)}$ and $\H^{(t)}$ are updated in the proposed framework. With the deep equilibrium CSC layers introduced in Section III-B, the iterative updates in Eq.~(\ref{eq:updatesh}) can be reformulated as two alternating layers, i,e., GIC layer and LSU layer:

\begin{equation}
\begin{split}
\S^{(t+1)} \!\!&= \!\!\text{Net}_1\left(\!\text{Soft}_{\theta_1} \left( \S^{(t)}\!\! + \!\!\W_K \otimes \!\!\left( \Y\!\! -\!\! \K \otimes \S^{(t)}\!\! -\!\! \D \star \H^{(t)} \right) \right) \right), \\
\H^{(t+1)} \!\!&= \!\!\text{Net}_2\left(\!\text{Soft}_{\theta_2} \left( \H^{(t)}\!\! + \!\!\W_D \star \!\! \left( \Y\!\! -\!\! \K \otimes \S^{(t+1)}\!\! -\!\! \D \star \H^{(t)} \right) \right) \right),
\end{split}
\end{equation}
where $\{\W_K, \K \W_D, \D, \theta_1, \theta_2\}$ denote the set of learnable parameters, with the following relationships: $\W_K = \K^{T}$, $\W_D = \D^{T}$.

Furthermore, we explicitly define an image reconstruction layer connected to GIC and LSU layers. This layer takes the equilibrium representations $\S^*$ and $\H^*$ as inputs and reconstructs the clean HSI as:
\begin{equation}
\widehat{\X} = \K \otimes \S^* + \D \star \H^*.
\end{equation}

\subsubsection{The Swin Transformer Module}

Since $\S^{(t)}$ captures the globally shared spatial structure across bands, we introduce multiple stacked swin transformer blocks within the GIC component to efficiently capture its non-local properties, corresponding to the regularization term $\mathcal{R}(\S)$ in Eq.~(\ref{eq:opt_problem2}). The Swin transformer is adopted.   Specifically,  we divide $\S^{(t)}$ into multiple windows. For the tokens $\P$ within a window, the query, key, and value of the $i$-th head are computed by linear projections using parameters $\W^q_i$, $\W^k_i$ ,and $\W^v_i$, respectively. Afterwards, the attention output for the $i$-th head is computed as follows:
\begin{equation}
    \text{head}_i = \text{Softmax}\left( \frac{(\W^q_i\P)(\W^k_i\P)^{T}}{\sqrt{d_i}} \right) \W^v_i\P,
\end{equation} where $d_i$ representing the feature dimension of the $i$-th head. The outputs from all heads are concatenated and further projected to obtain the final result:
\begin{equation}
    \text{MSA}(\mathbf{P}) = \text{Concat}(\text{head}_1, \ldots, \text{head}_h) \W^o.
\end{equation} where $h$ is the number of heads and $W^o$ is the projection matrix.

\subsubsection{The Detail Enhancement Module}

\begin{figure}
    \centering
    \includegraphics[width=1\linewidth]{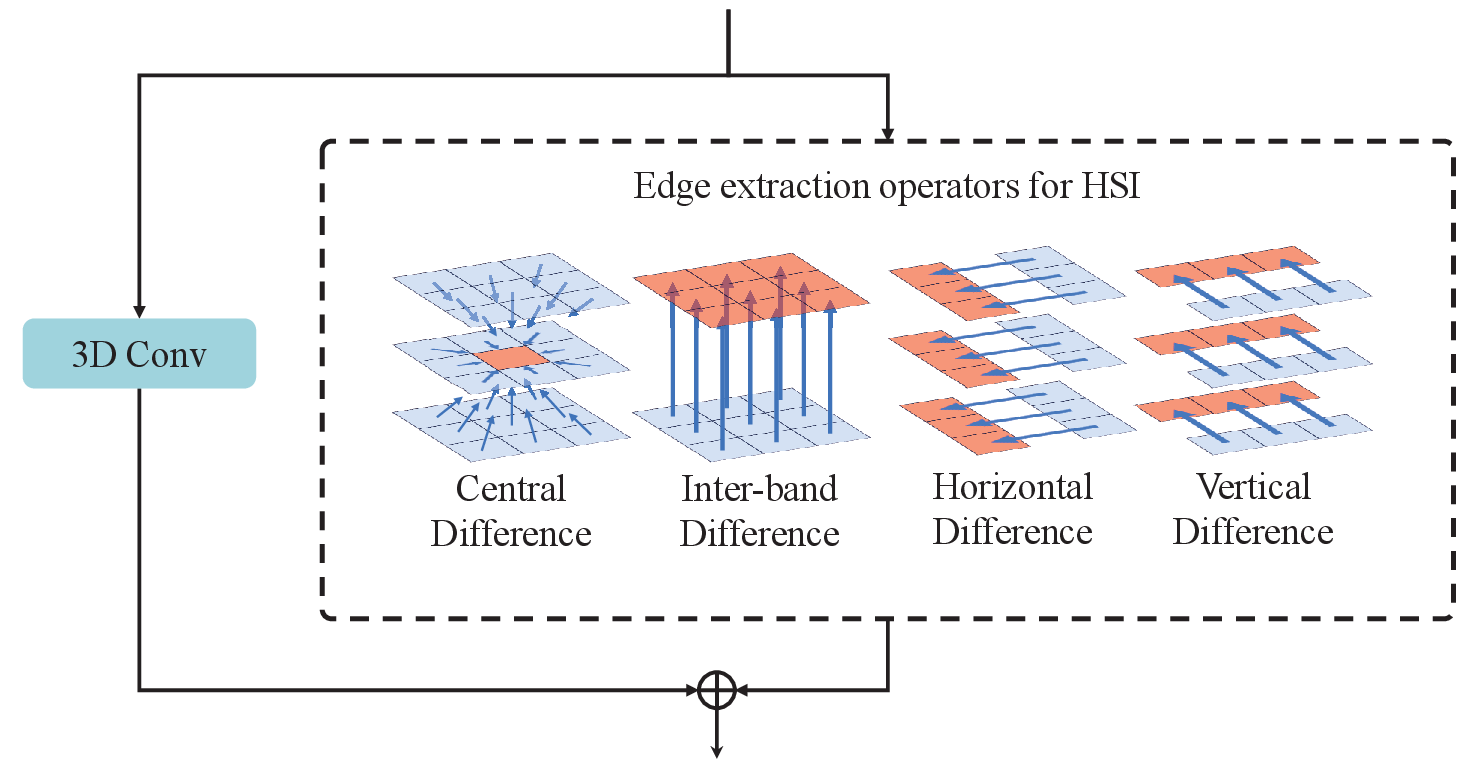}
    \caption{The difference convolution module consists of a 3D convolution and four HSI-specific edge extraction operators.}\label{fig:DConv}
\end{figure}

The detail enhancement module, composed of a difference convolution and a spatial attention, is designed to strengthen the preservation of fine image details in the LSU component and corresponds regularization term $\mathcal{R}(\H)$  in Eq.~(\ref{eq:opt_problem2}). It is formulated as:

\begin{equation}
\begin{split}
\S &= \S + \text{DConv}(\S), \\
\S &= \S + \S \bigodot \text{Conv}_1(\text{Conv}_1(\S)),
\end{split}
\end{equation}
where $\mathrm{Conv}_1$  is the $3\times 3 \times 3$ 3D convolution and $\text{DConv}$ denotes the difference convolution operator~\cite{yu2020searching}. Due to the fact that conventional CNNs are optimized from random initialization, they often struggle to effectively focus on extracting image edges and fine details during training. In contrast, traditional edge detection operators characterize abrupt intensity changes and fine structural features by leveraging differential information from the image. Therefore, $\text{DConv}$\cite{yu2020searching} combines conventional edge detection operators with CNNs, enabling more precise capture of image gradient information while fully exploiting the powerful representational capability of deep learning. As illustrated in Fig.\ref{fig:DConv}, we extend conventional edge detection operators to design a DConv module tailored for HSI. It consists of a standard 3D convolution for extracting intensity information, along with four edge extraction operators specifically designed for HSI to capture gradient information. In particular, the central difference operator enhances image sharpness, the inter-band difference operator captures variations along the spectral dimension, and the horizontal difference and vertical difference operators in the spatial domain extract horizontal and vertical edge information, respectively. This intermediate output is then passed through two cascaded 3D convolutional layers, which serve as an attention  that adaptively emphasizes regions rich in image detail information.

\subsubsection{Training Loss}
The loss function of our DEQCSC is  the Euclidean distance between the estimated clean HSI and the ground truth,
\begin{equation}
        \mathcal{L}_{\Theta}=\frac{1}{N}\sum_{i}^{N} {\parallel \widehat{\X}_i-\X_i\parallel}^2_F,
\end{equation}
where $\Theta$ represents the network parameters, $N$ is the total number of training samples, and $\parallel \cdot \parallel^2_F$ denotes the Frobenius norm.

\section{Experiment}\label{experiment}

In this section, we first present the experiment settings and implementation details, followed by an analysis of the results for synthetic noise and real-world noise removal experiments. Finally, we present a series of ablation studies.
\subsection{Experiment Settings and Implementation Details}

\subsubsection{Benchmarked Models}
The comparison involves 12 methods, comprising 7 model-driven methods, 3 data-driven methods, and 2 hybrid-driven methods. The model-driven methods include BM4D~\cite{Maggioni2013BM4D}, MTSNMF~\cite{Ye2015MTSNMF}, LLRT~\cite{Chang2017LLRT}, NGMeet~\cite{He2022NGMeet}, LRMR~\cite{Zhang2014LRMR}, E-3DTV~\cite{Peng2020E-3DTV}, and 3DlogTNN~\cite{Zheng20203DlogTNN}. The data-driven methods include SST~\cite{li2022spatialspectral}, TRQ3D~\cite{Pang2022TRQ3DNet} and SERT~\cite{li2023spectral}, while the hybrid-driven methods based on sparse priors include T3SC~\cite{bodrito2021T3SC} and MTSNN++~\cite{xiong2023multitask}.
\subsubsection{Metrics}
To quantitatively evaluate the performance of the models, we adopt standard metrics including Peak Signal-to-Noise Ratio (PSNR), Structural Similarity Index Measure (SSIM), and Spectral Angle Mapper (SAM), where higher PSNR and SSIM values, along with lower SAM values, reflect better denoising performance.

\subsubsection{Noise Patterns}
Non-independent and identically distributed (non-i.i.d.) Gaussian noise and mixture noise were considered in the synthetic experiment. In addition, we adopted the "noise with spectrally correlated variance" model proposed by Bodrito~\emph{et al.}~\cite{bodrito2021T3SC}. The detailed noise settings are described as follows:
\begin{itemize}
    \item \textbf{Non-i.i.d. Gaussian Noise:} Each band is contaminated with Gaussian noise, where the standard deviation $\sigma$ is uniformly sampled from  fixed intervals, i.e., [0, 15], [0, 55], and [0, 95].
    \item \textbf{Mixture Noise:} In addition to being contaminated by non-i.i.d. distributed Gaussian noise with intensity [0-95], every one-third of the bands is further degraded by the additional noise types: impulse noise with intensities ranging from 0.1 to 0.7, strip noise affecting 5\%-15\% of columns, and deadline noise.
    \item \textbf{Noise with Spectrally Correlated Variance:} Each band is contaminated with Gaussian noise whose standard deviation $\sigma$ varies continuously across bands following a Gaussian distribution. Specifically, for each band $i \in [0, B{-}1]$, $\sigma$ is defined as:
    \begin{equation}
            \sigma_i=\beta exp[-\frac{1}{4\eta^2}{(\frac{i}{c}-\frac{1}{2})}^2],    \label{eq:scv}
    \end{equation} where $\beta = 23.08$ and $ \eta = 0.157$ as in~\cite{bodrito2021T3SC}.
\end{itemize}
\begin{table*}[!t]
    \caption{Comparison of Different Methods on 50 Testing HSIs from ICVL Dataset. The Top Three Values Are Marked as \1{Red}, \2{Blue}, And \3{Green}.}\label{tab:icvl}
    % \vspace{-0.3cm}
    \centering
    \resizebox{\linewidth}{!}{
            \tablesize{
    \begin{tabular}{c|c|c|c|c|c|c|c|c|c|c|c|c|c|c|c}
            \Xhline{1.2pt}
    \multirow{3}*{\makebox[0.0588\textwidth][c]{$\sigma$}}&\multirow{3}*{\makebox[0.0588\textwidth][c]{Index}}&\multirow{3}*{\makebox[0.0588\textwidth][c]{Noisy}}
    &\multicolumn{7}{c|}{\textbf{Model-driven}}&\multicolumn{3}{c|}{\textbf{Data-driven}}&\multicolumn{3}{c}{\textbf{Hybrid-driven}}\\
    \cline{4-16}
    &&&\multirow{1}*{\makebox[0.0588\textwidth][c]{BM4D}}&\multirow{1}*{\makebox[0.0588\textwidth][c]{MTSNMF}}&\multirow{1}*{\makebox[0.0588\textwidth][c]{LLRT}}&\multirow{1}*{\makebox[0.0588\textwidth][c]{NGMeet}}&\multirow{1}*{\makebox[0.0588\textwidth][c]{LRMR}}&\multirow{1}*{\makebox[0.0588\textwidth][c]{E-3DTV}}&\multirow{1}*{\makebox[0.0588\textwidth][c]{3DlogTNN}}&\multirow{1}*{\makebox[0.0588\textwidth][c]{SST}}&\multirow{1}*{\makebox[0.0588\textwidth][c]{TRQ3D}}&\multirow{1}*{\makebox[0.0588\textwidth][c]{SERT}}&\multirow{1}*{\makebox[0.0588\textwidth][c]{T3SC}}&\multirow{1}*{\makebox[0.0588\textwidth][c]{MTSNN++}}&\makebox[0.063\textwidth][c]{\textbf{DECSC}}\\
    &&&\multirow{1}*{\makebox[0.0588\textwidth][c]{\cite{Maggioni2013BM4D}}}&\multirow{1}*{\makebox[0.0588\textwidth][c]{\cite{Ye2015MTSNMF}}}&\multirow{1}*{\makebox[0.0588\textwidth][c]{\cite{Chang2017LLRT}}}&\multirow{1}*{\makebox[0.0588\textwidth][c]{\cite{He2022NGMeet}}}&\multirow{1}*{\makebox[0.0588\textwidth][c]{\cite{Zhang2014LRMR}}}&\multirow{1}*{\makebox[0.0588\textwidth][c]{\cite{Peng2020E-3DTV}}}&\multirow{1}*{\makebox[0.0588\textwidth][c]{\cite{Zheng20203DlogTNN}}}&\multirow{1}*{\makebox[0.0588\textwidth][c]{\cite{li2022spatialspectral}}}&\multirow{1}*{\makebox[0.0588\textwidth][c]{\cite{Pang2022TRQ3DNet}}}&\multirow{1}*{\makebox[0.0588\textwidth][c]{\cite{li2023spectral}}}&\multirow{1}*{\makebox[0.0588\textwidth][c]{\cite{bodrito2021T3SC}}}&\multirow{1}*{\makebox[0.0588\textwidth][c]{\cite{xiong2023multitask}}}&\multirow{1}*{\makebox[0.07\textwidth][c]{\textbf{(Ours)}}}\\
    \Xhline{1.2pt}  		
    \multirow{3}*{\textbf{[0,15]}}
    & PSNR$\uparrow$  & 33.18 & 44.39 & 45.39 & 45.74 & 39.63 & 41.50 & 46.05 & 43.89      & \1{50.87} & 46.43 & \3{50.17} & 49.68 & 48.86 & \2{50.63} \\
    & SSIM$\uparrow$  & .6168 & .9683 & .9592 & .9657 & .8612 & .9356 & .9811 & .9902      & \2{.9938} & .9878 & \1{.9976} & .9912 & .9917 & \3{.9936} \\
    & SAM$\downarrow$ & .3368 & .0692 & .0845 & .0832 & .2144 & .1289 & .0560 & \1{.0150}  & .0298     & .0437 & \3{.0277} & .0486 & .0346 & \2{.0265} \\
    \hline 		
    \multirow{3}*{\textbf{[0,55]}}
    & PSNR$\uparrow$  & 21.72 & 37.63 & 38.02 & 36.80 & 31.53 & 31.50 & 40.20 & 33.37      & \2{46.39} & 44.64  & \3{46.33} & 45.15 & 43.88 & \1{46.92} \\
    & SSIM$\uparrow$  & .2339 & .9008 & .8586 & .8285 & .6785 & .6233 & .9505 & .6892      & \3{.9872} & .9840  & \1{.9950} & .9810 & .9794 & \2{.9876} \\
    & SAM$\downarrow$ & .7012 & .1397 & .2340 & .2316 & .4787 & .3583 & .0993 & .2766      & \3{.0457} & .0487  & \2{.0372} & .0652 & .0528 & \1{.0348} \\
    \hline\
    \multirow{3}*{\textbf{[0,95]}}\
    & PSNR$\uparrow$  & 17.43 & 34.71 & 34.81 & 31.89 & 27.62 & 27.00 & 37.80 & 24.53      & \2{44.83} & 43.54 & \3{44.47} & 43.10 & 42.15 & \1{45.64} \\
    & SSIM$\uparrow$  & .1540 & .8402 & .7997 & .6885 & .5363 & .4208 & .9279 & .4251      & \3{.9838} & .9806 & \1{.9929} & .9734 & .9720 & \2{.9848} \\
    & SAM$\downarrow$ & .8893 & .1906 & .3266 & .3444 & .6420 & .5142 & .1317 & .6087      & \3{.0513} & .0523 & \2{.0446} & .0747 & .0665 & \1{.0387} \\
    \Xhline{1.2pt}  		  \
    \multirow{3}*{\textbf{Mixture}}  \
    & PSNR$\uparrow$  & 13.21 & 23.36 & 27.55 & 18.23 & 23.61 & 23.10 & 34.90 & 17.52      & \3{39.22} & \2{39.73} & 39.13     & 34.09 & 38.90 & \1{42.67} \\
    & SSIM$\uparrow$  & .0841 & .4275 & .6743 & .1731 & .4448 & .3463 & .9041 & .2389      & \3{.9626} & .9491     & \2{.9678} & .9052 & .9531 & \1{.9756} \\
    & SAM$\downarrow$ & .9124 & .5476 & .5326 & .6873 & .6252 & .5144 & .1468 & .6905      & \2{.0743} & \3{.0869} & .0963     & .2340 & .0885 & \1{.0625} \\
    \Xhline{1.2pt}  		\
    \multirow{3}*{\textbf{Corr}}\
    & PSNR$\uparrow$  & 28.22 & 41.15 & 42.44 & 41.92 & 35.82 & 39.32 & 43.58 & 41.49      & \3{47.59} & 46.26 & \1{48.66} & 47.33     & 46.83 & \2{48.06} \\
    & SSIM$\uparrow$  & .4640 & .8963 & .9221 & .9080 & .7891 & .9081 & .9733 & .9709      & \2{.9904} & .9870 & \1{.9966} & .9858     & .9871 & \3{.9891} \\
    & SAM$\downarrow$ & .4601 & .1582 & .1121 & .1547 & .3113 & .1212 & .0601 & .0574      & \1{.0258} & .0403 & \3{.0351} & .0524     & .0399 & \2{.0298} \\
    \Xhline{1.2pt} 		
    \end{tabular}}}
            %  \vspace{-0.1cm}
\end{table*}

\begin{figure*}[!t]
    \centering
    \subfloat[]{\label{fig:Master5000K_clean_}\includegraphics[width=0.1240\linewidth]{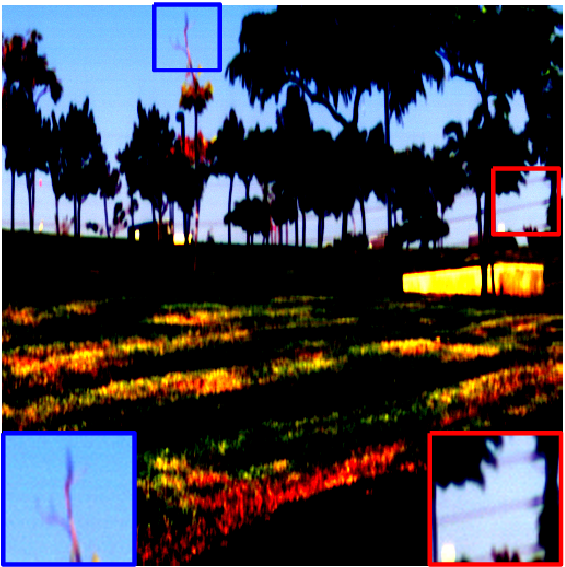}}
    \hspace{-1.1mm}
    \subfloat[]{\label{fig:Master5000K_noise_}\includegraphics[width=0.1240\linewidth]{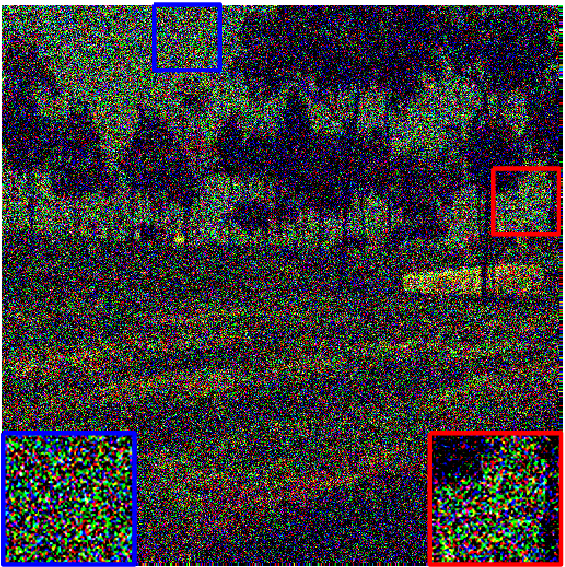}}
    \hspace{-1.1mm}
    \subfloat[]{\label{fig:Master5000K_BM4D}\includegraphics[width=0.1240\linewidth]{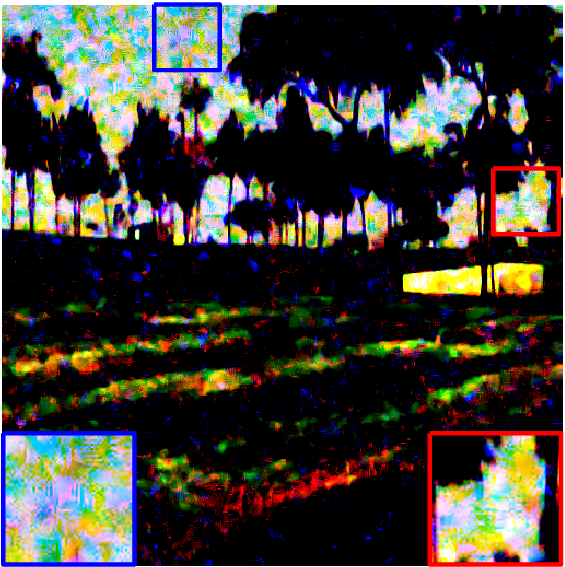}}
    \hspace{-1.1mm}
    \subfloat[]{\label{fig:Master5000K_MTSNMF}\includegraphics[width=0.1240\linewidth]{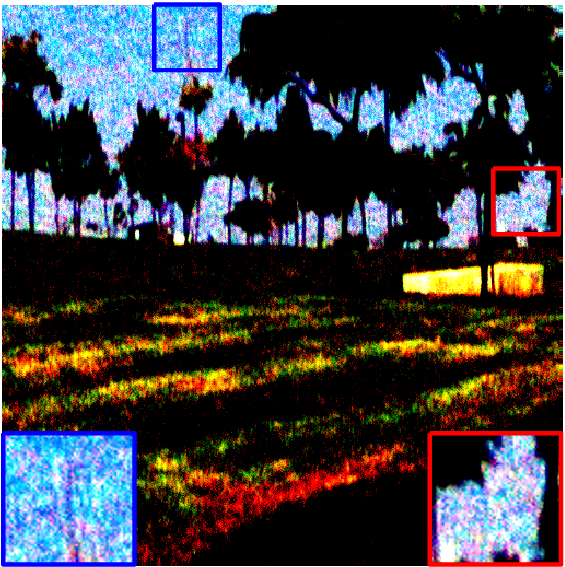}}
    \hspace{-1.1mm}
    \subfloat[]{\label{fig:Master5000K_LLRT}\includegraphics[width=0.1240\linewidth]{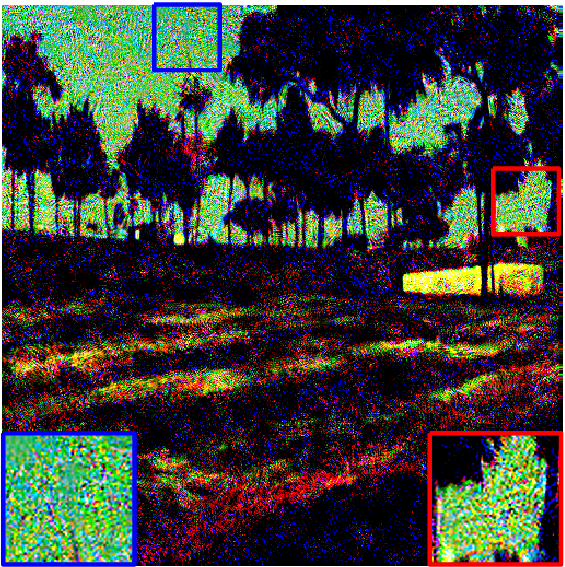}}
    \hspace{-1.1mm}
    \subfloat[]{\label{fig:Master5000K_NGMeet}\includegraphics[width=0.1240\linewidth]{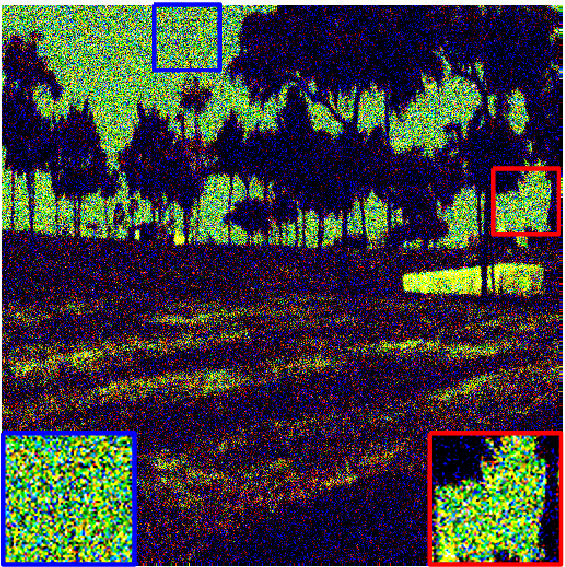}}
    \hspace{-1.1mm}
    \subfloat[]{\label{fig:Master5000K_LRMR}\includegraphics[width=0.1240\linewidth]{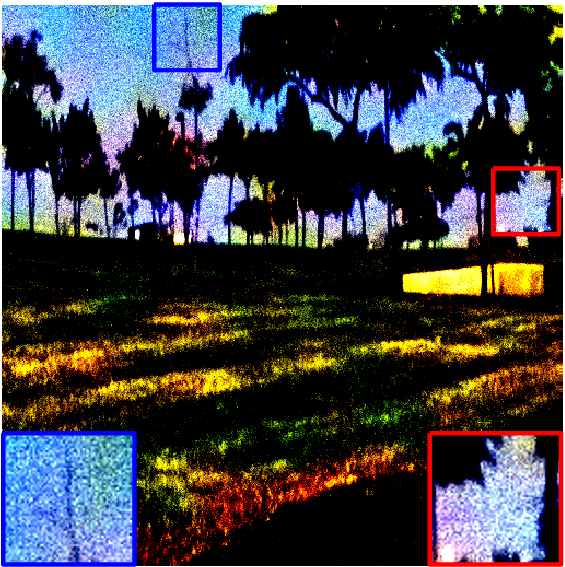}}
    \hspace{-1.1mm}
    \subfloat[]{\label{fig:Master5000K_E-3DTV}\includegraphics[width=0.1240\linewidth]{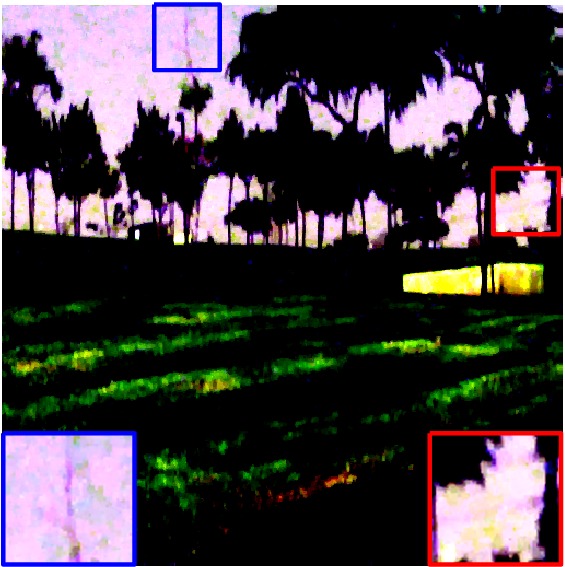}}
    \hspace{-1.1mm}
    \subfloat[]{\label{fig:Master5000K_3DlogTNN}\includegraphics[width=0.1240\linewidth]{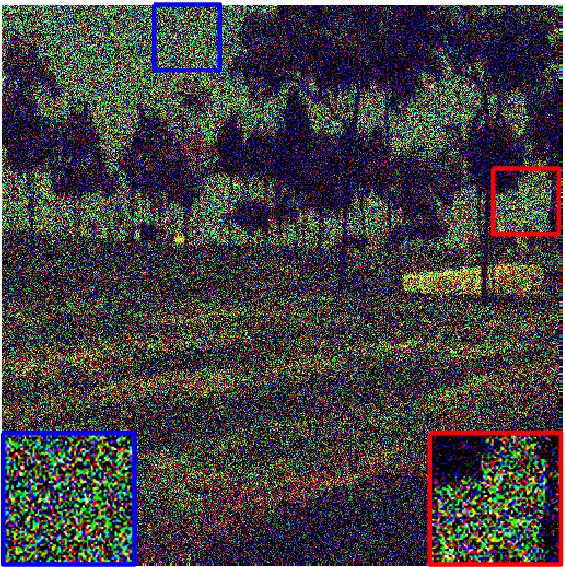}}
    \hspace{-1.1mm}
    \subfloat[]{\label{fig:Master5000K_SST}\includegraphics[width=0.1240\linewidth]{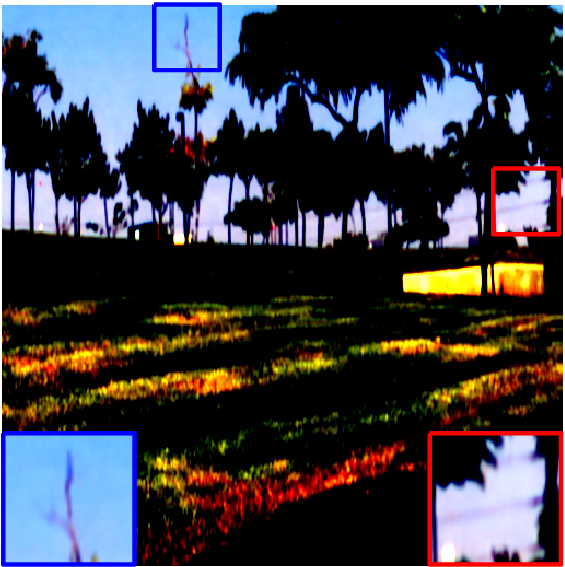}}
    \hspace{-1.1mm}
    \subfloat[]{\label{fig:Master5000K_TRQ3D}\includegraphics[width=0.1240\linewidth]{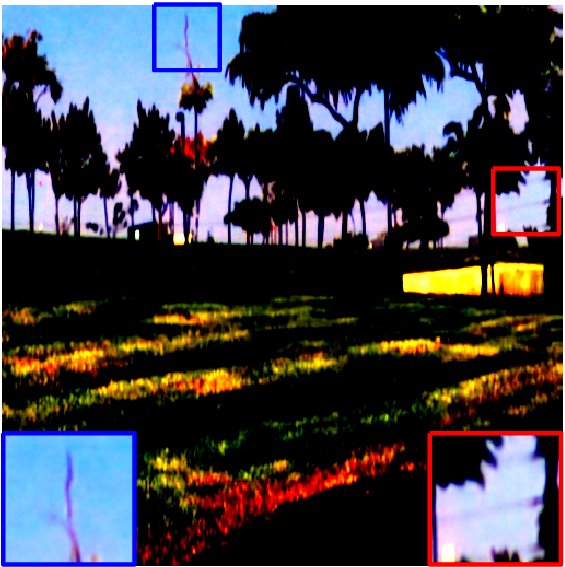}}
    \hspace{-1.1mm}
    \subfloat[]{\label{fig:Master5000K_SERT}\includegraphics[width=0.1240\linewidth]{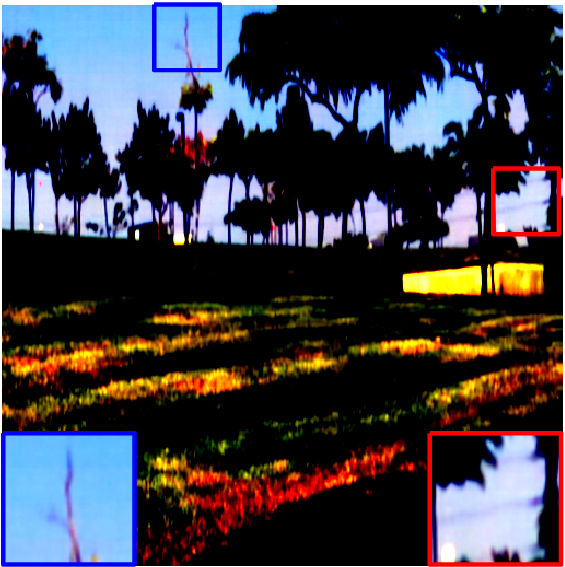}}
    \hspace{-1.1mm}
    \subfloat[]{\label{fig:Master5000K_T3SC}\includegraphics[width=0.1240\linewidth]{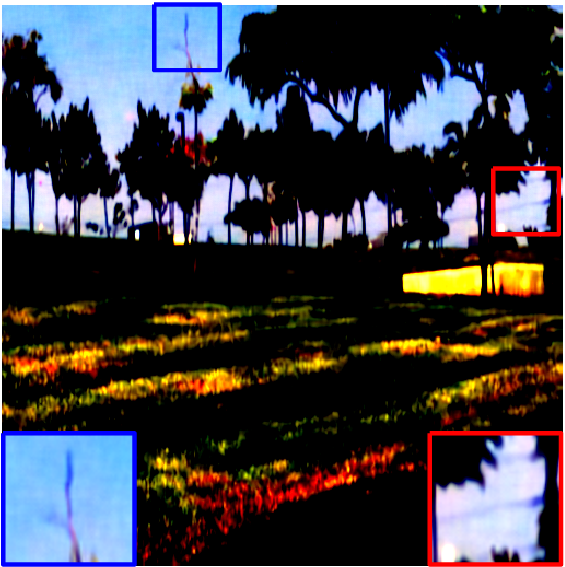}}
    \hspace{-1.1mm}
    \subfloat[]{\label{fig:Master5000K_MTSNN++}\includegraphics[width=0.1240\linewidth]{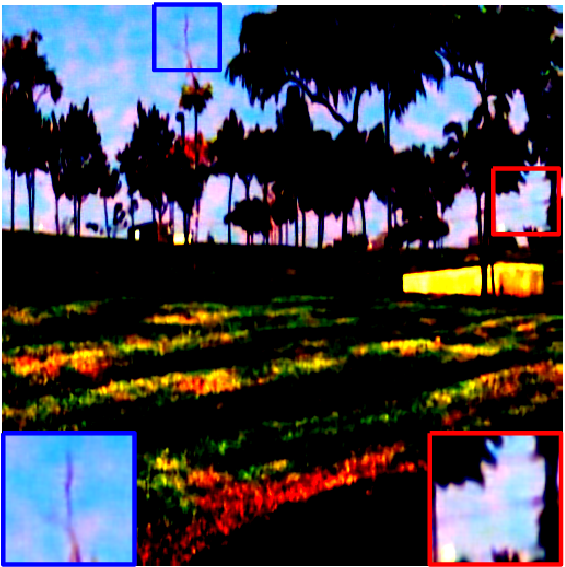}}
    \hspace{-1.1mm}
    \subfloat[]{\label{fig:Master5000K_DEQ-CSC}\includegraphics[width=0.1240\linewidth]{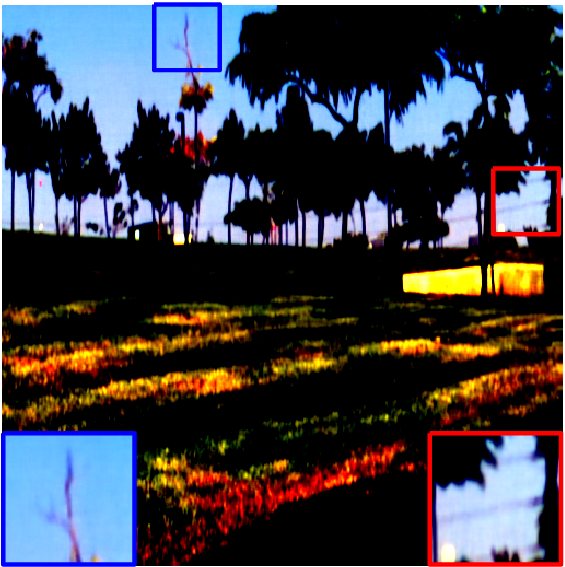}}
    %   \vspace{-0.2cm}
    \caption{Denoising results on the $Nachal\_0823-1214$ HSI from the ICVL dataset under the non-i.i.d. Gaussian noise with $\sigma \in [0,95]$. The false-color images are generated by combining bands 31, 17, and 2. (a) Clean. (b) Noisy. (c) BM4D~\cite{Maggioni2013BM4D}. (d) MTSNMF~\cite{Ye2015MTSNMF}. (e) LLRT~\cite{Chang2017LLRT}. (f) NGMeet~\cite{He2022NGMeet}. (g) LRMR~\cite{Zhang2014LRMR}. (h) E-3DTV~\cite{Peng2020E-3DTV}. (i) 3DlogTNN~\cite{Zheng20203DlogTNN}. (j) SST~\cite{li2022spatialspectral}. (k) TRQ3D~\cite{Pang2022TRQ3DNet}. (l) SERT~\cite{li2023spectral} (m) T3SC~\cite{bodrito2021T3SC}. (n) MTSNN++~\cite{xiong2023multitask}. (o) \textbf{DECSC}.} \label{fig:ICVL_visual}
 %    \vspace{-0.5cm}
 %   \vspace{-0.3cm}
\end{figure*}
\subsubsection{Implementation Details}
The proposed DECSC model was implemented in PyTorch and trained using the Adam optimizer, starting with an initial learning rate of 0.0001 and a batch size of 8. The learning rate was reduced by half every 10 epochs, and training was conducted for 30 epochs on on a Linux machine equipped with an Intel(R) Xeon(R) E5-2650 v4 CPU @ 2.20 GHz and four NVIDIA GeForce RTX 4090 GPUs. For the GIC component, the dictionary size was set to 192 with a convolutional filter size of $9\times 9$, and the Swin Transformer consists of four stacked stages with a window size of $ 4\times 4$. For the LSU component, the dictionary size was set to 96, also with a convolutional filter size of $9\times 9\times  3$. Anderson acceleration was employed to efficiently compute the equilibrium point (fixed point), while a rolling-based phantom gradient strategy with a truncation length of $L=5$ was used to approximate gradients during backpropagation. For synthetic noise removal, all models were pre-trained on the ICVL dataset and subsequently fine-tuned on the corresponding test datasets. In real noise scenarios, where ground truth is unavailable, band-splitting was applied directly during inference.

\subsection{Synthetic Noise Removal}
To comprehensively evaluate the DECSC's ability to remove synthetic noise, we conducted denoising experiments on both close-range HSIs, i.e., the ICVL dataset, and remote-range HSI, i.e., the Houston 2018 HSI, using the synthetic noise patterns described earlier.
\subsubsection{ICVL Dataset}

\begin{figure*}[!t]
    \centering
    \subfloat[]{\label{fig:Master5000K_noise_reflectance}\includegraphics[width=0.140\linewidth]{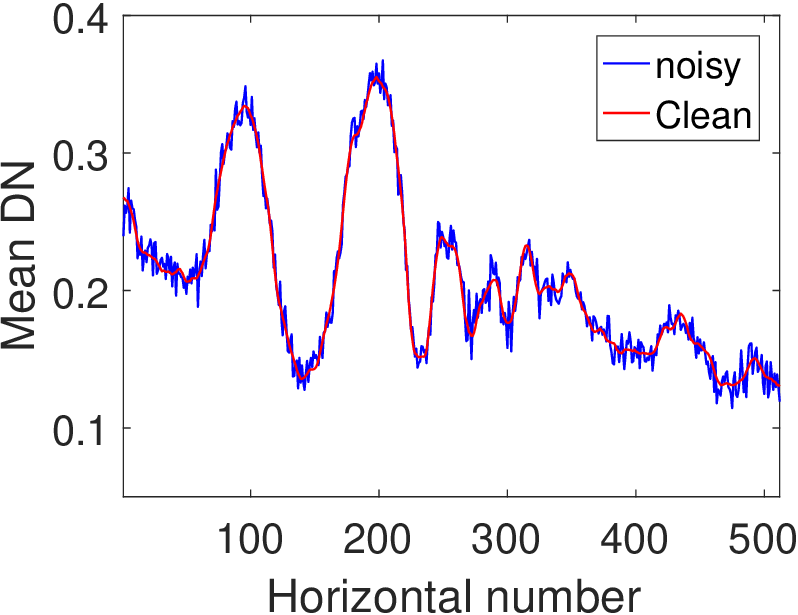}}
    \hspace{-1.1mm}
    \subfloat[]{\label{fig:Master5000K_BM4D_reflectance}\includegraphics[width=0.140\linewidth]{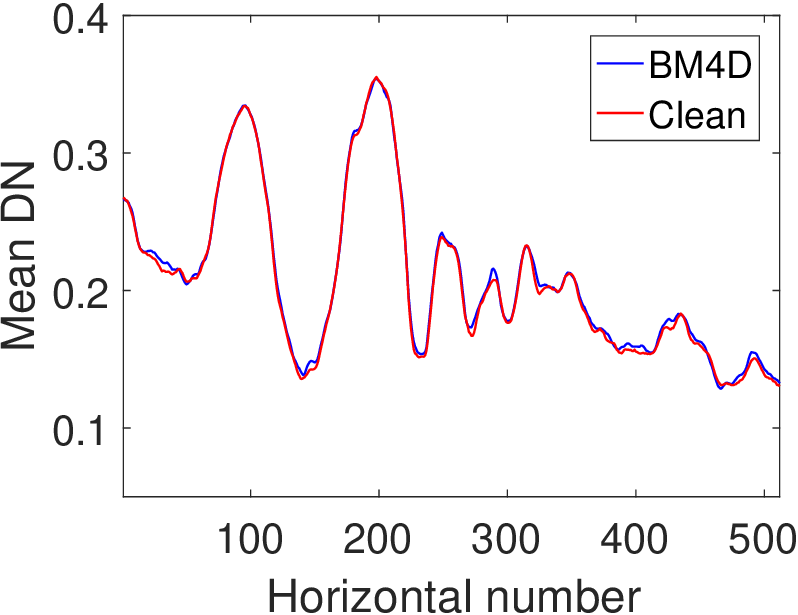}}
    \hspace{-1.1mm}
    \subfloat[]{\label{fig:Master5000K_MTSNMF_reflectance}\includegraphics[width=0.140\linewidth]{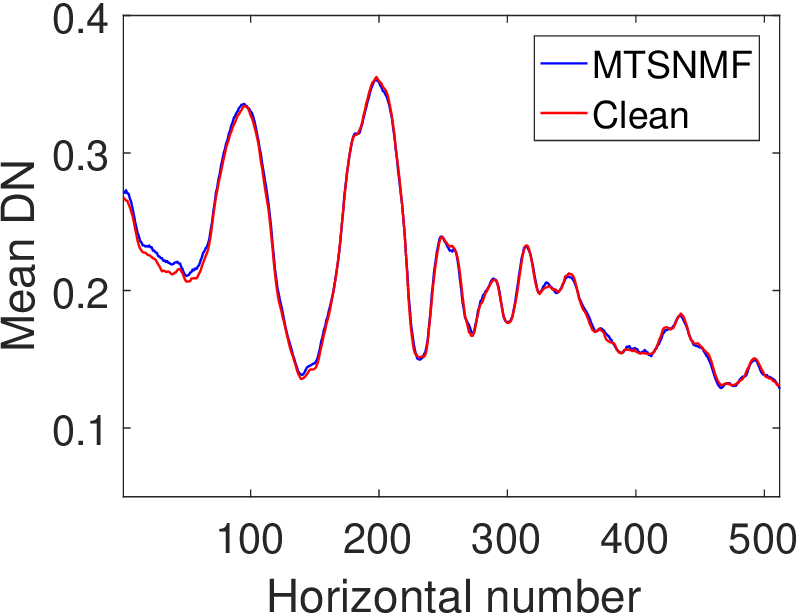}}
    \hspace{-1.1mm}
    \subfloat[]{\label{fig:Master5000K_LLRT_reflectance}\includegraphics[width=0.140\linewidth]{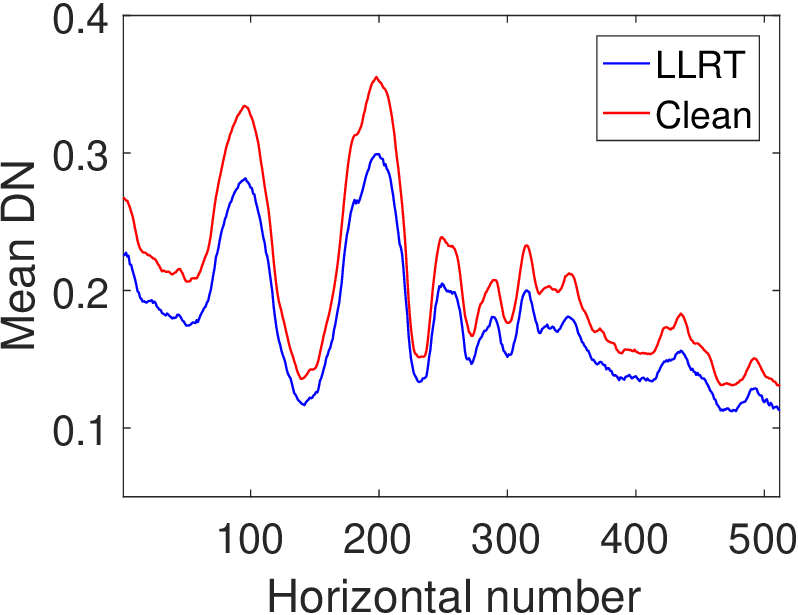}}
    \hspace{-1.1mm}
    \subfloat[]{\label{fig:Master5000K_NGMeet_reflectance}\includegraphics[width=0.140\linewidth]{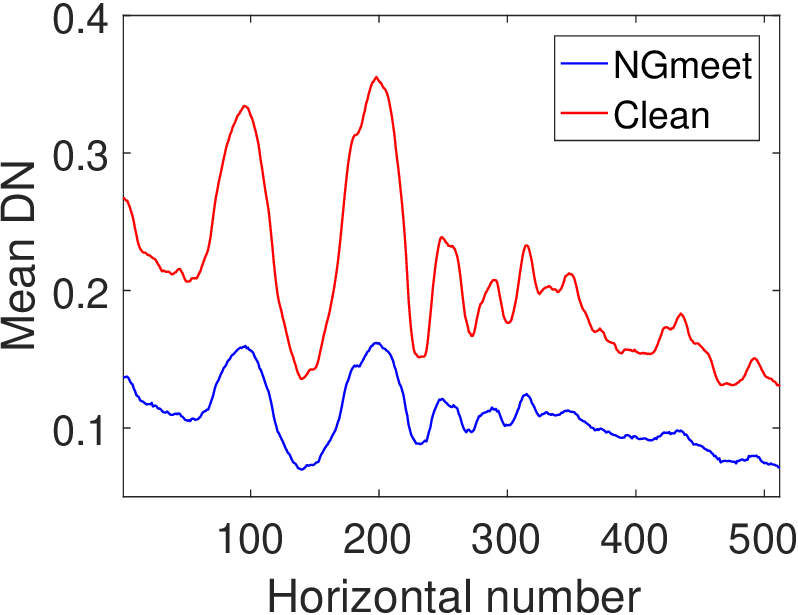}}
    \hspace{-1.1mm}
    \subfloat[]{\label{fig:Master5000K_LRMR_reflectance}\includegraphics[width=0.140\linewidth]{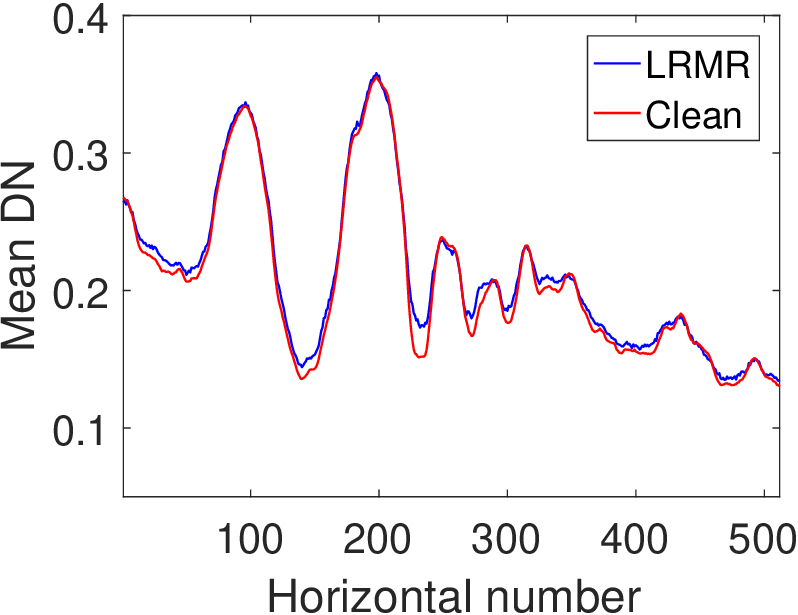}}
    \hspace{-1.1mm}
    \subfloat[]{\label{fig:Master5000K_E-3DTV_reflectance}\includegraphics[width=0.140\linewidth]{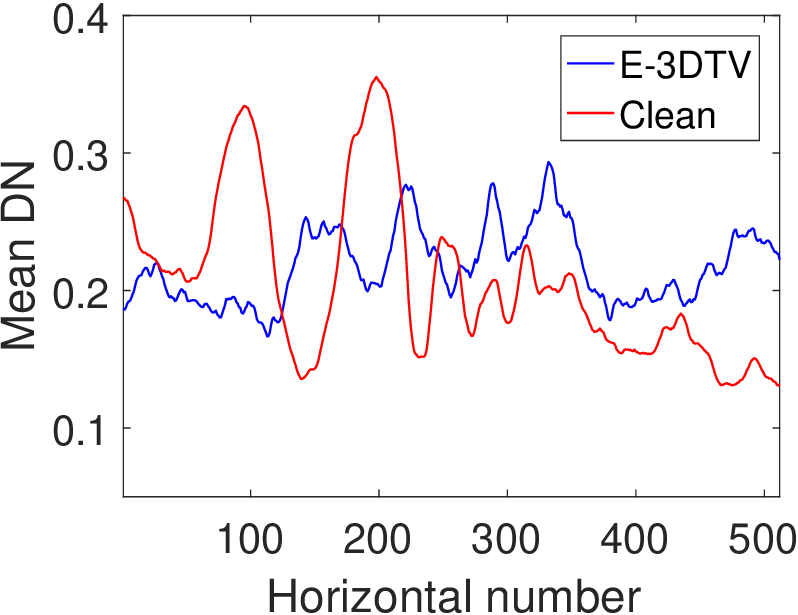}}
    \hspace{-1.1mm}
    \subfloat[]{\label{fig:Master5000K_3DlogTNN_reflectance}\includegraphics[width=0.140\linewidth]{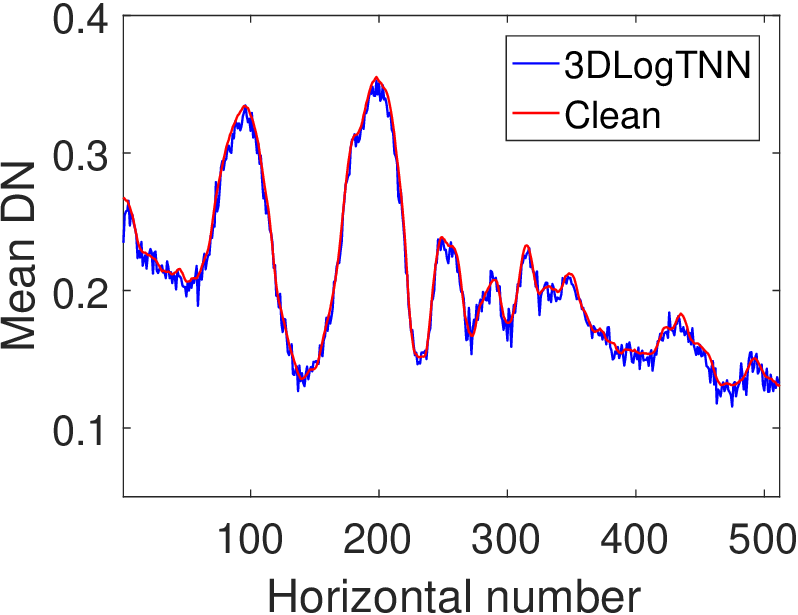}}
    \hspace{-1.1mm}
    \subfloat[]{\label{fig:Master5000K_SST_reflectance}\includegraphics[width=0.140\linewidth]{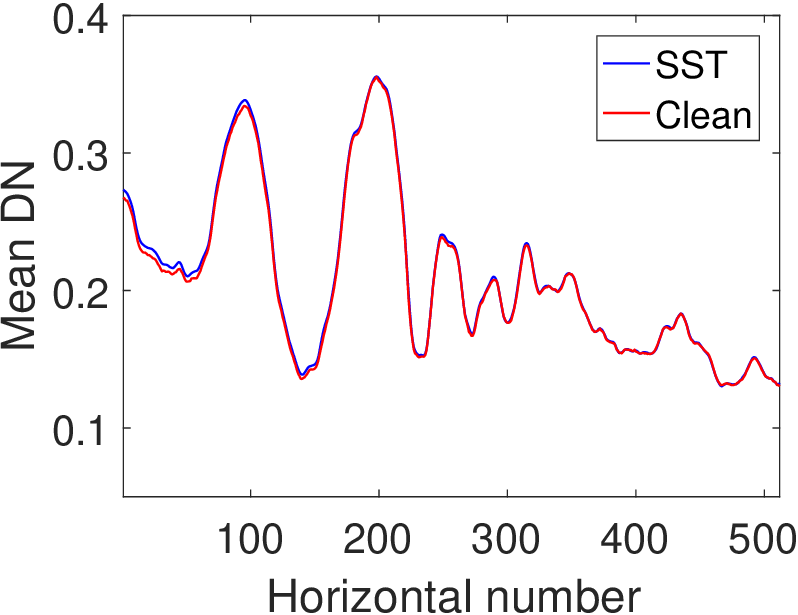}}
    \hspace{-1.1mm}
    \subfloat[]{\label{fig:Master5000K_TRQ3D_reflectance}\includegraphics[width=0.140\linewidth]{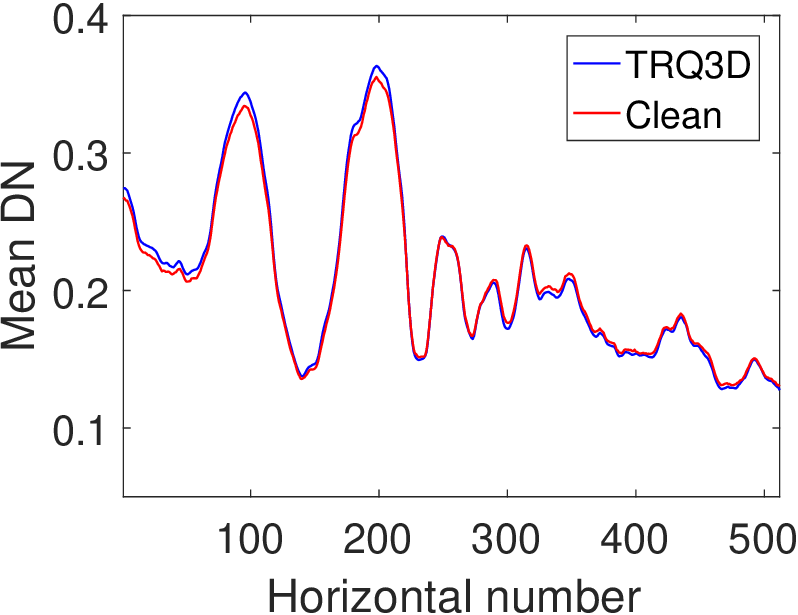}}
    \hspace{-1.1mm}
    \subfloat[]{\label{fig:Master5000K_SERT_reflectance}\includegraphics[width=0.140\linewidth]{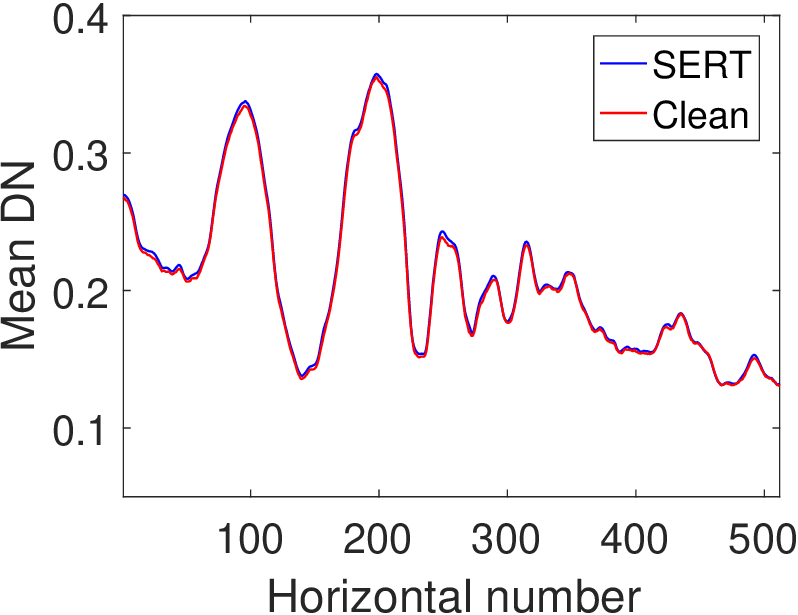}}
    \hspace{-1.1mm}
    \subfloat[]{\label{fig:Master5000K_T3SC_reflectance}\includegraphics[width=0.140\linewidth]{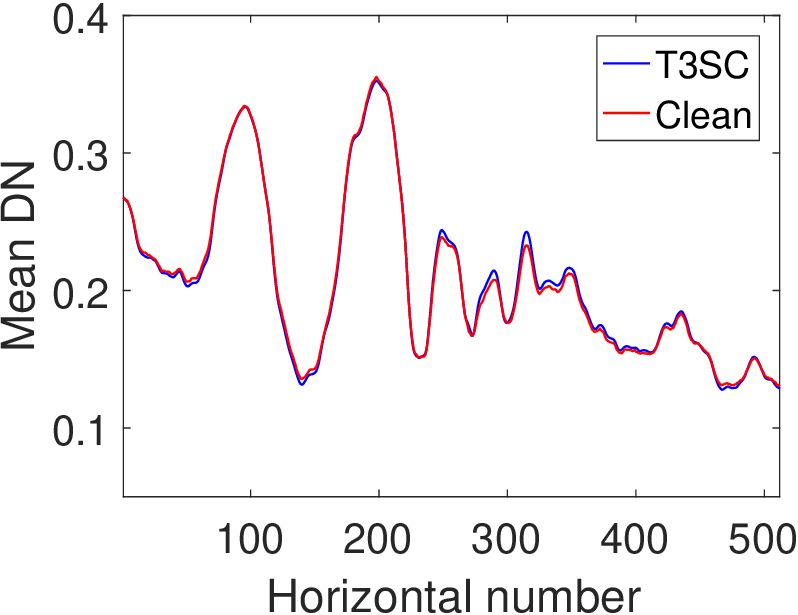}}
    \hspace{-1.1mm}
    \subfloat[]{\label{fig:Master5000K_MTSNN++_reflectance}\includegraphics[width=0.140\linewidth]{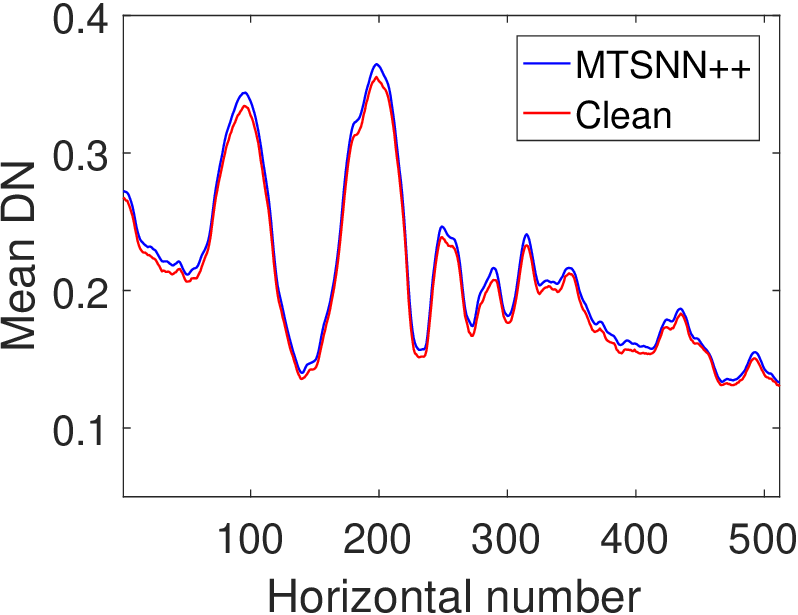}}
    \hspace{-1.1mm}
    \subfloat[]{\label{fig:Master5000K_DEQ-CSC_reflectance}\includegraphics[width=0.140\linewidth]{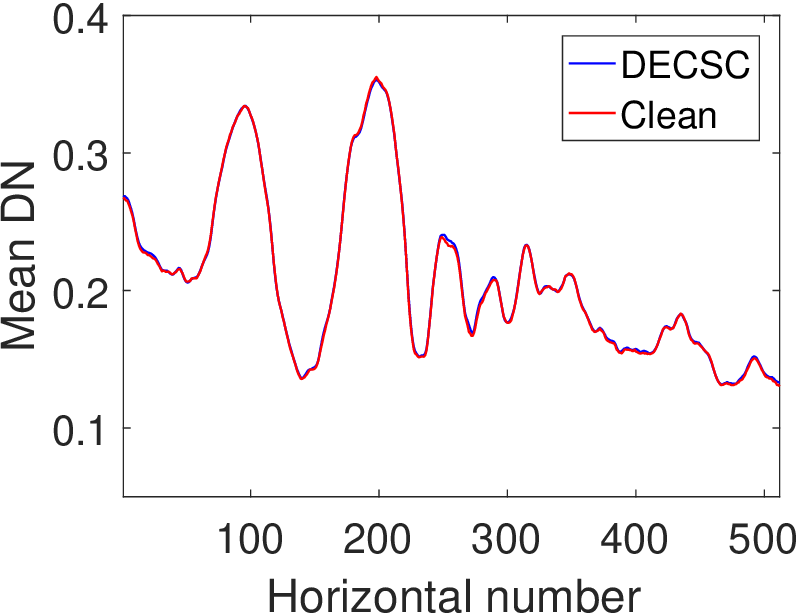}}
    %   \vspace{-0.2cm}
    \caption{Row mean profiles of band 28 for the $Nachal\_0823-1214$ HSI from the ICVL dataset under the non-i.i.d. Gaussian noise with $\sigma \in [0,95]$. (a) Noisy. (b) BM4D~\cite{Maggioni2013BM4D}. (c) MTSNMF~\cite{Ye2015MTSNMF}. (d) LLRT~\cite{Chang2017LLRT}. (e) NGMeet~\cite{He2022NGMeet}. (f) LRMR~\cite{Zhang2014LRMR}. (g) E-3DTV~\cite{Peng2020E-3DTV}. (h) 3DlogTNN~\cite{Zheng20203DlogTNN}. (i) SST~\cite{li2022spatialspectral}. (j) TRQ3D~\cite{Pang2022TRQ3DNet}. (k) SERT~\cite{li2023spectral}. (l) T3SC~\cite{bodrito2021T3SC}. (m) MTSNN++~\cite{xiong2023multitask}. (n) \textbf{DECSC}.} \label{fig:ICVL_row_mean_profiles}
 %    \vspace{-0.5cm}
 %   \vspace{-0.3cm}
\end{figure*}
\begin{table*}[!t]
    \caption{Comparison of Different Methods on Houston 2018 HSI. The Top Three Values Are Marked as \1{Red}, \2{Blue}, And \3{Green}.}\label{tab:houston}
    % \vspace{-0.3cm}
    \centering
    \resizebox{\linewidth}{!}{
            \tablesize{
    \begin{tabular}{c|c|c|c|c|c|c|c|c|c|c|c|c|c|c|c}
            \Xhline{1.2pt}
    \multirow{3}*{\makebox[0.0588\textwidth][c]{$\sigma$}}&\multirow{3}*{\makebox[0.0588\textwidth][c]{Index}}&\multirow{3}*{\makebox[0.0588\textwidth][c]{Noisy}}
    &\multicolumn{7}{c|}{\textbf{Model-driven}}&\multicolumn{3}{c|}{\textbf{Data-driven}}&\multicolumn{3}{c}{\textbf{Hybrid-driven}}\\
    \cline{4-16}
    &&&\multirow{1}*{\makebox[0.0588\textwidth][c]{BM4D}}&\multirow{1}*{\makebox[0.0588\textwidth][c]{MTSNMF}}&\multirow{1}*{\makebox[0.0588\textwidth][c]{LLRT}}&\multirow{1}*{\makebox[0.0588\textwidth][c]{NGMeet}}&\multirow{1}*{\makebox[0.0588\textwidth][c]{LRMR}}&\multirow{1}*{\makebox[0.0588\textwidth][c]{E-3DTV}}&\multirow{1}*{\makebox[0.0588\textwidth][c]{3DlogTNN}}&\multirow{1}*{\makebox[0.0588\textwidth][c]{SST}}&\multirow{1}*{\makebox[0.0588\textwidth][c]{TRQ3D}}&\multirow{1}*{\makebox[0.0588\textwidth][c]{SERT}}&\multirow{1}*{\makebox[0.0588\textwidth][c]{T3SC}}&\multirow{1}*{\makebox[0.0588\textwidth][c]{MTSNN++}}&\makebox[0.063\textwidth][c]{\textbf{DECSC}}\\
    &&&\multirow{1}*{\makebox[0.0588\textwidth][c]{\cite{Maggioni2013BM4D}}}&\multirow{1}*{\makebox[0.0588\textwidth][c]{\cite{Ye2015MTSNMF}}}&\multirow{1}*{\makebox[0.0588\textwidth][c]{\cite{Chang2017LLRT}}}&\multirow{1}*{\makebox[0.0588\textwidth][c]{\cite{He2022NGMeet}}}&\multirow{1}*{\makebox[0.0588\textwidth][c]{\cite{Zhang2014LRMR}}}&\multirow{1}*{\makebox[0.0588\textwidth][c]{\cite{Peng2020E-3DTV}}}&\multirow{1}*{\makebox[0.0588\textwidth][c]{\cite{Zheng20203DlogTNN}}}&\multirow{1}*{\makebox[0.0588\textwidth][c]{\cite{li2022spatialspectral}}}&\multirow{1}*{\makebox[0.0588\textwidth][c]{\cite{Pang2022TRQ3DNet}}}&\multirow{1}*{\makebox[0.0588\textwidth][c]{\cite{li2023spectral}}}&\multirow{1}*{\makebox[0.0588\textwidth][c]{\cite{bodrito2021T3SC}}}&\multirow{1}*{\makebox[0.0588\textwidth][c]{\cite{xiong2023multitask}}}&\multirow{1}*{\makebox[0.07\textwidth][c]{\textbf{(Ours)}}}\\
    \Xhline{1.2pt}		
    \multirow{3}*{\textbf{[0,15]}}
    & PSNR$\uparrow$  & 32.47 & 40.70 & 42.83 & 40.65 & 38.31 & 39.90 & 43.39 & 43.66   & \2{50.53} & 44.75 & \3{49.60} & 49.09 & 48.11 & \1{51.35} \\
    & SSIM$\uparrow$  & .6738 & .9667 & .9775 & .9468 & .8906 & .9601 & .9838 & .9869   & \3{.9959} & .9867 & \1{.9981} & .9947 & .9927 & \2{.9961} \\
    & SAM$\downarrow$ & .2562 & .0550 & .0357 & .0712 & .1279 & .0612 & .0427 & .0327   & \2{.0180} & .0284 & \3{.0200} & .0214 & .0232 & \1{.0173} \\
    \hline 		
    \multirow{3}*{\textbf{[0,55]}}
    & PSNR$\uparrow$  & 24.66 & 35.04 & 35.94 & 30.15 & 29.86 & 30.93 & 39.23 & 31.44   & \2{47.64} & 44.76 & \3{45.65} & 44.20 & 44.98 & \1{49.05} \\
    & SSIM$\uparrow$  & .3874 & .9002 & .8919 & .6390 & .6918 & .7624 & .9629 & .6480   & \3{.9926} & .9874 & \1{.9955} & .9872 & .9862 & \2{.9933} \\
    & SAM$\downarrow$ & .6484 & .1014 & .1387 & .3295 & .3707 & .1688 & .0690 & .4463   & \2{.0227} & .0307 & \3{.0263} & .0316 & .0292 & \1{.0209} \\
    \hline
    \multirow{3}*{\textbf{[0,95]}}
    & PSNR$\uparrow$  & 16.85 & 31.23 & 32.73 & 22.93 & 26.06 & 26.43 & 34.92 & 23.10   & \2{43.10} & 41.74 & \3{42.10} & 40.08 & 41.07 & \1{43.25} \\
    & SSIM$\uparrow$  & .2072 & .7766 & .8160 & .3724 & .5514 & .5662 & .9254 & .4339   & \2{.9834} & .9779 & \1{.9911} & .9699 & .9719 & \2{.9834} \\
    & SAM$\downarrow$ & .9201 & .1961 & .2160 & .5802 & .5400 & .3017 & .1117 & .8329   & \2{.0310} & .0389 & \3{.0356} & .0466 & .0434 & \1{.0305} \\
    \Xhline{1.2pt}  		
    \multirow{3}*{\textbf{Mixture}}
    & PSNR$\uparrow$  & 11.72 & 22.76 & 25.86 & 15.58 & 22.36 & 21.84 & 30.69 & 15.43   & \3{35.85} & \2{36.23} & 34.86     & 28.66 & 34.71 & \1{38.65} \\
    & SSIM$\uparrow$  & .0843 & .4762 & .6933 & .1386 & .5169 & .3914 & .8582 & .1965   & \3{.9449} & .9363     & \1{.9643} & .8471 & .9081 & \2{.9581} \\
    & SAM$\downarrow$ & .9778 & .5168 & .4977 & .7652 & .5728 & .4857 & .1316 & .9033   & \2{.0605} & \3{.0659} & .0842     & .2280 & .0817 & \1{.0480} \\
    \Xhline{1.2pt}  		
    \multirow{3}*{\textbf{Corr}}
    & PSNR$\uparrow$  & 28.21 & 37.28 & 40.22 & 36.04 & 35.69 & 37.12 & 40.65 & 41.01  & \2{46.54} & 45.36 & \1{48.32} & 45.90 & 45.42 & \3{46.32} \\
    & SSIM$\uparrow$  & .5631 & .8721 & .9554 & .8383 & .8634 & .9429 & .9706 & .9702  & \2{.9910} & .9886 & \1{.9973} & .9899 & .9882 & \3{.9904} \\
    & SAM$\downarrow$ & .3649 & .1466 & .0525 & .1597 & .1834 & .0761 & .0493 & .0423  & \2{.0221} & .0268 & \1{.0201} & .0251 & .0265 & \3{.0225} \\
    \Xhline{1.2pt} 		
    \end{tabular}}}
            %  \vspace{-0.1cm}
\end{table*}
\begin{figure*}[!t]
    \centering
    \subfloat[]{\label{fig:houston_clean_visual}\includegraphics[width=0.1240\linewidth]{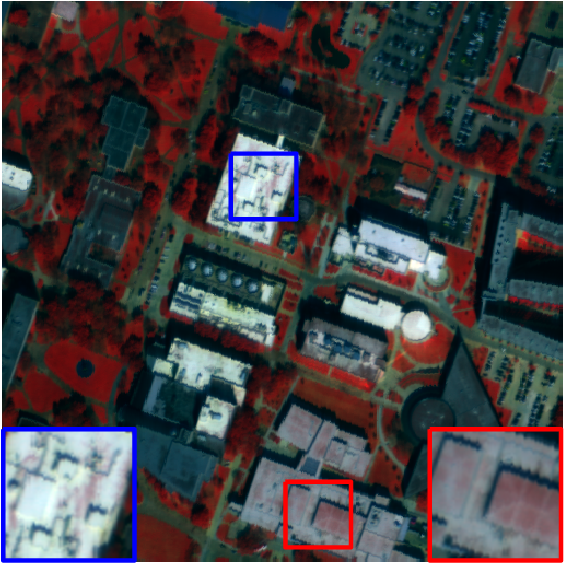}}
    \hspace{-1.1mm}
    \subfloat[]{\label{fig:houston_noise_visual}\includegraphics[width=0.1240\linewidth]{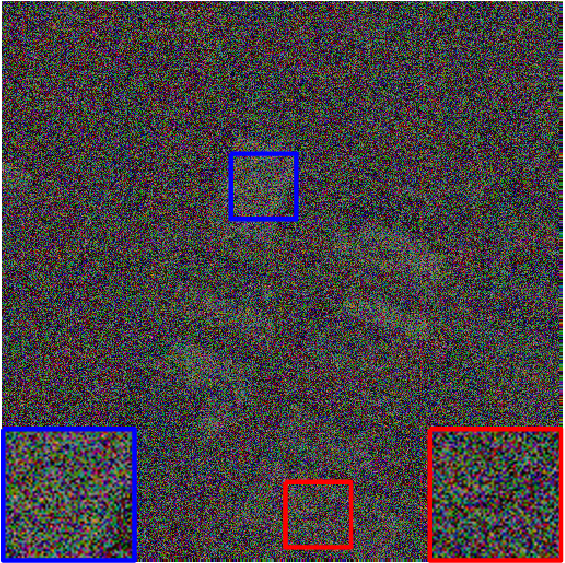}}
    \hspace{-1.1mm}
    \subfloat[]{\label{fig:houston_BM4D_visual}\includegraphics[width=0.1240\linewidth]{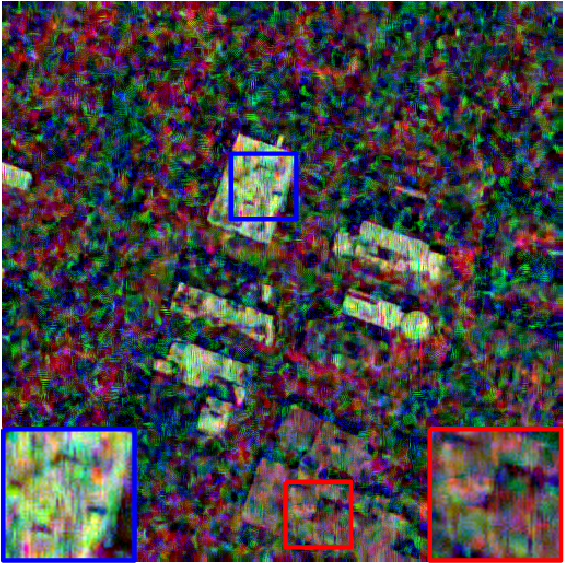}}
    \hspace{-1.1mm}
    \subfloat[]{\label{fig:houston_MTSNMF_visual}\includegraphics[width=0.1240\linewidth]{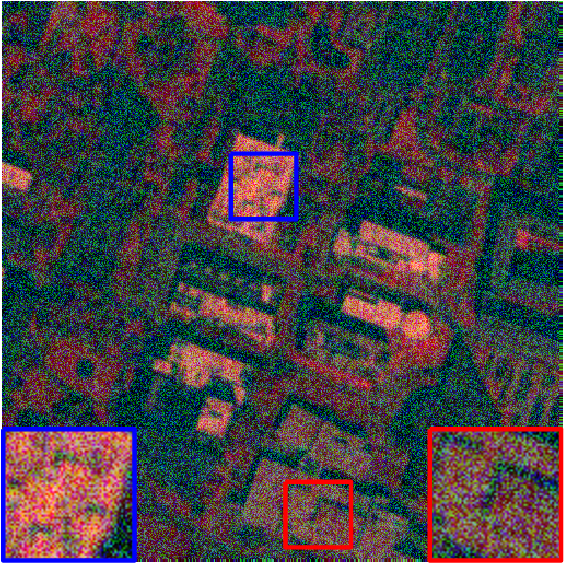}}
    \hspace{-1.1mm}
    \subfloat[]{\label{fig:houston_LLRT_visual}\includegraphics[width=0.1240\linewidth]{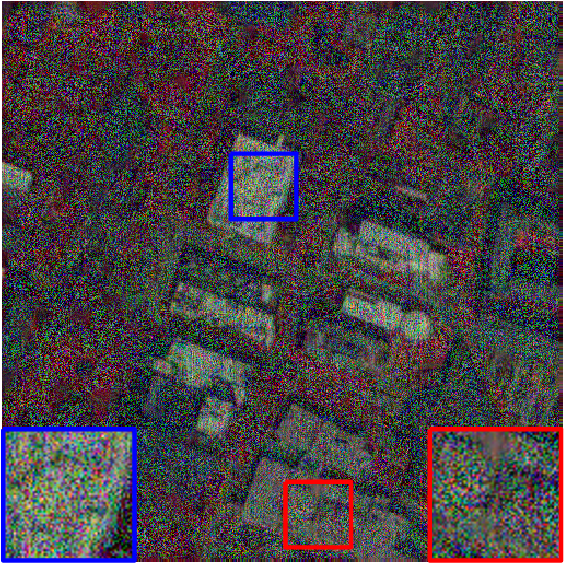}}
    \hspace{-1.1mm}
    \subfloat[]{\label{fig:houston_NGMeet_visual}\includegraphics[width=0.1240\linewidth]{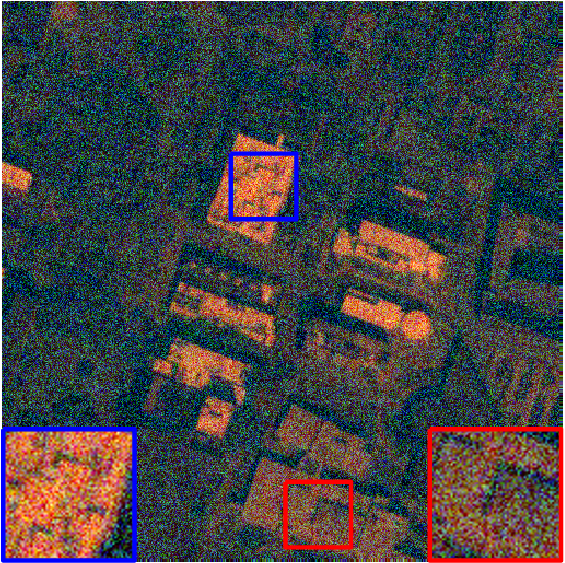}}
    \hspace{-1.1mm}
    \subfloat[]{\label{fig:houston_LRMR_visual}\includegraphics[width=0.1240\linewidth]{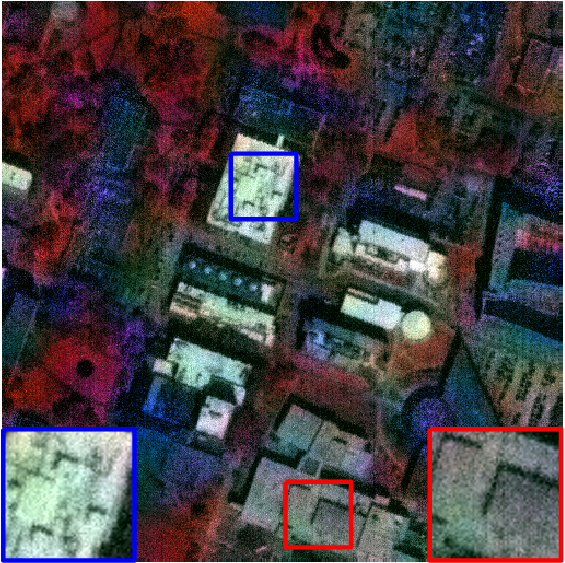}}
    \hspace{-1.1mm}
    \subfloat[]{\label{fig:houston_E-3DTV_visual}\includegraphics[width=0.1240\linewidth]{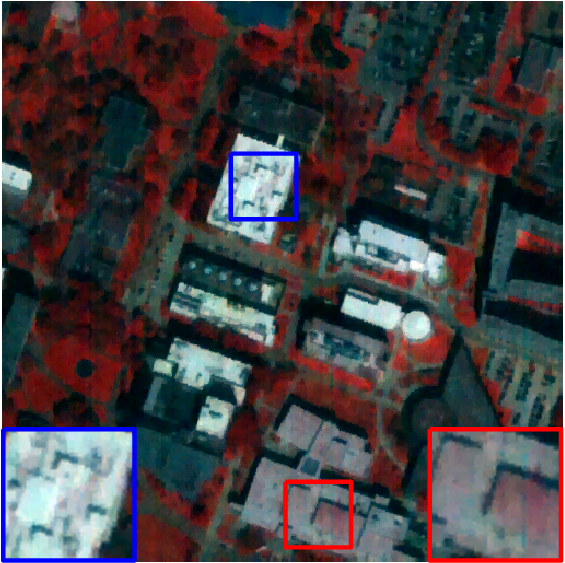}}
    \hspace{-1.1mm}
    \subfloat[]{\label{fig:houston_3DlogTNN_visual}\includegraphics[width=0.1240\linewidth]{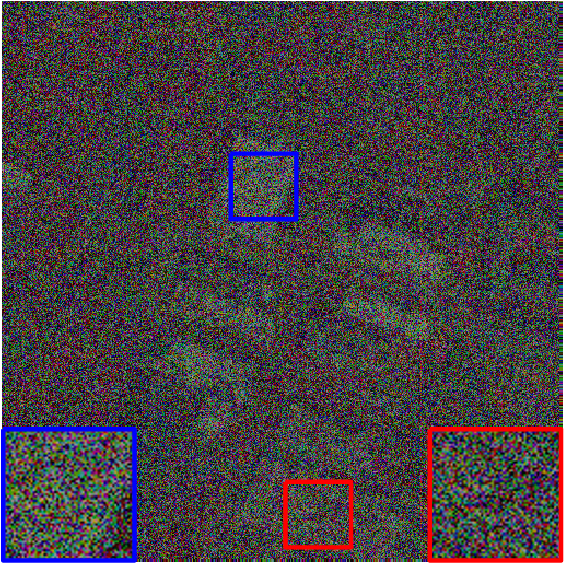}}
    \hspace{-1.1mm}
    \subfloat[]{\label{fig:houston_SST_visual}\includegraphics[width=0.1240\linewidth]{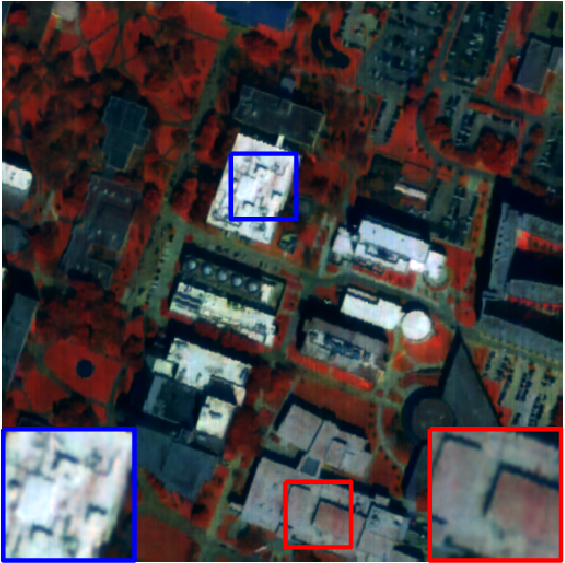}}
    \hspace{-1.1mm}
    \subfloat[]{\label{fig:houston_TRQ3D_visual}\includegraphics[width=0.1240\linewidth]{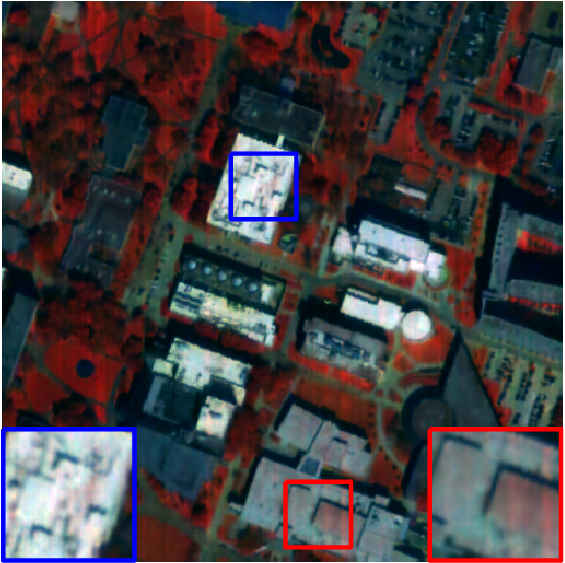}}
    \hspace{-1.1mm}
    \subfloat[]{\label{fig:houston_SERT_visual}\includegraphics[width=0.1240\linewidth]{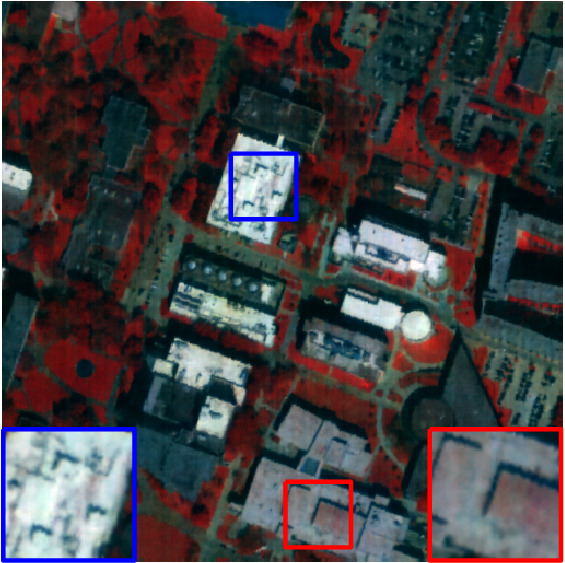}}
    \hspace{-1.1mm}
    \subfloat[]{\label{fig:houston_T3SC_visual}\includegraphics[width=0.1240\linewidth]{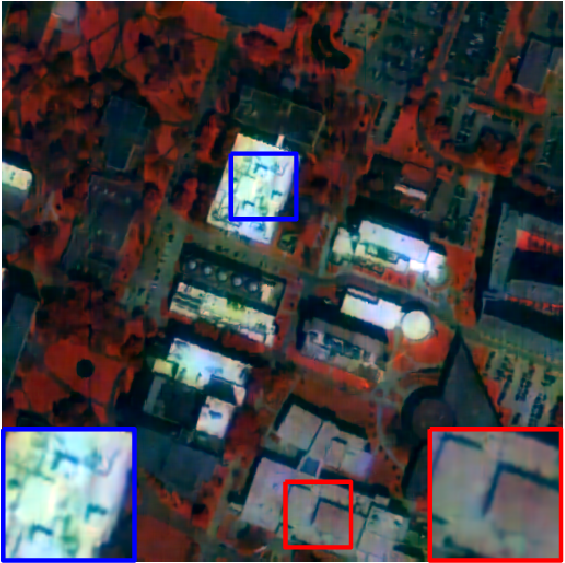}}
    \hspace{-1.1mm}
    \subfloat[]{\label{fig:houston_MTSNN++_visual}\includegraphics[width=0.1240\linewidth]{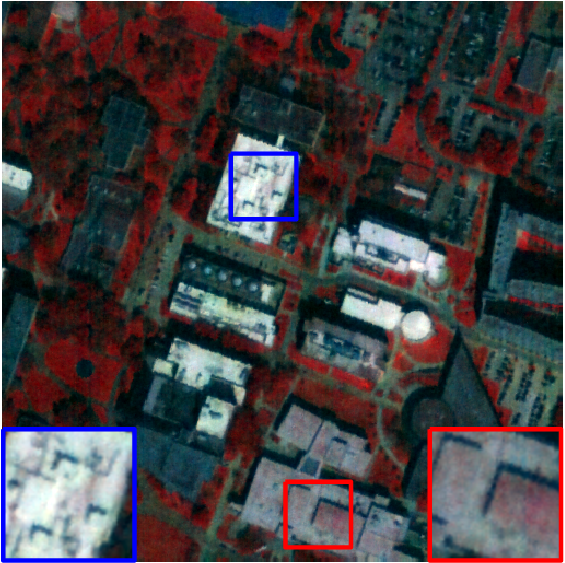}}
    \hspace{-1.1mm}
    \subfloat[]{\label{fig:houston_DECSC_visual}\includegraphics[width=0.1240\linewidth]{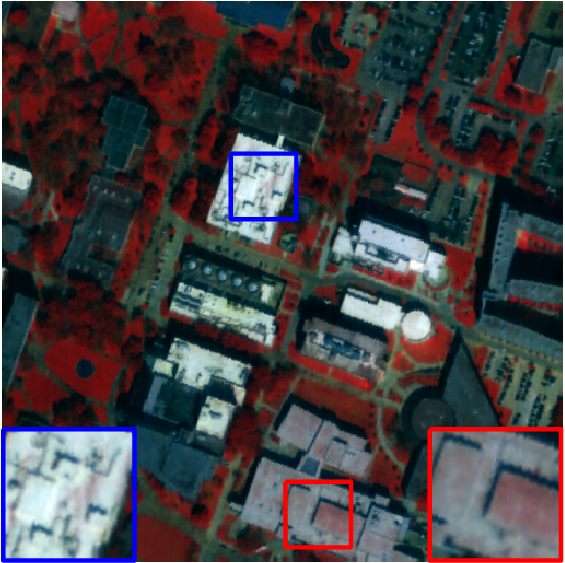}}
    %   \vspace{-0.2cm}
    \caption{Denoising results on the the Houston 2018 HSI under the mixture noise. The false-color images are generated by combining bands 39, 20, and 4. (a) Clean. (b) Noisy. (c) BM4D~\cite{Maggioni2013BM4D}. (d) MTSNMF~\cite{Ye2015MTSNMF}. (e) LLRT~\cite{Chang2017LLRT}. (f) NGMeet~\cite{He2022NGMeet}. (g) LRMR~\cite{Zhang2014LRMR}. (h) E-3DTV~\cite{Peng2020E-3DTV}. (i) 3DlogTNN~\cite{Zheng20203DlogTNN}. (j) SST~\cite{li2022spatialspectral}. (k) TRQ3D~\cite{Pang2022TRQ3DNet}. (l) SERT~\cite{li2023spectral} (m) T3SC~\cite{bodrito2021T3SC}. (n) MTSNN++~\cite{xiong2023multitask}. (o) \textbf{DECSC}.} \label{fig:houston_visual}
 %    \vspace{-0.5cm}
 %   \vspace{-0.3cm}
\end{figure*}
The ICVL dataset spans a spectral range of 400-700nm. For testing, each HSI is cropped to a size of $512\times512\times31$. Table~\ref{tab:icvl} presents a quantitative comparison of the denoising performance. As data-driven neural network methods learn mappings from noisy to clean HSIs using large-scale data, they are better suited to capture the intrinsic structures of HSIs than methods relying on hand-crafted priors, resulting in a notable performance advantage. Furthermore, the performance of model-driven approaches that depend on precise noise intensity estimation may degrade when noise levels vary independently across spectral bands. All data-driven methods evaluated in this study utilize transformer-based architectures. Among them, SST and SERT exhibit strong performance due to the incorporation of spectral attention mechanisms and low-rank memory units, respectively. The comparative hybrid-driven methods evaluated here are based on sparse priors. Although MTSNN++ shares a similar design philosophy with DECSC, it does not adequately account for nonlocal self-similarities. In contrast, the proposed DECSC outperforms  others by jointly modeling global inter-band structural consistency, local spatial-spectral correlations, and nonlocal self-similarities. Moreover, the DEQ framework further enhances the denoising performance compared to the deep unfolding-based MTSNN++. However, we also observe that our DECSC lags behind SERT in the Corr noise pattern, possibly due to SERT's superior ability to capture global spectral correlations through its low-rank memory unit.

Fig.~\ref{fig:ICVL_visual} presents a visual comparison of different methods. To better highlight the denoising performance, we selected three spectral bands with severe noise corruption. Consistent with the quantitative results, data-driven and hybrid-driven methods generally outperform purely model-driven approaches. Model-driven methods often leave residual noise, color distortions, and blurred details, while data-driven and hybrid-driven methods achieve more effective noise suppression. Nonetheless, T3SC, TRQ3D, and MTSNN++ still exhibit blur patterns, whereas SST and SERT produce sharper images but suffer from slight color shifts. In contrast, the proposed method demonstrates superior preservation of both image detail and color fidelity, benefiting from the guidance of low-noise bands and the effectiveness of the detail enhancement module.

Fig.~\ref{fig:ICVL_row_mean_profiles} compares the row mean profiles of the restoration results obtained by different methods. Except for some model-driven approaches, most methods are able to recover results that are globally similar to the clean HSI, though local deviations still exist. It is worth noting that the row mean profile, which represents the global trend along the vertical spatial dimension by averaging, inherently suppresses local spatial variations. As a result, it can appear close to the ground truth even when spatial details are degraded, leading to discrepancies with pixel-level metrics such as PSNR and with visual quality, both of which are sensitive to fine spatial distortions. Methods with strong global regularization often maintain accurate row mean profiles but may suffer from excessive spatial smoothing or insufficient noise removal, thus lowering PSNR and perceived quality. In contrast, the proposed DECSC achieves results that are noticeably closer to the clean HSI in both global trends and fine details, demonstrating its strong denoising ability.

\subsubsection{Houston 2018 HSI}

\begin{figure*}[!t]
    \centering
    \subfloat[]{\label{fig:houston_noise_reflectance}\includegraphics[width=0.140\linewidth]{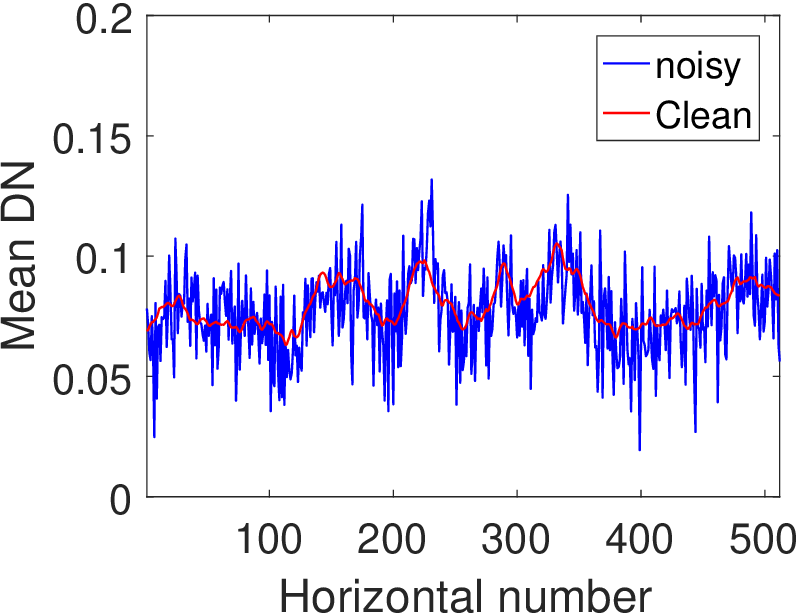}}
    \hspace{-1.1mm}
    \subfloat[]{\label{fig:houston_BM4D_reflectance}\includegraphics[width=0.140\linewidth]{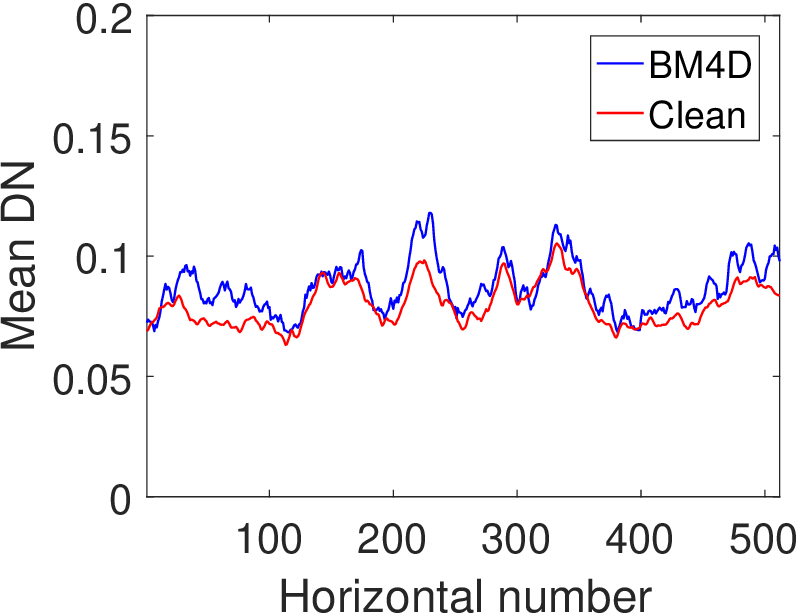}}
    \hspace{-1.1mm}
    \subfloat[]{\label{fig:houston_MTSNMF_reflectance}\includegraphics[width=0.140\linewidth]{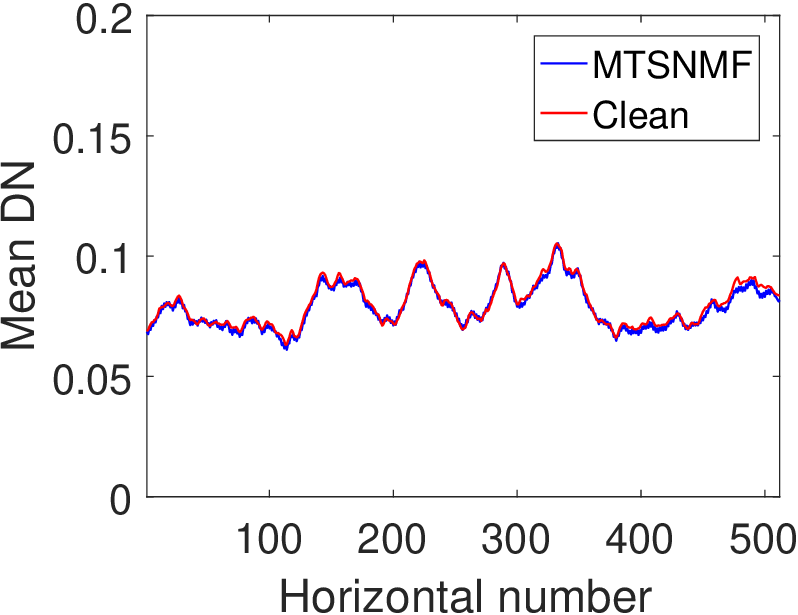}}
    \hspace{-1.1mm}
    \subfloat[]{\label{fig:houston_LLRT_reflectance}\includegraphics[width=0.140\linewidth]{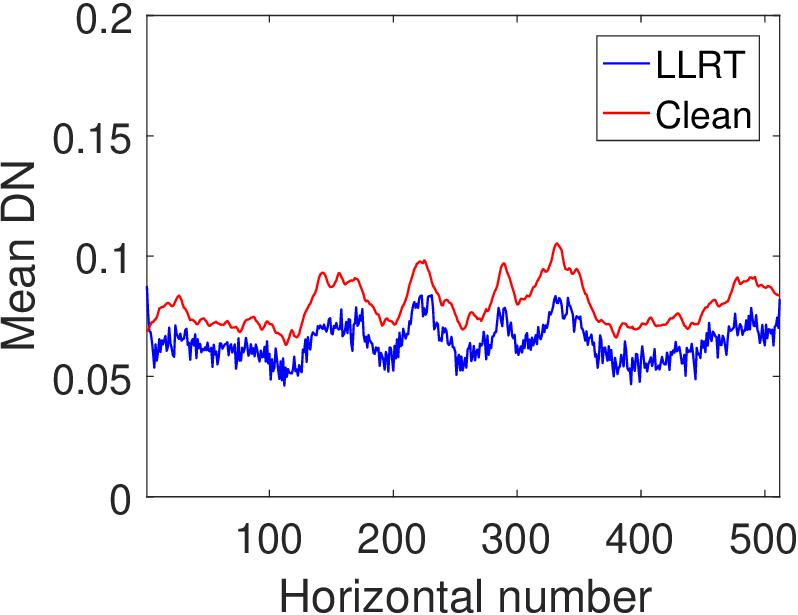}}
    \hspace{-1.1mm}
    \subfloat[]{\label{fig:houston_NGMeet_reflectance}\includegraphics[width=0.140\linewidth]{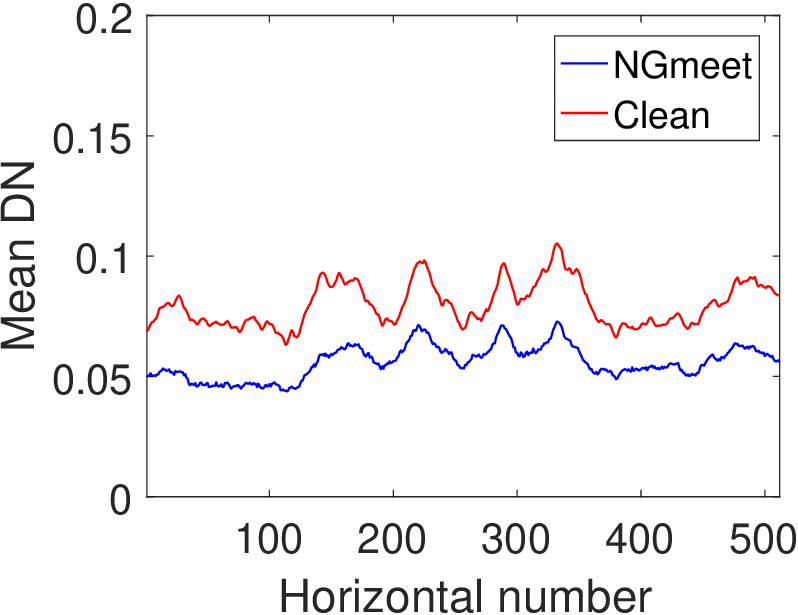}}
    \hspace{-1.1mm}
    \subfloat[]{\label{fig:houston_LRMR_reflectance}\includegraphics[width=0.140\linewidth]{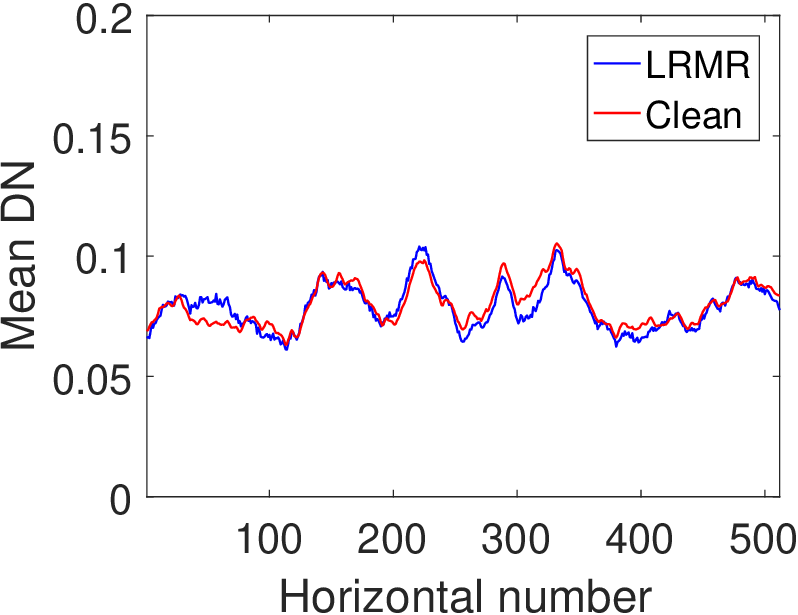}}
    \hspace{-1.1mm}
    \subfloat[]{\label{fig:houston_E-3DTV_reflectance}\includegraphics[width=0.140\linewidth]{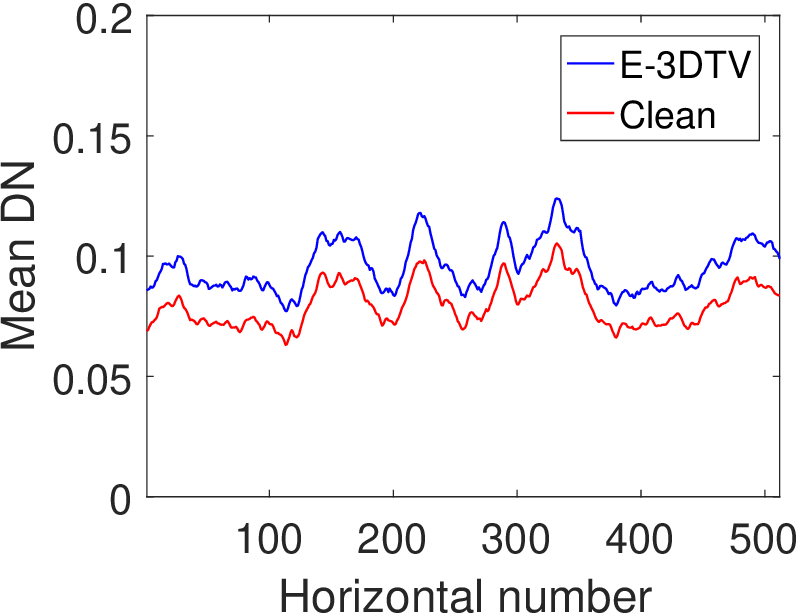}}
    \hspace{-1.1mm}
    \subfloat[]{\label{fig:houston_3DlogTNN_reflectance}\includegraphics[width=0.140\linewidth]{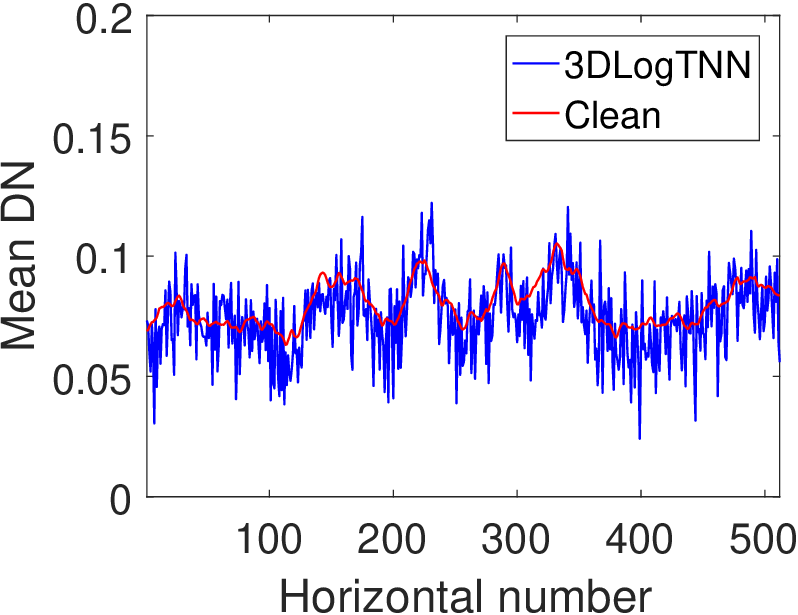}}
    \hspace{-1.1mm}
    \subfloat[]{\label{fig:houston_SST_reflectance}\includegraphics[width=0.140\linewidth]{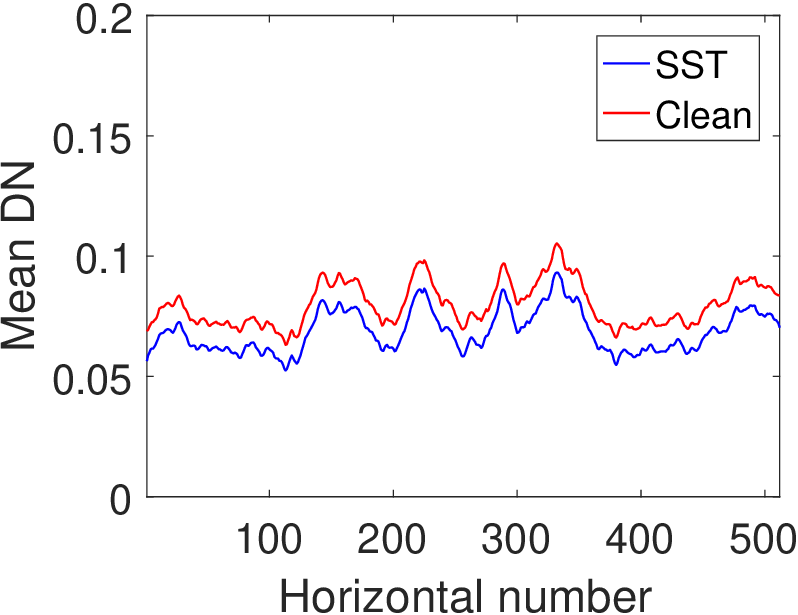}}
    \hspace{-1.1mm}
    \subfloat[]{\label{fig:houston_TRQ3D_reflectance}\includegraphics[width=0.140\linewidth]{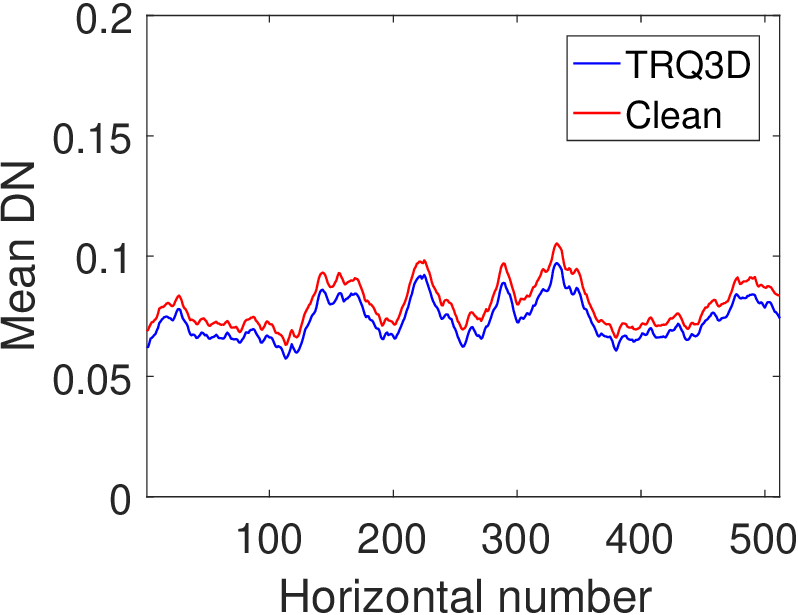}}
    \hspace{-1.1mm}
    \subfloat[]{\label{fig:houston_SERT_reflectance}\includegraphics[width=0.140\linewidth]{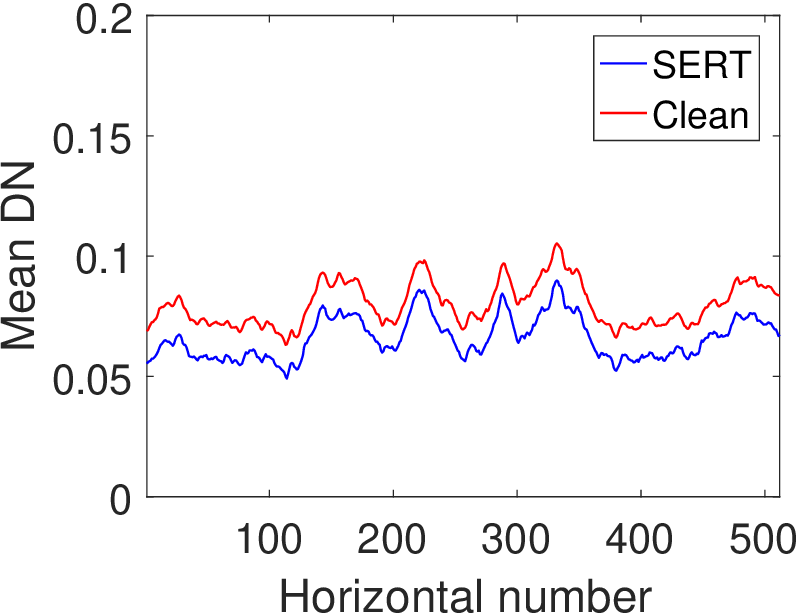}}
    \hspace{-1.1mm}
    \subfloat[]{\label{fig:houston_T3SC_reflectance}\includegraphics[width=0.140\linewidth]{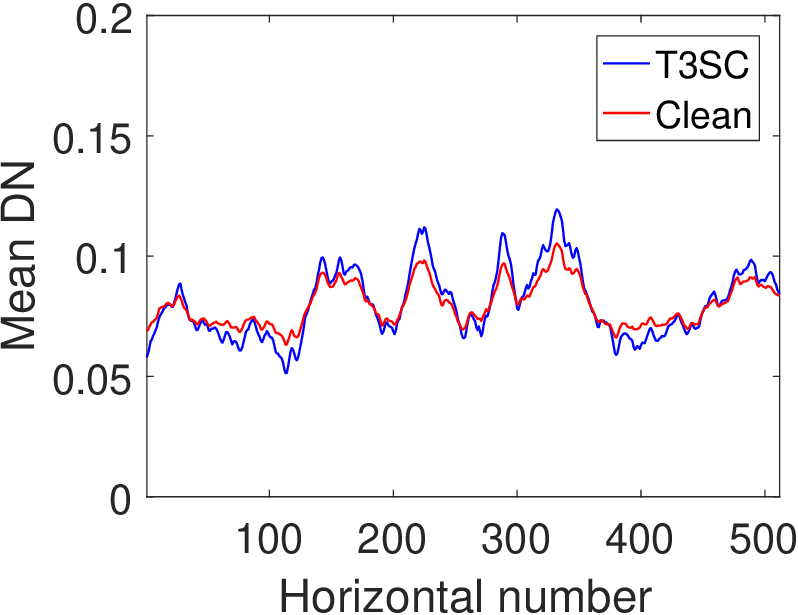}}
    \hspace{-1.1mm}
    \subfloat[]{\label{fig:houston_MTSNN++_reflectance}\includegraphics[width=0.140\linewidth]{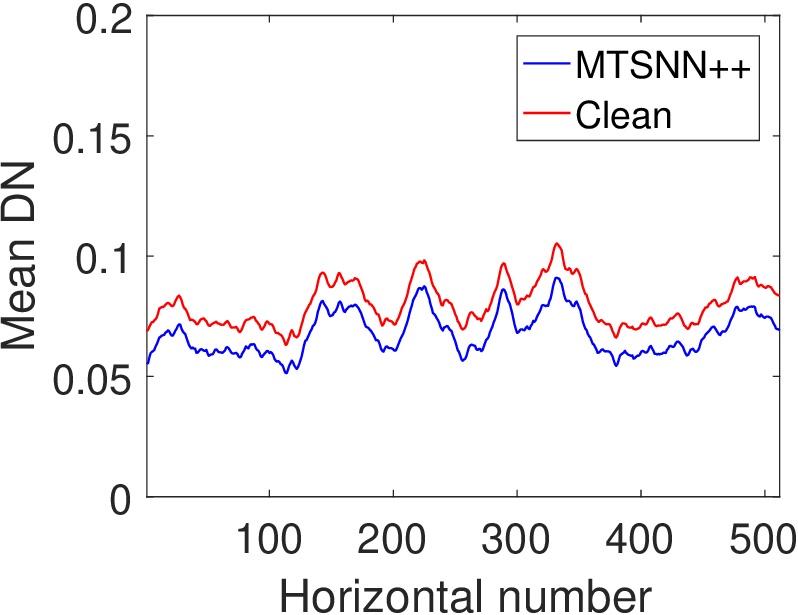}}
    \hspace{-1.1mm}
    \subfloat[]{\label{fig:houston_DECSC_reflectance}\includegraphics[width=0.140\linewidth]{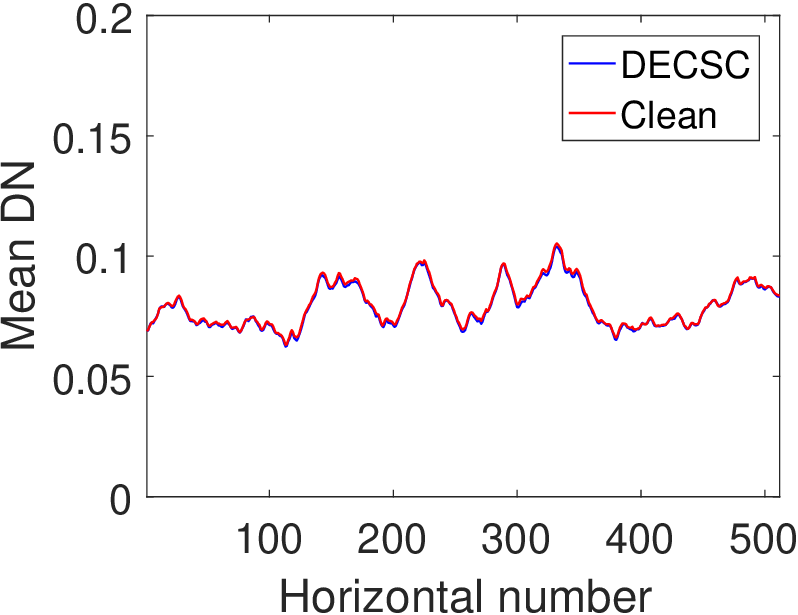}}
    %   \vspace{-0.2cm}
    \caption{Row mean profiles of band 39 for the Houston 2018 HSI under the mixture noise. (a) Noisy. (b) BM4D~\cite{Maggioni2013BM4D}. (c) MTSNMF~\cite{Ye2015MTSNMF}. (d) LLRT~\cite{Chang2017LLRT}. (e) NGMeet~\cite{He2022NGMeet}. (f) LRMR~\cite{Zhang2014LRMR}. (g) E-3DTV~\cite{Peng2020E-3DTV}. (h) 3DlogTNN~\cite{Zheng20203DlogTNN}. (i) SST~\cite{li2022spatialspectral}. (j) TRQ3D~\cite{Pang2022TRQ3DNet}. (k) SERT~\cite{li2023spectral}. (l) T3SC~\cite{bodrito2021T3SC}. (m) MTSNN++~\cite{xiong2023multitask}. (n) \textbf{DECSC}.} \label{fig:houston_row_mean_profiles}
 %    \vspace{-0.5cm}
 %   \vspace{-0.3cm}
\end{figure*}
\begin{figure*}[!t]
    \centering
    \subfloat[]{\label{fig:eo1_clean_visual}\includegraphics[width=0.1420\linewidth]{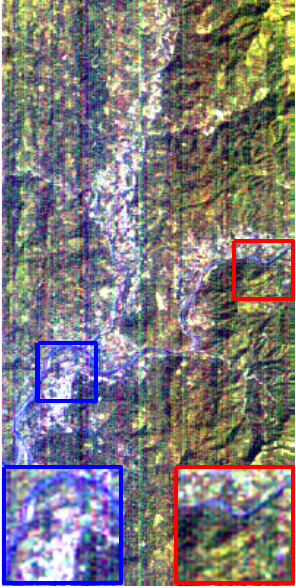}}
    \hspace{-1.1mm}
    \subfloat[]{\label{fig:eo1_BM4D_visual}\includegraphics[width=0.1420\linewidth]{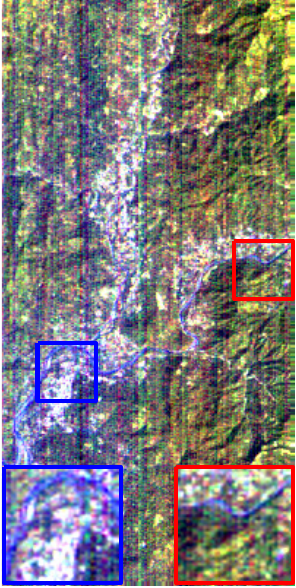}}
    \hspace{-1.1mm}
    \subfloat[]{\label{fig:eo1_MTSNMF_visual}\includegraphics[width=0.1420\linewidth]{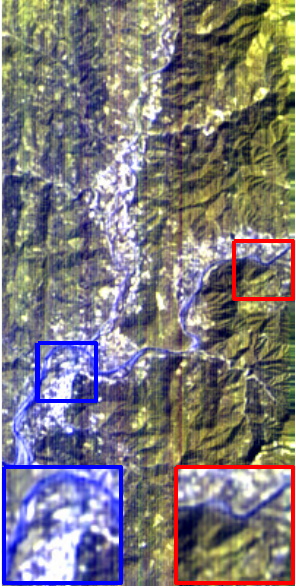}}
    \hspace{-1.1mm}
    \subfloat[]{\label{fig:eo1_LLRT_visual}\includegraphics[width=0.1420\linewidth]{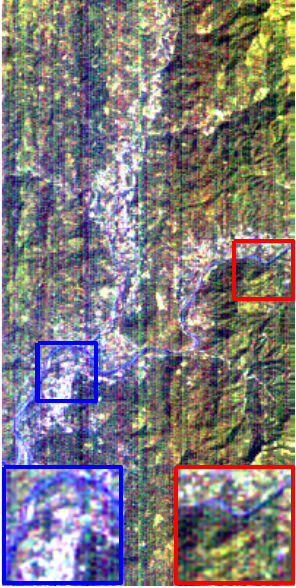}}
    \hspace{-1.1mm}
    \subfloat[]{\label{fig:eo1_NGMeet_visual}\includegraphics[width=0.1420\linewidth]{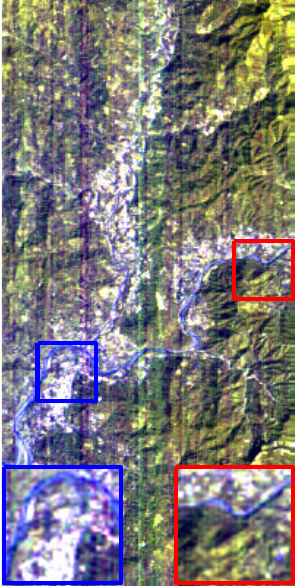}}
    \hspace{-1.1mm}
    \subfloat[]{\label{fig:eo1_LRMR_visual}\includegraphics[width=0.1420\linewidth]{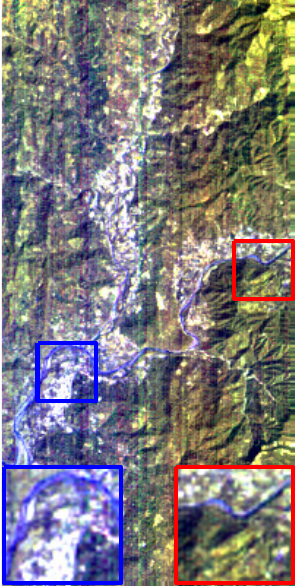}}
    \hspace{-1.1mm}
    \subfloat[]{\label{fig:eo1_E-3DTV_visual}\includegraphics[width=0.1420\linewidth]{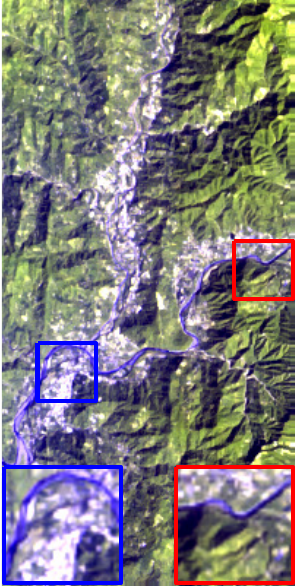}}
    \hspace{-1.1mm}
    \subfloat[]{\label{fig:eo1_3DlogTNN_visual}\includegraphics[width=0.1420\linewidth]{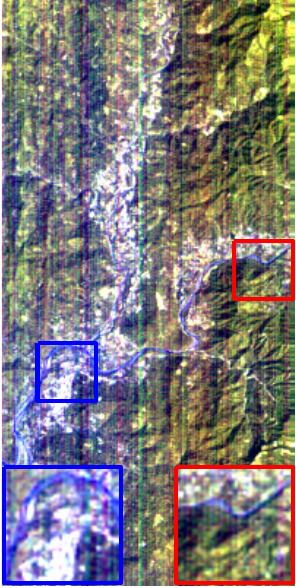}}
    \hspace{-1.1mm}
    \subfloat[]{\label{fig:eo1_SST_visual}\includegraphics[width=0.1420\linewidth]{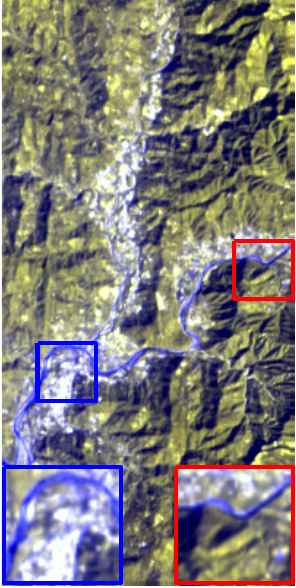}}
    \hspace{-1.1mm}
    \subfloat[]{\label{fig:eo1_TRQ3D_visual}\includegraphics[width=0.1420\linewidth]{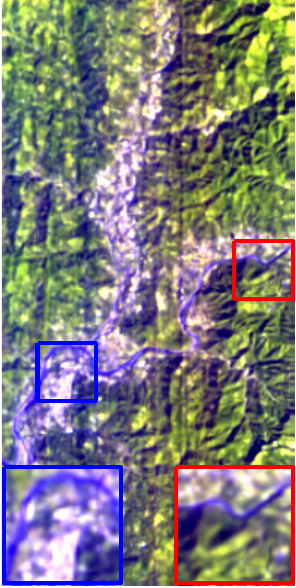}}
    \hspace{-1.1mm}
    \subfloat[]{\label{fig:eo1_SERT_visual}\includegraphics[width=0.1420\linewidth]{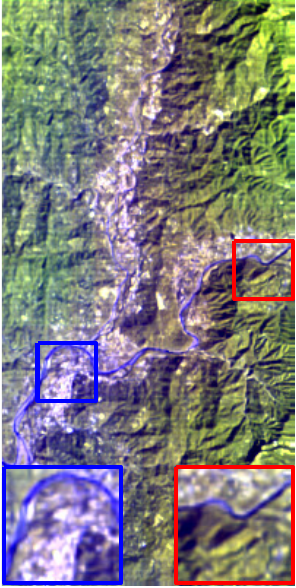}}
    \hspace{-1.1mm}
    \subfloat[]{\label{fig:eo1_T3SC_visual}\includegraphics[width=0.1420\linewidth]{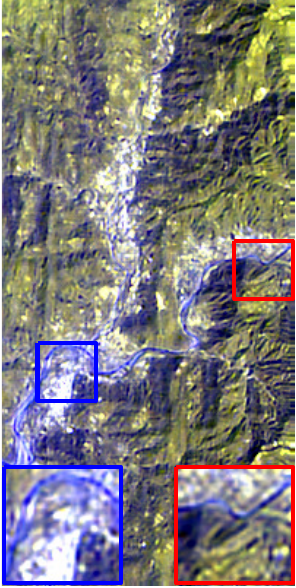}}
    \hspace{-1.1mm}
    \subfloat[]{\label{fig:eo1_MTSNN++_visual}\includegraphics[width=0.1420\linewidth]{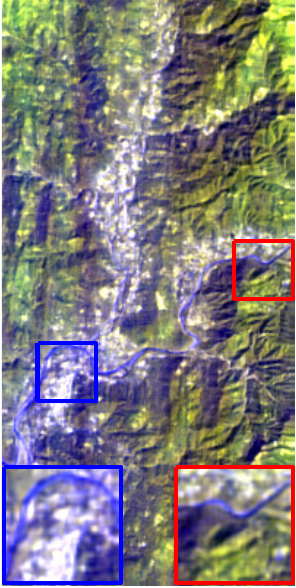}}
    \hspace{-1.1mm}
    \subfloat[]{\label{fig:eo1_DECSC_visual}\includegraphics[width=0.1420\linewidth]{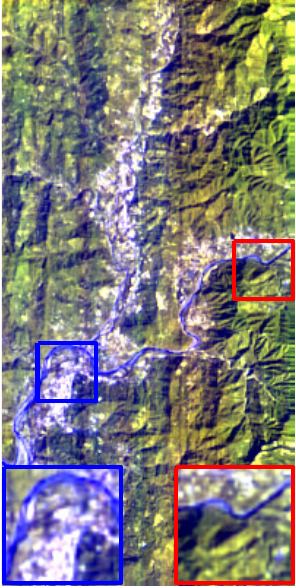}}
    %   \vspace{-0.2cm}
    \caption{Denoising results on the EO-1 HSI. The false-color images are generated by combining bands 97, 95, and 1. (a) Noisy. (b) BM4D~\cite{Maggioni2013BM4D}. (c) MTSNMF~\cite{Ye2015MTSNMF}. (d) LLRT~\cite{Chang2017LLRT}. (e) NGMeet~\cite{He2022NGMeet}. (f) LRMR~\cite{Zhang2014LRMR}. (g) E-3DTV~\cite{Peng2020E-3DTV}. (h) 3DlogTNN~\cite{Zheng20203DlogTNN}. (i) SST~\cite{li2022spatialspectral}. (j) TRQ3D~\cite{Pang2022TRQ3DNet}. (k) SERT~\cite{li2023spectral}. (l) T3SC~\cite{bodrito2021T3SC}. (m) MTSNN++~\cite{xiong2023multitask}. (n) \textbf{DECSC}.} \label{fig:eo1_visual}
 %    \vspace{-0.5cm}
 %   \vspace{-0.3cm}
\end{figure*}
The Houston 2018 HSI contains $1202 \times 4172$ pixels and spans 48 spectral bands ranging from 380 to 1050 nm. For testing, we selected the last 46 relatively clean bands and cropped a $512 \times 512$ patch from the central region as the test sample. The remaining regions were divided into $64 \times 64$ patches for model fine-tuning. Table~\ref{tab:houston} presents a quantitative comparison of denoising performance on the Houston 2018 HSI under various noise patterns. The results are largely consistent with those reported in Table~\ref{tab:icvl}. Despite differences in imaging conditions across datasets, fine-tuning enables both data-driven and hybrid-driven methods to outperform purely model-driven approaches. The proposed DECSC achieves the best performance under non-i.i.d. Gaussian and Mixture noise patterns, which we attribute to the robustness of its convolutional sparse dictionary specifically designed to capture GIC and LSU structures. Moreover, the DEQ model provides an infinite-depth network that inherits the robustness of the physical model while guaranteeing convergence.

\begin{figure*}[!t]
    \centering
    \subfloat[]{\label{fig:eo1_noise_reflectance}\includegraphics[width=0.1420\linewidth]{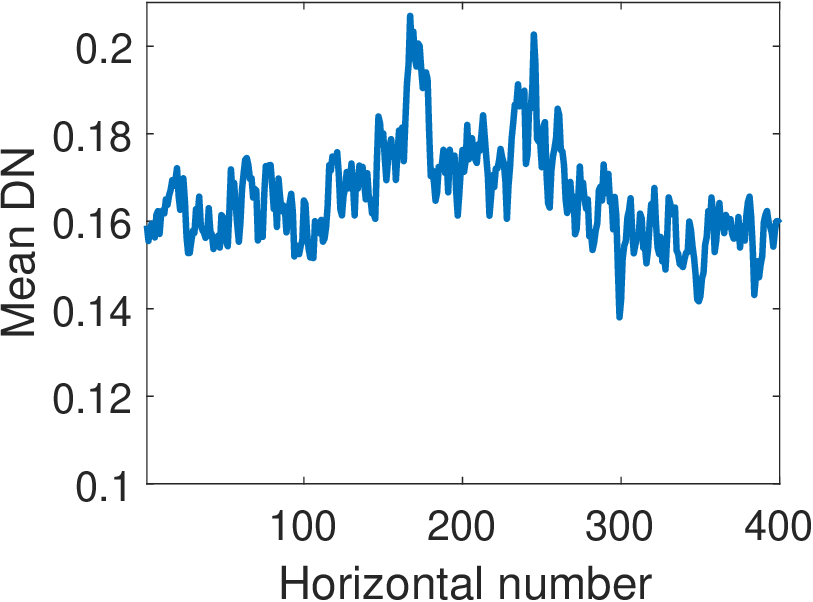}}
    \hspace{-1.1mm}
    \subfloat[]{\label{fig:eo1_BM4D_reflectance}\includegraphics[width=0.1420\linewidth]{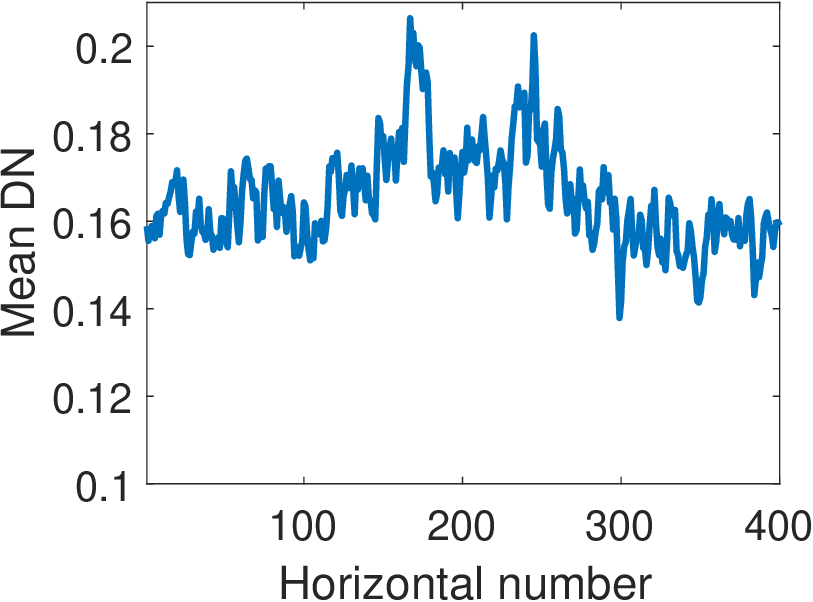}}
    \hspace{-1.1mm}
    \subfloat[]{\label{fig:eo1_MTSNMF_reflectance}\includegraphics[width=0.1420\linewidth]{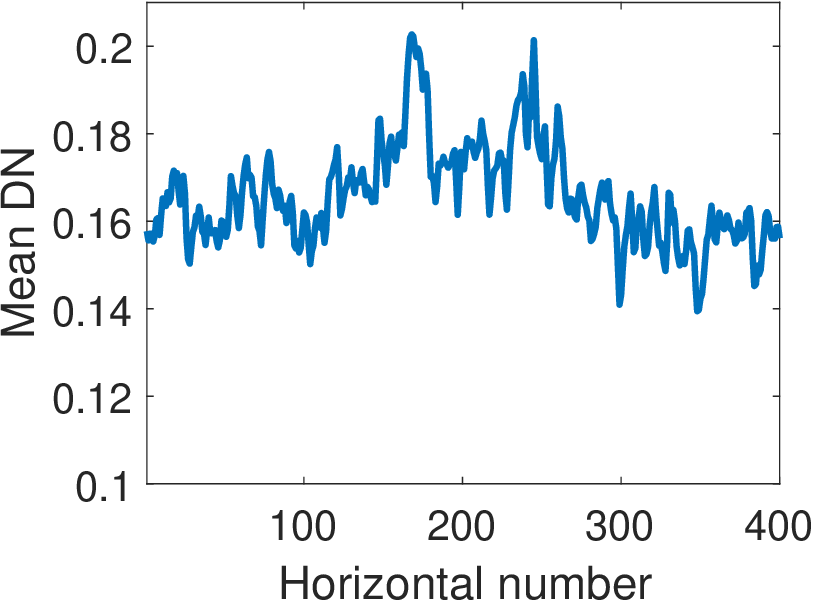}}
    \hspace{-1.1mm}
    \subfloat[]{\label{fig:eo1_LLRT_reflectance}\includegraphics[width=0.1420\linewidth]{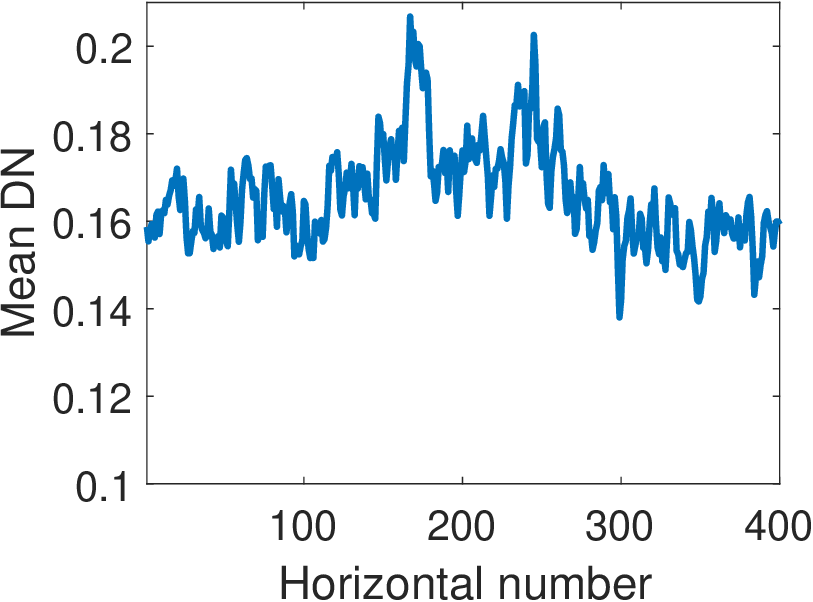}}
    \hspace{-1.1mm}
    \subfloat[]{\label{fig:eo1_NGMeet_reflectance}\includegraphics[width=0.1420\linewidth]{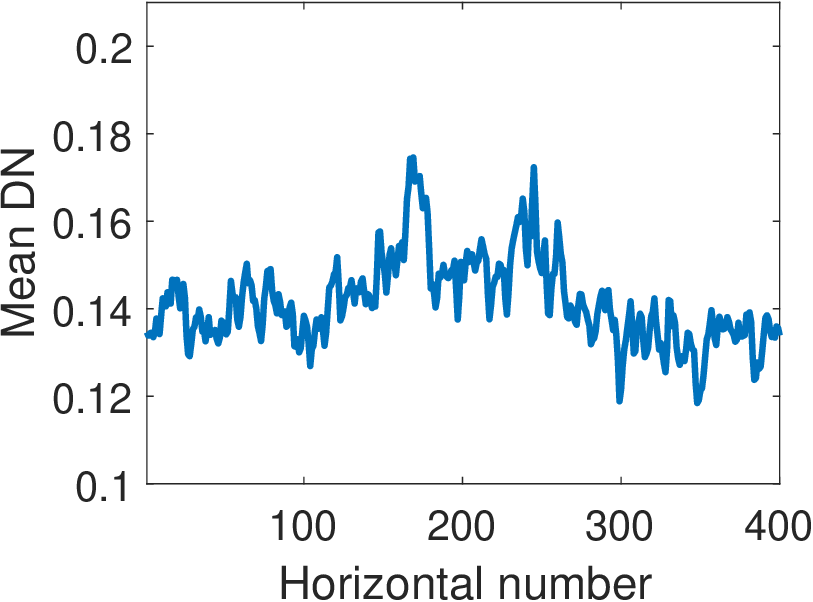}}
    \hspace{-1.1mm}
    \subfloat[]{\label{fig:eo1_LRMR_reflectance}\includegraphics[width=0.1420\linewidth]{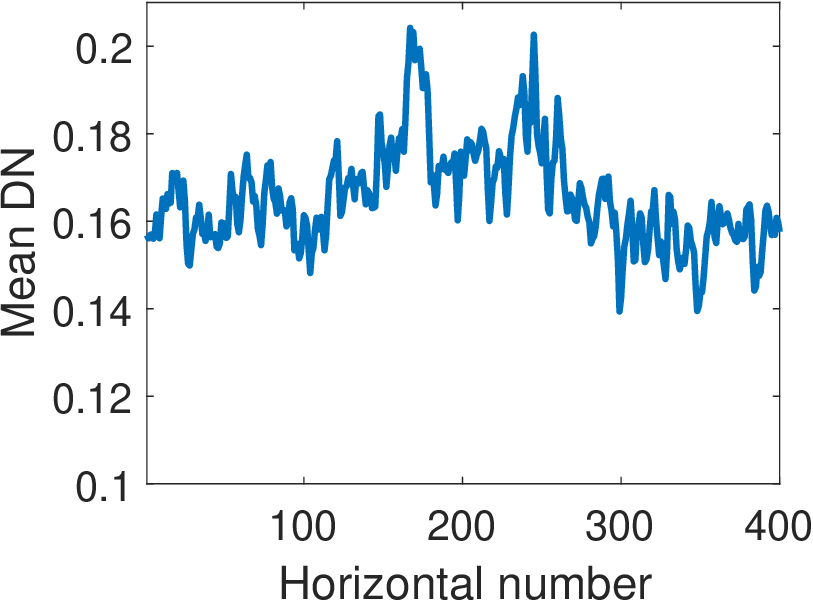}}
    \hspace{-1.1mm}
    \subfloat[]{\label{fig:eo1_E-3DTV_reflectance}\includegraphics[width=0.1420\linewidth]{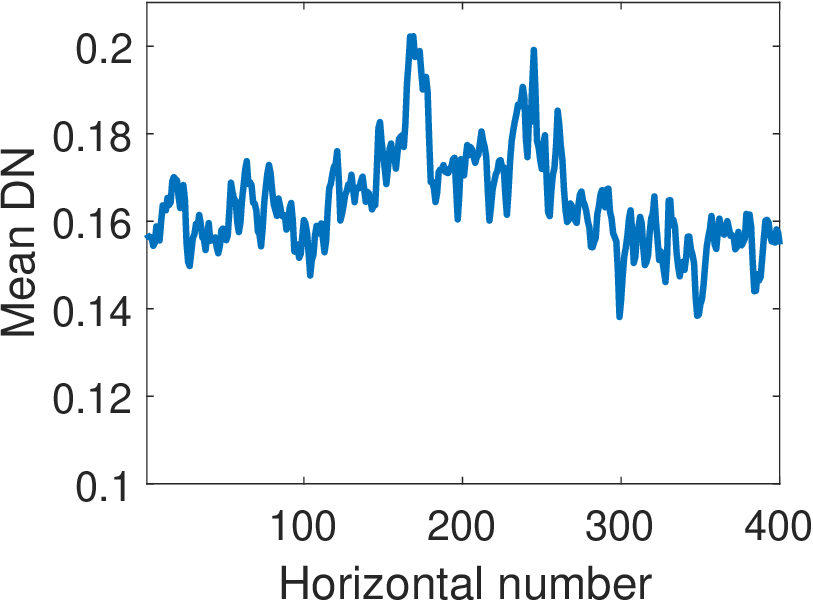}}
    \hspace{-1.1mm}
    \subfloat[]{\label{fig:eo1_3DlogTNN_reflectance}\includegraphics[width=0.1420\linewidth]{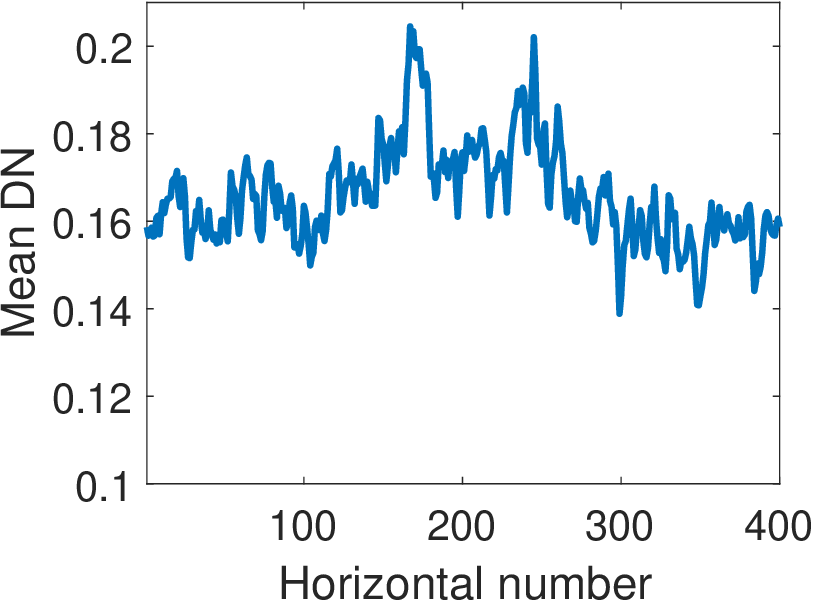}}
    \hspace{-1.1mm}
    \subfloat[]{\label{fig:eo1_SST_reflectance}\includegraphics[width=0.1420\linewidth]{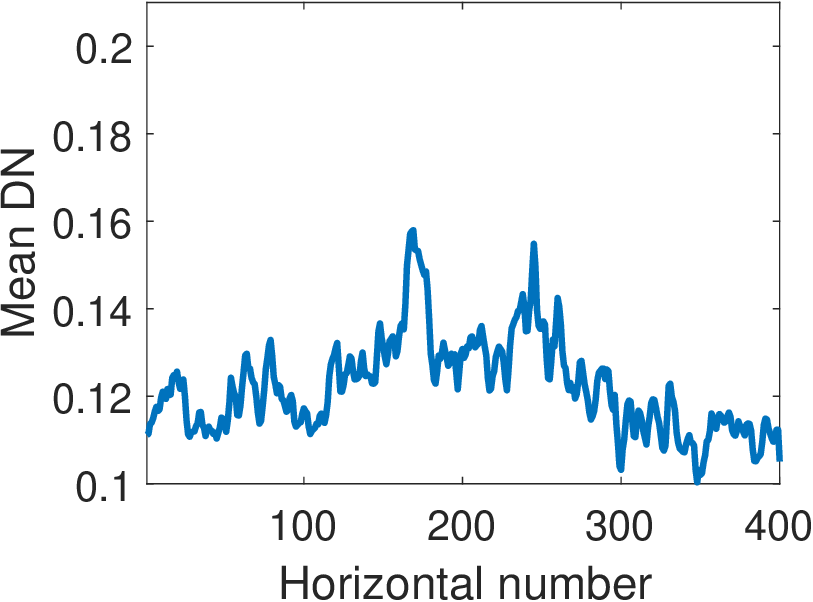}}
    \hspace{-1.1mm}
    \subfloat[]{\label{fig:eo1_TRQ3D_reflectance}\includegraphics[width=0.1420\linewidth]{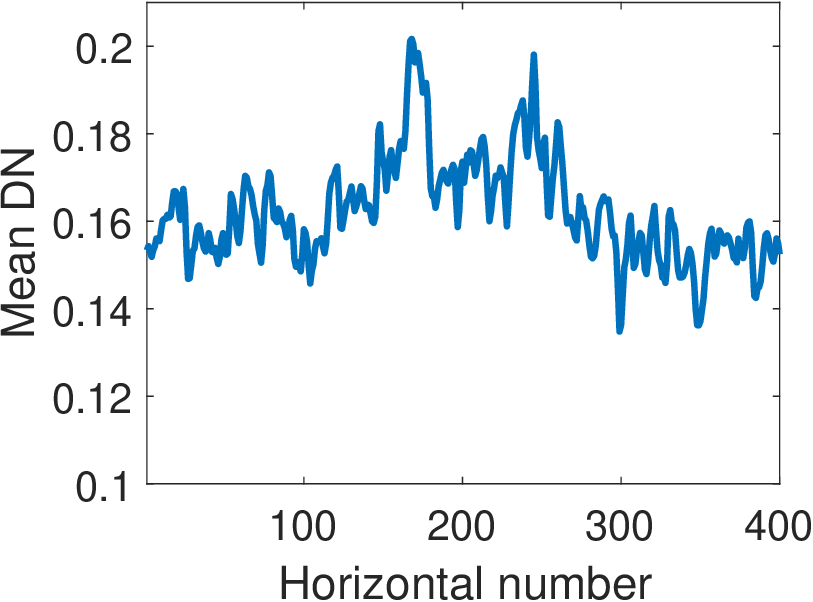}}
    \hspace{-1.1mm}
    \subfloat[]{\label{fig:eo1_SERT_reflectance}\includegraphics[width=0.1420\linewidth]{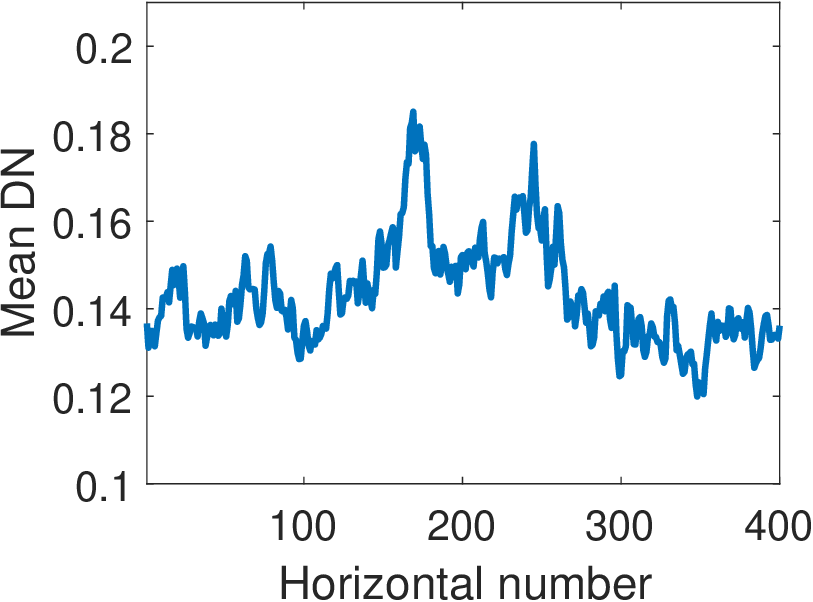}}
    \hspace{-1.1mm}
    \subfloat[]{\label{fig:eo1_T3SC_reflectance}\includegraphics[width=0.1420\linewidth]{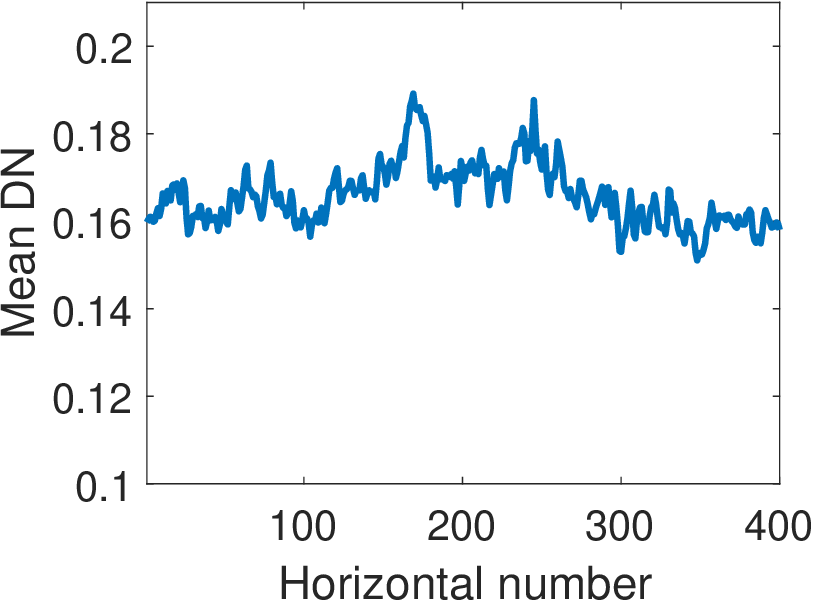}}
    \hspace{-1.1mm}
    \subfloat[]{\label{fig:eo1_MTSNN++_reflectance}\includegraphics[width=0.1420\linewidth]{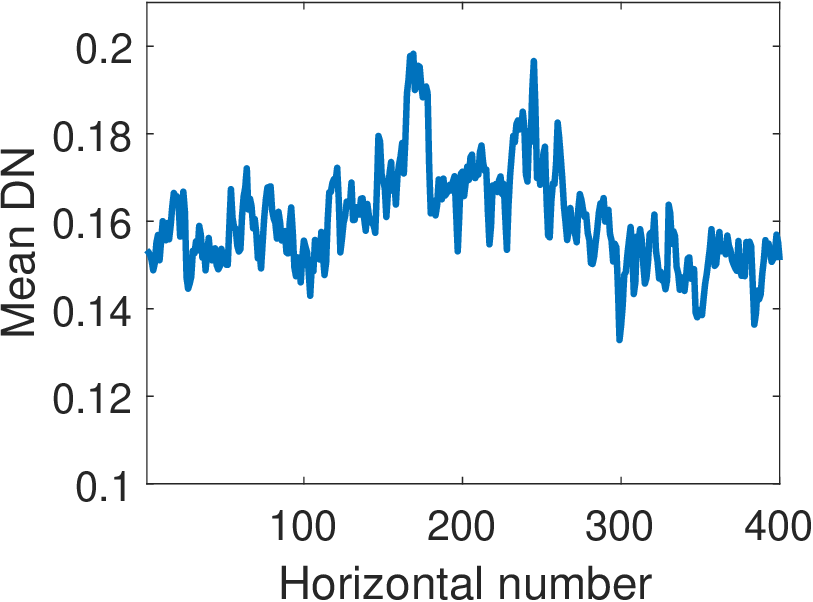}}
    \hspace{-1.1mm}
    \subfloat[]{\label{fig:eo1_DECSC_reflectance}\includegraphics[width=0.1420\linewidth]{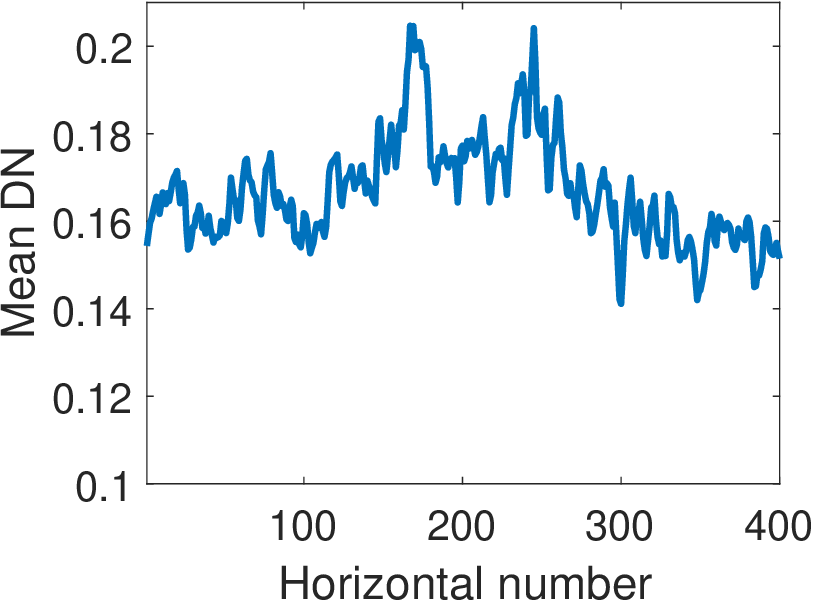}}
    %   \vspace{-0.2cm}
    \caption{Row mean profiles of band 159 for the EO-1 HSI. (a) Noisy. (b) BM4D~\cite{Maggioni2013BM4D}. (c) MTSNMF~\cite{Ye2015MTSNMF}. (d) LLRT~\cite{Chang2017LLRT}. (e) NGMeet~\cite{He2022NGMeet}. (f) LRMR~\cite{Zhang2014LRMR}. (g) E-3DTV~\cite{Peng2020E-3DTV}. (h) 3DlogTNN~\cite{Zheng20203DlogTNN}. (i) SST~\cite{li2022spatialspectral}. (j) TRQ3D~\cite{Pang2022TRQ3DNet}. (k) SERT~\cite{li2023spectral}. (l) T3SC~\cite{bodrito2021T3SC}. (m) MTSNN++~\cite{xiong2023multitask}. (n) \textbf{DECSC}.} \label{fig:eo1_row_mean_profiles}
 %    \vspace{-0.5cm}
 %   \vspace{-0.3cm}
\end{figure*}

\begin{figure*}[!t]
    \centering
    \subfloat[]{\label{fig:capitalairport_noise_visual}\includegraphics[width=0.1420\linewidth]{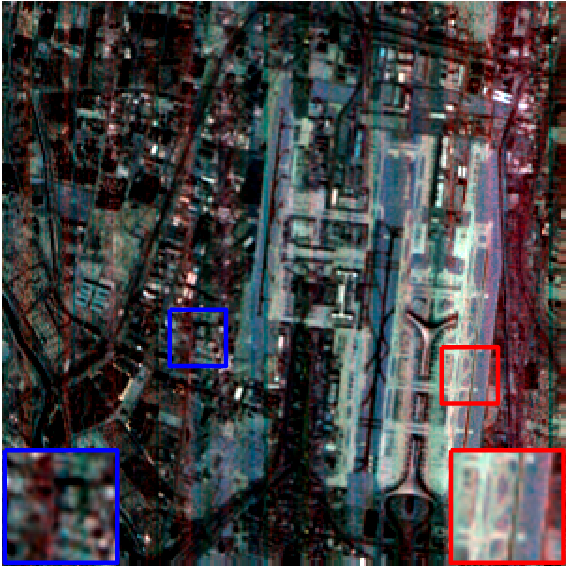}}
    \hspace{-1.1mm}
    \subfloat[]{\label{fig:capitalairport_BM4D_visual}\includegraphics[width=0.1420\linewidth]{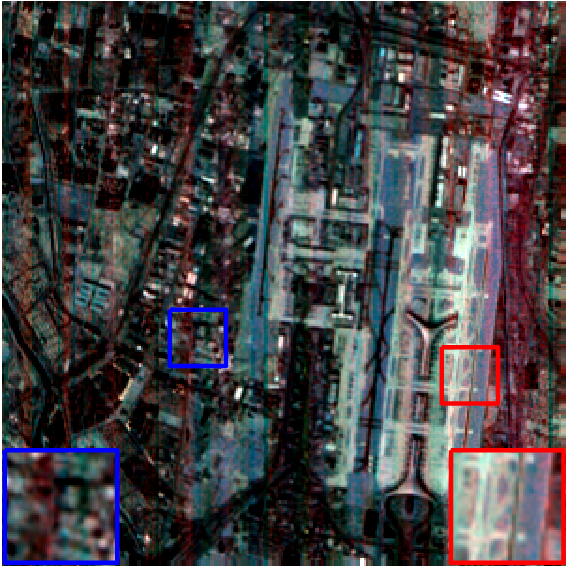}}
    \hspace{-1.1mm}
    \subfloat[]{\label{fig:capitalairport_MTSNMF_visual}\includegraphics[width=0.1420\linewidth]{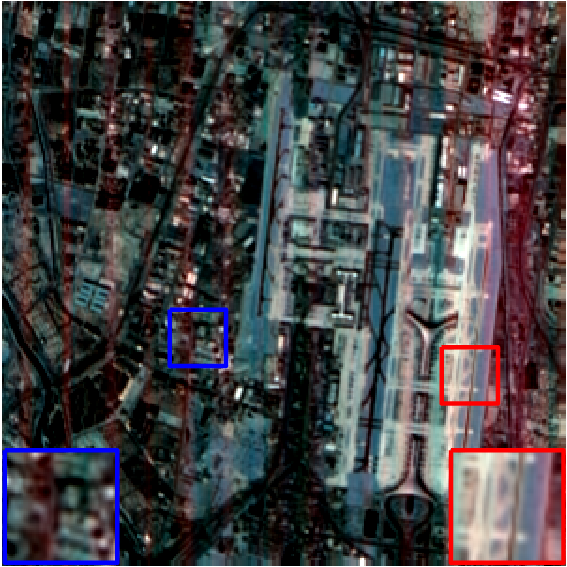}}
    \hspace{-1.1mm}
    \subfloat[]{\label{fig:capitalairport_LLRT_visual}\includegraphics[width=0.1420\linewidth]{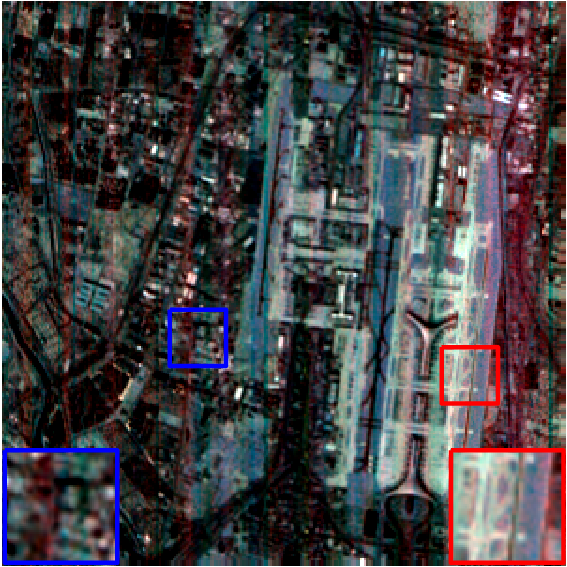}}
    \hspace{-1.1mm}
    \subfloat[]{\label{fig:capitalairport_NGMeet_visual}\includegraphics[width=0.1420\linewidth]{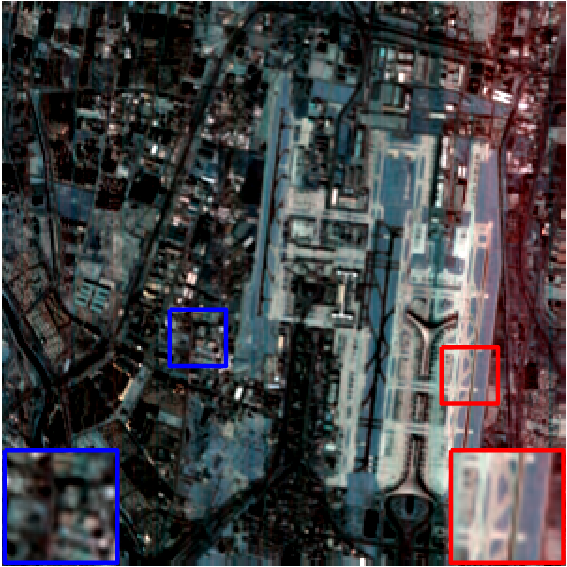}}
    \hspace{-1.1mm}
    \subfloat[]{\label{fig:capitalairport_LRMR_visual}\includegraphics[width=0.1420\linewidth]{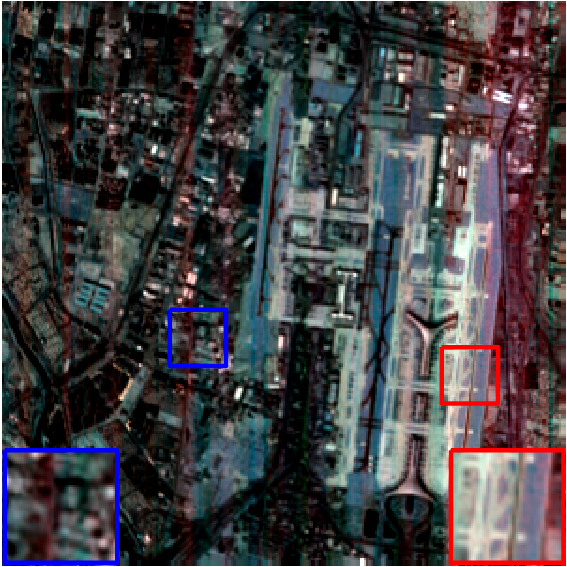}}
    \hspace{-1.1mm}
    \subfloat[]{\label{fig:capitalairport_E-3DTV_visual}\includegraphics[width=0.1420\linewidth]{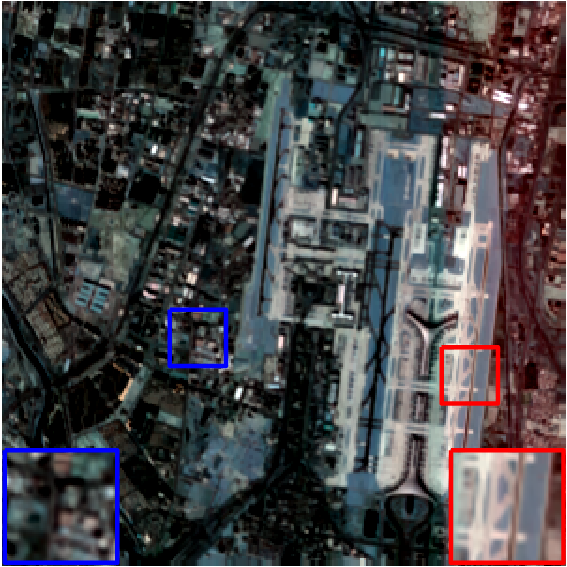}}
    \hspace{-1.1mm}
    \subfloat[]{\label{fig:capitalairport_3DlogTNN_visual}\includegraphics[width=0.1420\linewidth]{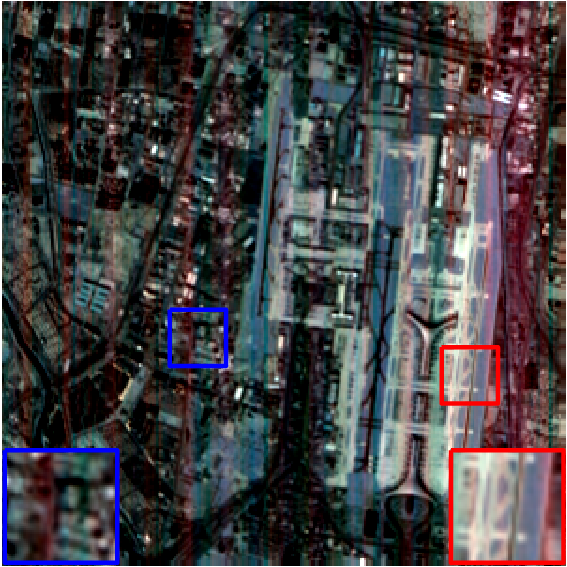}}
    \hspace{-1.1mm}
    \subfloat[]{\label{fig:capitalairport_SST_visual}\includegraphics[width=0.1420\linewidth]{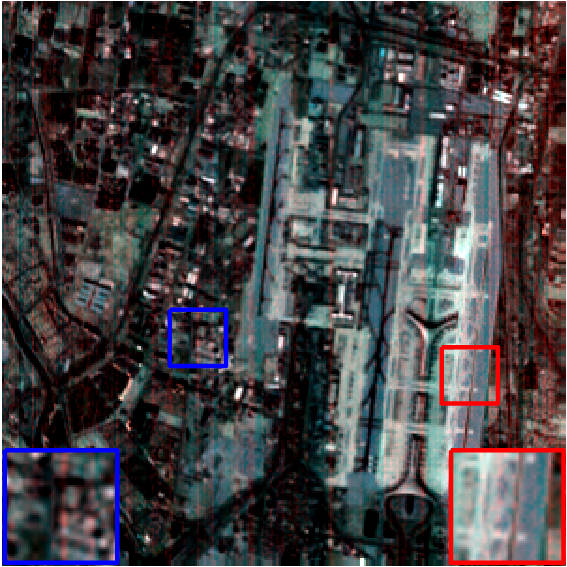}}
    \hspace{-1.1mm}
    \subfloat[]{\label{fig:capitalairport_TRQ3D_visual}\includegraphics[width=0.1420\linewidth]{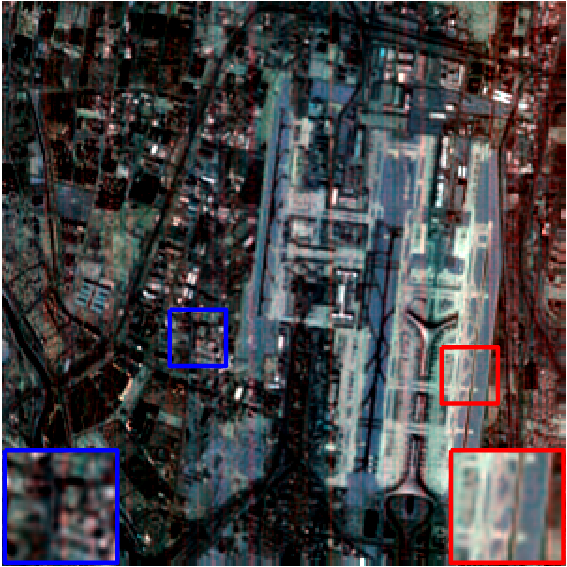}}
    \hspace{-1.1mm}
    \subfloat[]{\label{fig:capitalairport_SERT_visual}\includegraphics[width=0.1420\linewidth]{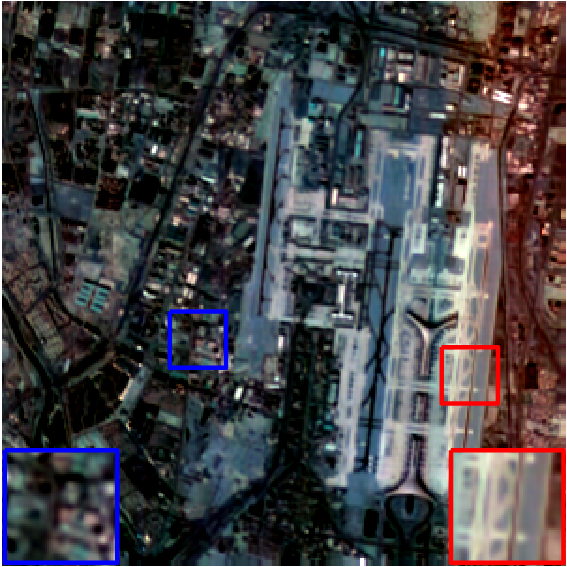}}
    \hspace{-1.1mm}
    \subfloat[]{\label{fig:capitalairport_T3SC_visual}\includegraphics[width=0.1420\linewidth]{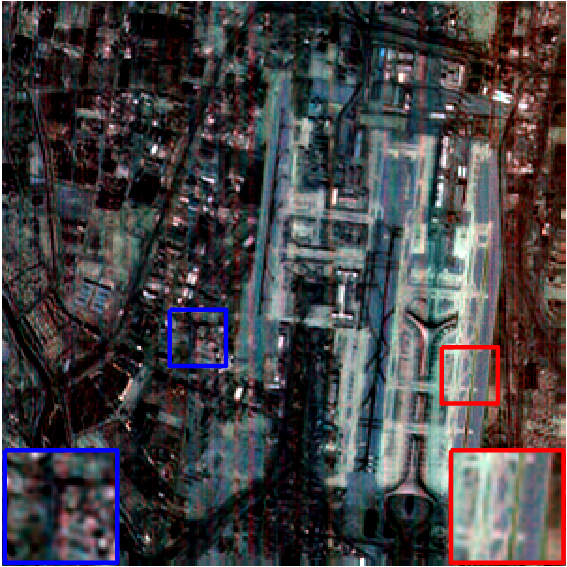}}
    \hspace{-1.1mm}
    \subfloat[]{\label{fig:capitalairport_MTSNN++_visual}\includegraphics[width=0.1420\linewidth]{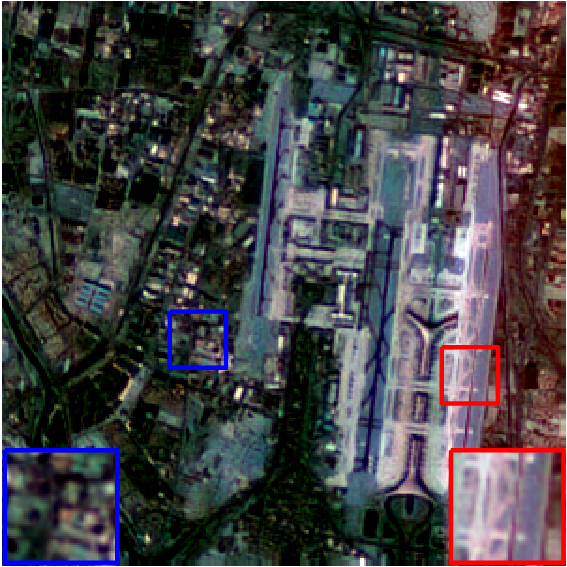}}
    \hspace{-1.1mm}
    \subfloat[]{\label{fig:capitalairport_DECSC_visual}\includegraphics[width=0.1420\linewidth]{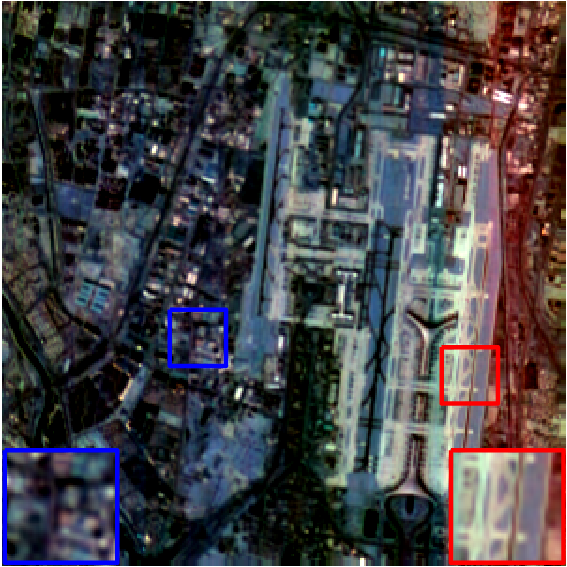}}
    %   \vspace{-0.2cm}
    \caption{Denoising results of CapitalAirport HSI collected from the GF-5 satellite. The false-color images are generated by combining bands 153, 107, and 94. (a) Noisy. (b) BM4D~\cite{Maggioni2013BM4D}. (c) MTSNMF~\cite{Ye2015MTSNMF}. (d) LLRT~\cite{Chang2017LLRT}. (e) NGMeet~\cite{He2022NGMeet}. (f) LRMR~\cite{Zhang2014LRMR}. (g) E-3DTV~\cite{Peng2020E-3DTV}. (h) 3DlogTNN~\cite{Zheng20203DlogTNN}. (i) SST~\cite{li2022spatialspectral}. (j) TRQ3D~\cite{Pang2022TRQ3DNet}. (k) SERT~\cite{li2023spectral}. (l) T3SC~\cite{bodrito2021T3SC}. (m) MTSNN++~\cite{xiong2023multitask}. (n) \textbf{DECSC}.} \label{fig:capitalairport_visual}
 %    \vspace{-0.5cm}
 %   \vspace{-0.3cm}
\end{figure*}

Fig.~\ref{fig:houston_visual} compares the restoration results of different methods on the Houston 2018 HSI. As a remote-sensing dataset, the Houston HSI is typically affected by mixture noise; thus, Fig.~\ref{fig:houston_visual} focuses on restoration performance under the Mixture noise pattern. Most model-driven methods, which are primarily designed for Gaussian noise, struggle with complex noise types, leading to lower visual quality. In contrast, data-driven and hybrid-driven methods generally yield better visual results due to the strong representational capacity of DNNs. Nevertheless, color distortions and detail over-smoothing remain common issues in many of these methods. Among all compared approaches, SERT and the proposed DECSC demonstrate visual results that are most consistent with the ground truth.  Furthermore, Fig.~\ref{fig:houston_row_mean_profiles} compares the row mean profiles of the restoration results produced by different methods. The mixture noise poses a significant challenge for most methods, leading to evident overall shifts or residual noise in the restored results. Although MTSNMF yields reconstructions that are relatively closer to the clean HSI in terms of overall structure, noticeable local deviations still persist. In contrast, DECSC consistently produces profiles that are more aligned with the clean HSI.
% Furthermore, Fig.~\ref{fig:houston_reflectance} compares the spectral reflectance curves recovered by different methods, revealing that only SERT and DECSC accurately restore the spectral information, which corroborates the superior visual quality observed in Fig.~\ref{fig:houston_visual}.

\subsection{Real-world Noise Removal}

To further evaluate the denoising performance of DECSC in real-world scenarios, we selected two real-world remote sensing HSIs for experiments involving real-world noise removal. Due to the absence of corresponding ground truth, the evaluation was limited to a qualitative analysis of the restoration results.
% To further assess the denoising capability of the DECSC in real-world scenarios, we selected two real-world remote sensing HSIs for experiments on real-world noise removal. Due to the lack of corresponding ground truth, we conducted only a qualitative analysis of the restoration results.

\subsubsection{Earth Observing-1 (EO-1) HSI}

We selected an HSI captured by the EO-1 satellite, which covers a spectral range of 400 to 2500 nm. Following the experimental setup of Zhang~\emph{et al.}~\cite{Zhang2014LRMR}, a sub-image of size $200 \times 400 \times 166$ was used for testing. Fig.~\ref{fig:eo1_visual} compares the visual quality of restoration results produced by different methods. Most model-driven methods struggle to effectively remove stripe noise. E-3DTV, benefiting from its ability to capture correlations and differences across bands, produces relatively clear results even under severe noise. The differences between synthetic and real-world noise pose challenges for data-driven and hybrid-driven methods, leading to blurring in the results of methods like SST, TRQ3D, T3SC, and MTSNN++. In contrast, SERT and DECSC not only achieve effective denoising but also preserve well-defined edges and structural integrity in the reconstructions. The superior performance of SERT can be attributed to its rectangle self-attention mechanism, while DECSC's advantage lies in the difference convolution's ability to capture edge information, complemented by an attention mechanism that adaptively enhances critical features. Fig.~\ref{fig:eo1_row_mean_profiles} shows the row mean profiles of band 159 for the EO-1 HSI. Although clean ground truth is unavailable in this real-world scenario, the profiles still provide insight into the smoothness and consistency of the restored results. E-3DTV and DECSC achieve a favorable balance between smoothness and consistency with the value range of the noisy input, validating their robustness and effectiveness in practical scenarios.

% As shown in Fig.~\ref{fig:eo1_reflectance}, the spectral curves recovered by DECSC strike a good balance between smoothness and the preservation of fine spectral details, consistent with the visual observations discussed above.

\subsubsection{Gaofen-5 (GF-5) CapitalAirport HSI}

\begin{figure*}[!t]
    \centering
    \subfloat[]{\label{fig:capitalairport_noise_reflectance}\includegraphics[width=0.1420\linewidth]{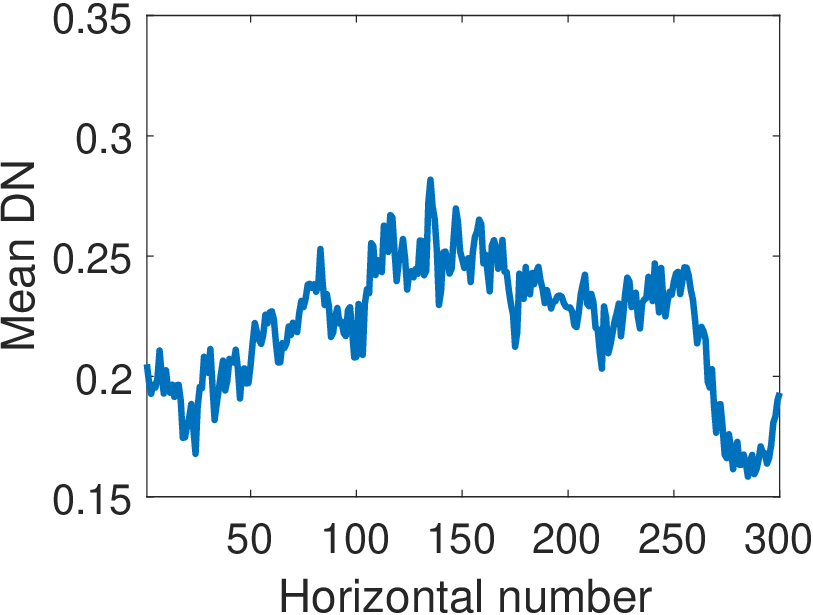}}
    \hspace{-1.1mm}
    \subfloat[]{\label{fig:capitalairport_BM4D_reflectance}\includegraphics[width=0.1420\linewidth]{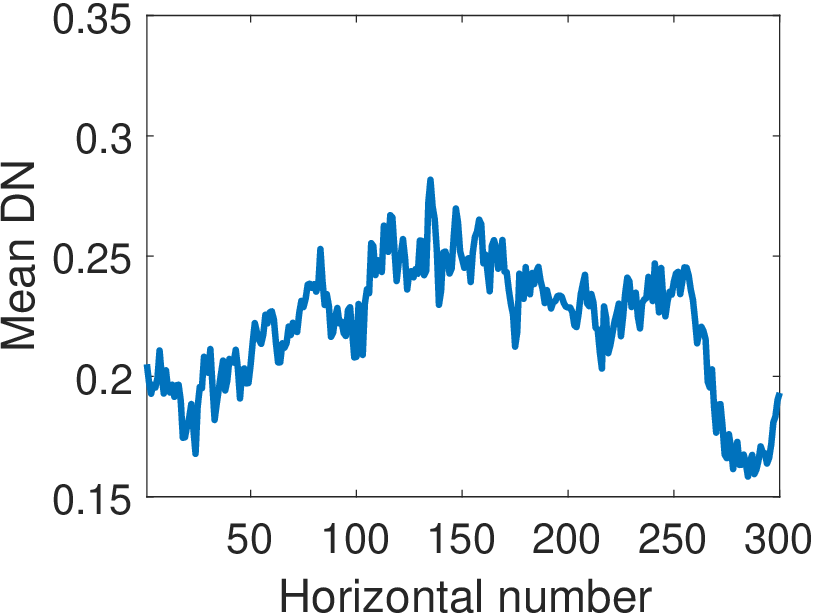}}
    \hspace{-1.1mm}
    \subfloat[]{\label{fig:capitalairport_MTSNMF_reflectance}\includegraphics[width=0.1420\linewidth]{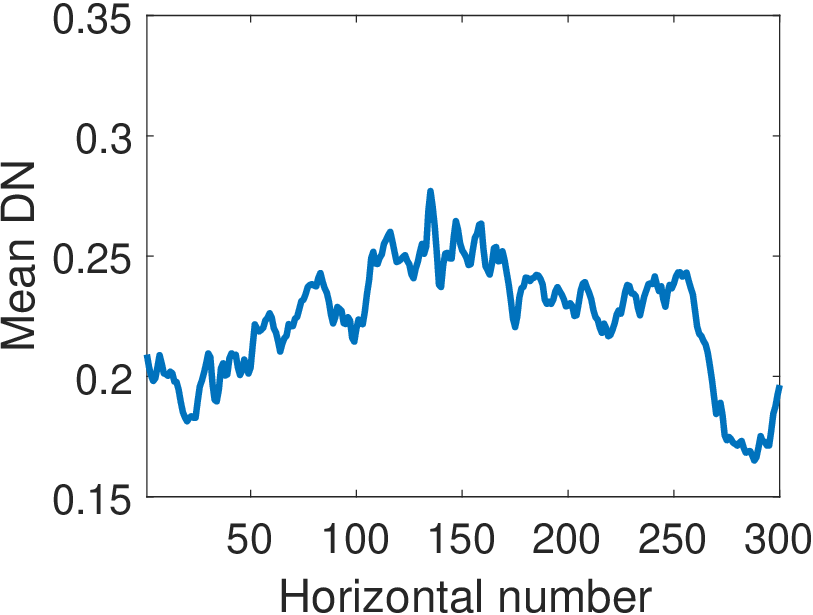}}
    \hspace{-1.1mm}
    \subfloat[]{\label{fig:capitalairport_LLRT_reflectance}\includegraphics[width=0.1420\linewidth]{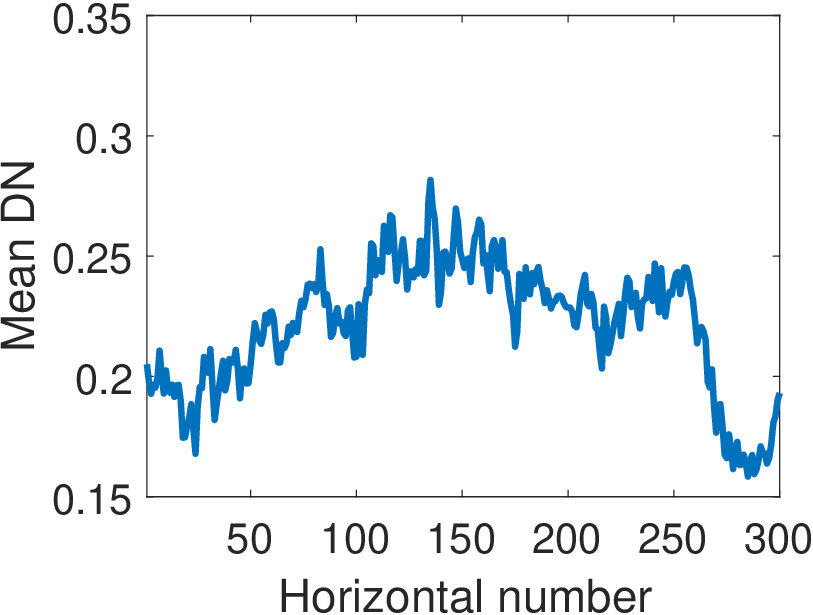}}
    \hspace{-1.1mm}
    \subfloat[]{\label{fig:capitalairport_NGMeet_reflectance}\includegraphics[width=0.1420\linewidth]{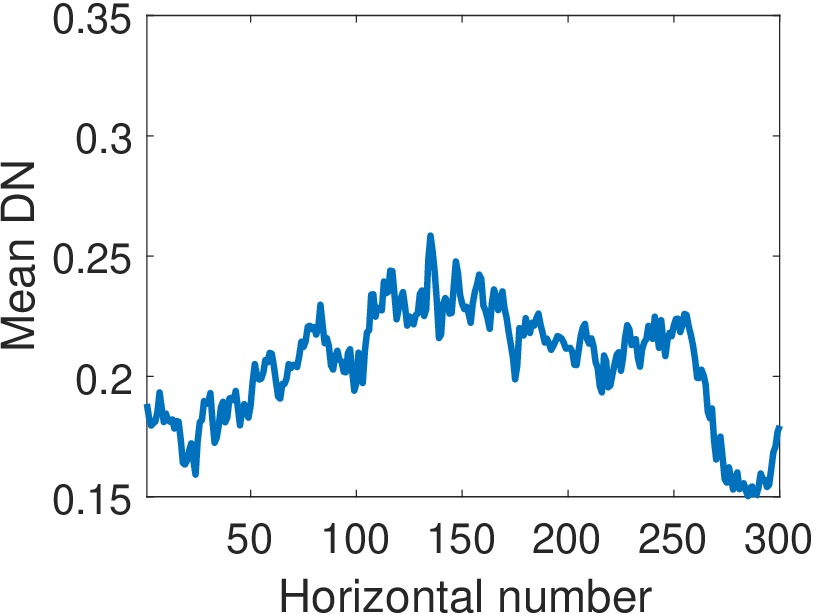}}
    \hspace{-1.1mm}
    \subfloat[]{\label{fig:capitalairport_LRMR_reflectance}\includegraphics[width=0.1420\linewidth]{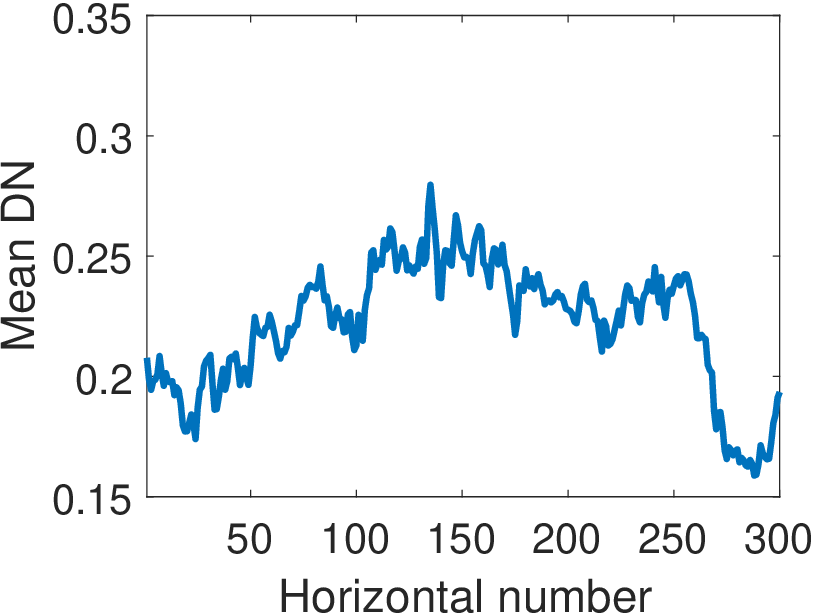}}
    \hspace{-1.1mm}
    \subfloat[]{\label{fig:capitalairport_E-3DTV_reflectance}\includegraphics[width=0.1420\linewidth]{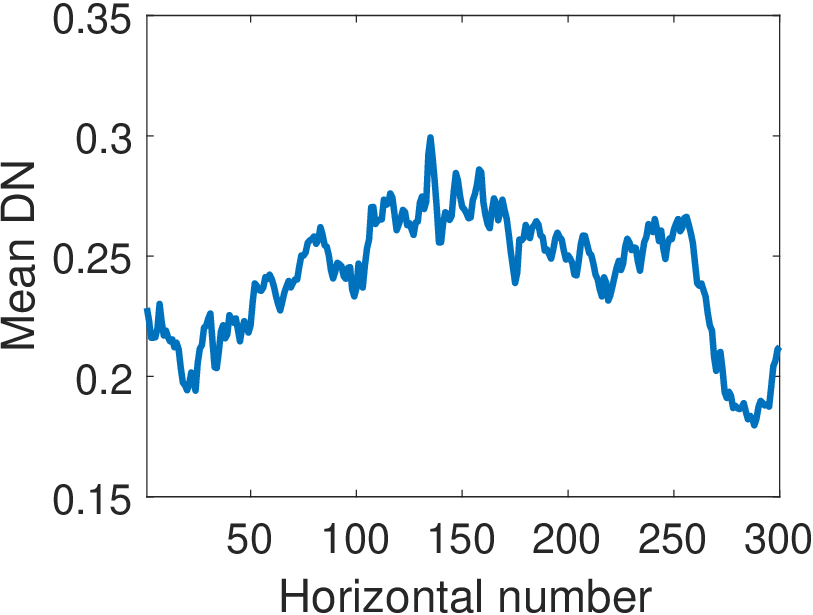}}
    \hspace{-1.1mm}
    \subfloat[]{\label{fig:capitalairport_3DlogTNN_reflectance}\includegraphics[width=0.1420\linewidth]{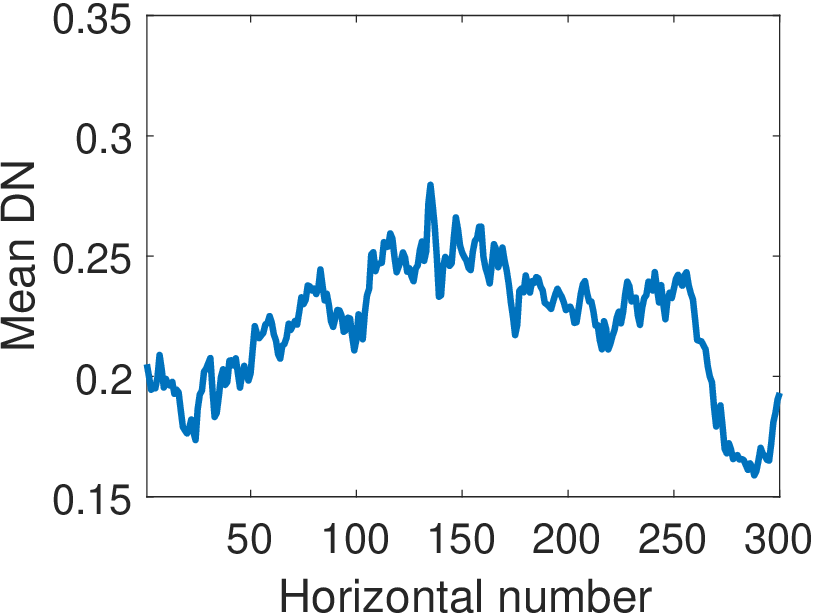}}
    \hspace{-1.1mm}
    \subfloat[]{\label{fig:capitalairport_SST_reflectance}\includegraphics[width=0.1420\linewidth]{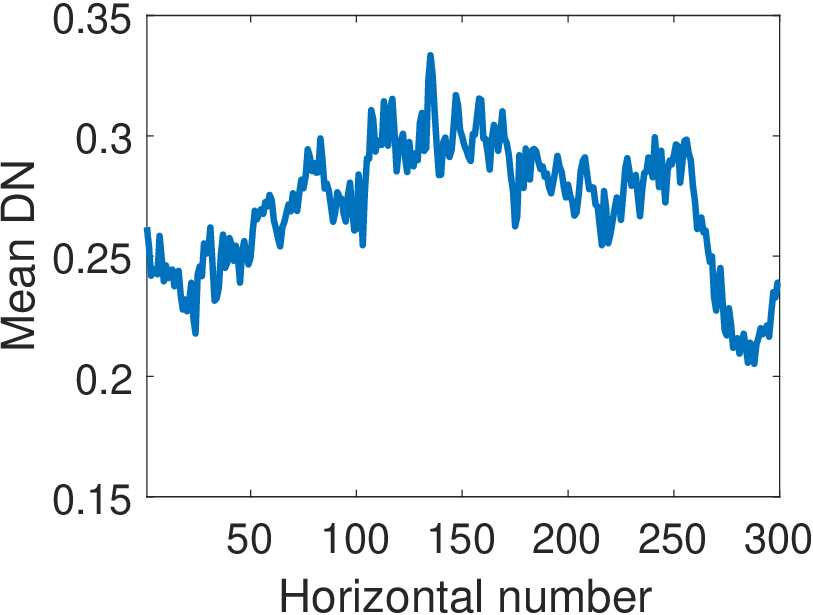}}
    \hspace{-1.1mm}
    \subfloat[]{\label{fig:capitalairport_TRQ3D_reflectance}\includegraphics[width=0.1420\linewidth]{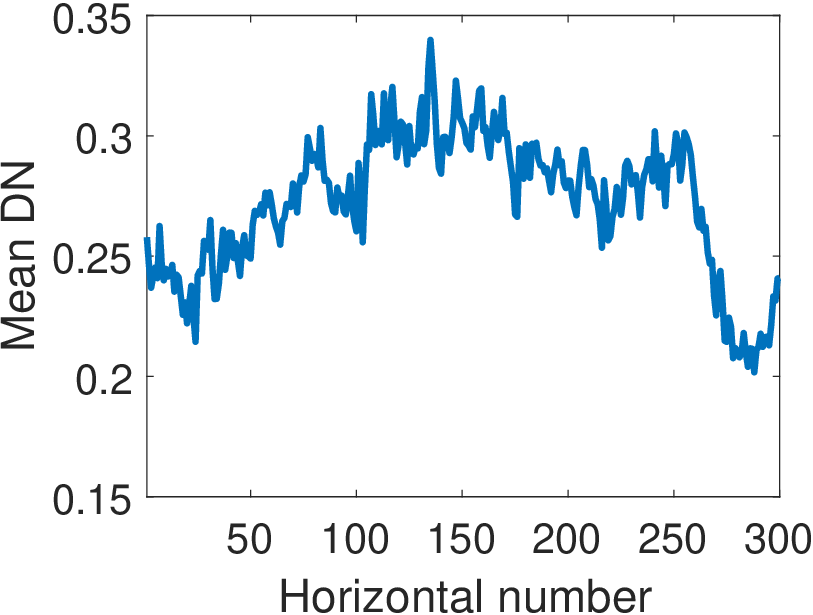}}
    \hspace{-1.1mm}
    \subfloat[]{\label{fig:capitalairport_SERT_reflectance}\includegraphics[width=0.1420\linewidth]{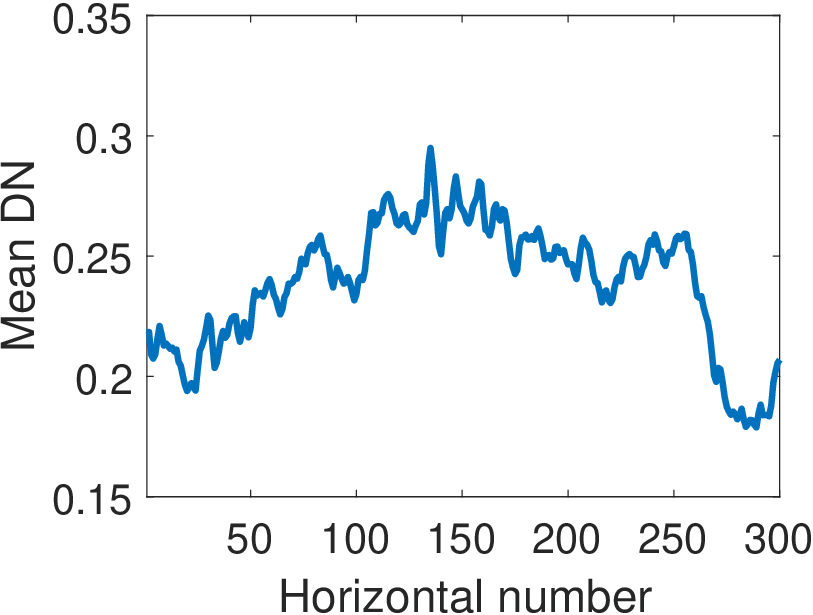}}
    \hspace{-1.1mm}
    \subfloat[]{\label{fig:capitalairport_T3SC_reflectance}\includegraphics[width=0.1420\linewidth]{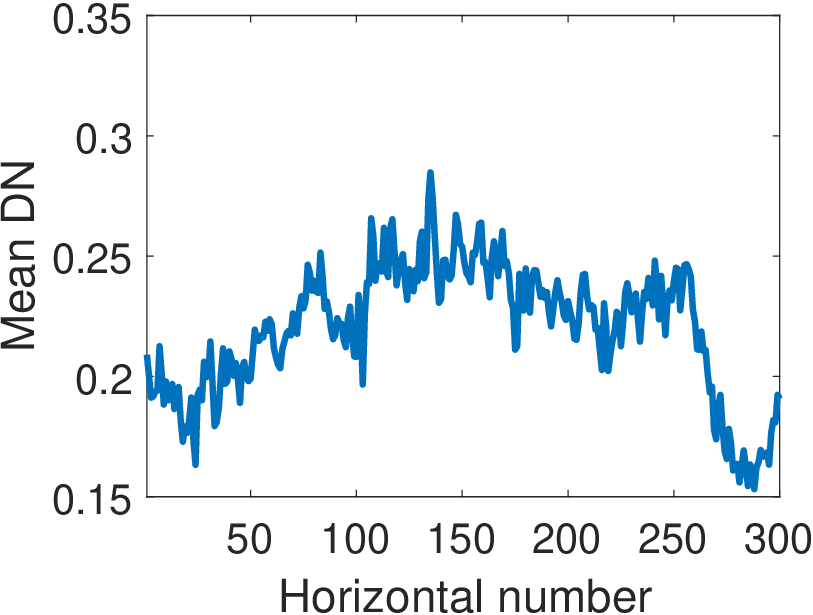}}
    \hspace{-1.1mm}
    \subfloat[]{\label{fig:capitalairport_MTSNN++_reflectance}\includegraphics[width=0.1420\linewidth]{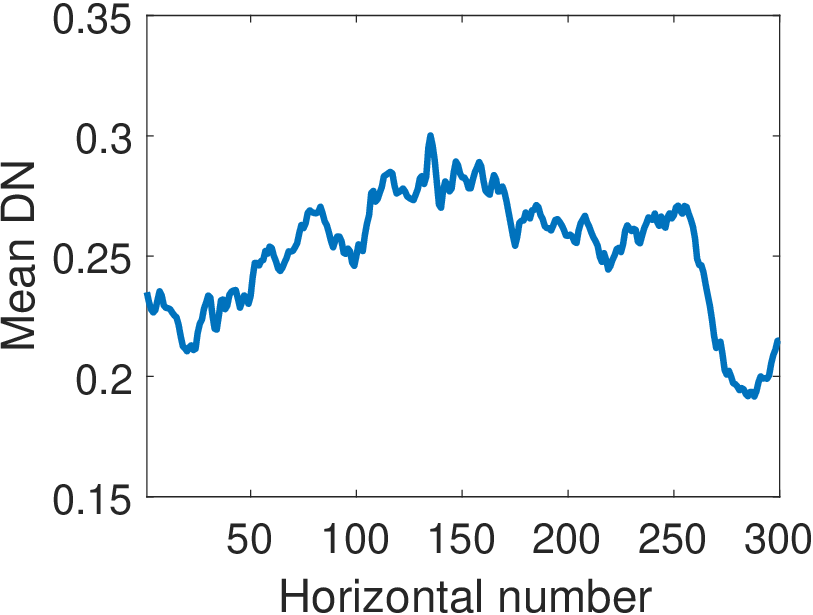}}
    \hspace{-1.1mm}
    \subfloat[]{\label{fig:capitalairport_DECSC_reflectance}\includegraphics[width=0.1420\linewidth]{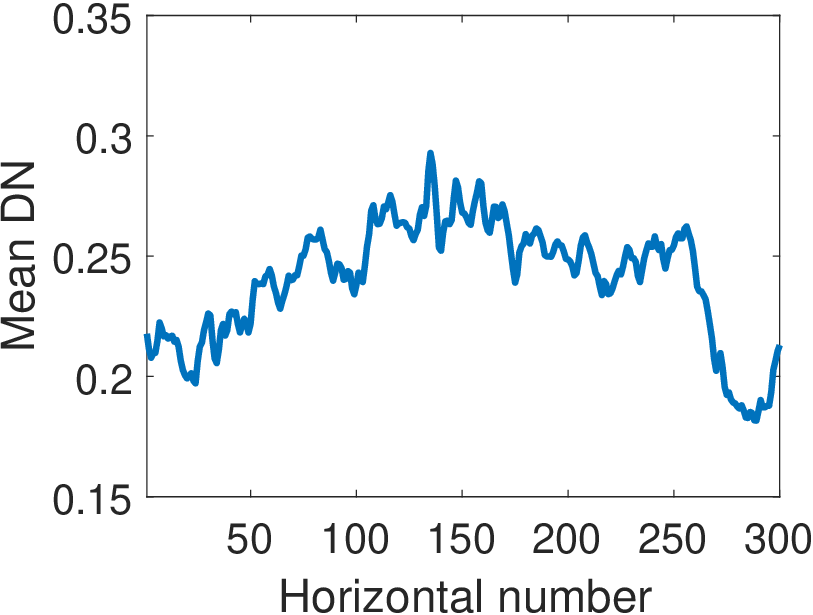}}
    %   \vspace{-0.2cm}
    \caption{Row mean profiles of band 154 for the CapitalAirport HSI collected from the GF-5 satellite. (a) Noisy. (b) BM4D~\cite{Maggioni2013BM4D}. (c) MTSNMF~\cite{Ye2015MTSNMF}. (d) LLRT~\cite{Chang2017LLRT}. (e) NGMeet~\cite{He2022NGMeet}. (f) LRMR~\cite{Zhang2014LRMR}. (g) E-3DTV~\cite{Peng2020E-3DTV}. (h) 3DlogTNN~\cite{Zheng20203DlogTNN}. (i) SST~\cite{li2022spatialspectral}. (j) TRQ3D~\cite{Pang2022TRQ3DNet}. (k) SERT~\cite{li2023spectral}. (l) T3SC~\cite{bodrito2021T3SC}. (m) MTSNN++~\cite{xiong2023multitask}. (n) \textbf{DECSC}.} \label{fig:capitalairport_row_mean_profiles}
 %    \vspace{-0.5cm}
 %   \vspace{-0.3cm}
\end{figure*}

We selected an HSI captured by the GF-5 satellite, covering a spectral range of 400 to 2500 nm, for real-world noise removal experiments. In this study, a sub-image of size $300 \times 300 \times 166$ was extracted for testing. Shown in Fig.~\ref{fig:capitalairport_visual} and~\ref{fig:capitalairport_row_mean_profiles},   MTSNMF, SERT, and DECSC yield results whose value ranges are more consistent with the noisy input.

% As shown in the spectral curves in Fig.~\ref{fig:capitalairport_reflectance}, the results reconstructed by E-3DTV and several data-driven or hybrid-driven methods tend to be overly smooth, whereas those produced by most model-driven methods exhibit noticeable fluctuations. In contrast, DECSC achieves a better balance between smoothness and the preservation of fine spectral variations in its restoration results.

\subsection{Network Analysis}
In this section, we conduct a further analysis of the DEQCSC components and hyperparameter settings. All experiments are performed on the ICVL dataset under the Non-i.i.d. Gaussian noise setting with noise levels in the range [0,95].
\subsubsection{Ablation Study}
We conduct ablation studies on key components of our network to validate the effectiveness of the proposed design choices. Specifically, we evaluate the contributions of the Swin Transformer and the detail enhancement module by independently removing the Swin Transformer, the difference convolution, and the attention mechanism. Each variant was trained independently under identical settings. As shown in Table~\ref{tab:breakdown}, all three components significantly contribute to the model's overall performance. The results indicate that the Swin Transformer effectively reinforces nonlocal spatial self-similarities across  bands, thereby enhancing denoising performance. The difference convolution preserves fine image details, improving the model's denoising capacity. Meanwhile, the attention mechanism guides the network to focus on important regions, further boosting denoising effectiveness.
\subsubsection{Convergence Property}
\begin{figure}
    \centering
    \includegraphics[width=0.4\linewidth]{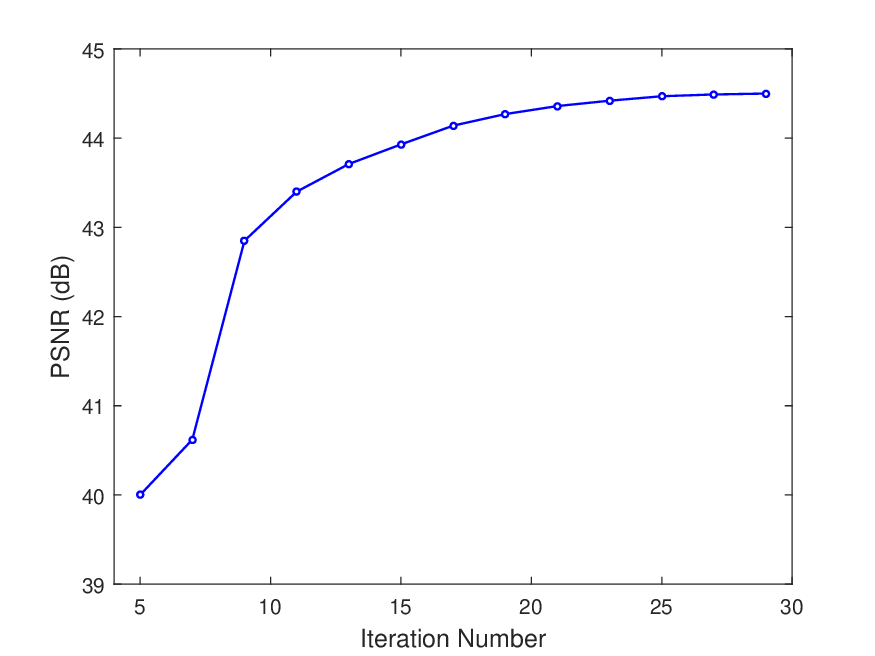}
    \caption{As the number of iterations increases, the PSNR of the fixed-point solution improves steadily and asymptotically converges.}\label{fig:conver}
\end{figure}

To further validate the relationship between our network and numerical optimization, we plot the changes in PSNR with respect to the number of layers during the forward process. As shown in Fig.~\ref{fig:conver}, the PSNR value gradually reaches an optimal level and then stabilizes, with minimal changes as the number of layers increases. This phenomenon clearly demonstrates the promising relationship between the network and the physical model.

\begin{table*}[!t]
    \caption{Comparison of Model Complexity and Efficiency Across All Methods.}\label{tab:params}
    % \vspace{-0.3cm}
    \centering
    \resizebox{\linewidth}{!}{
            \tablesize{
    \begin{tabular}{c|c|c|c|c|c|c|c|c|c|c|c|c|c|c}
            \Xhline{1.2pt}
    \multirow{3}*{\makebox[0.0588\textwidth][c]{Index}}
    &\multicolumn{7}{c|}{\textbf{Model-driven}}&\multicolumn{3}{c|}{\textbf{Data-driven}}&\multicolumn{4}{c}{\textbf{Hybrid-driven}}\\
    \cline{2-15}
    &\multirow{1}*{\makebox[0.0588\textwidth][c]{BM4D}}&\multirow{1}*{\makebox[0.0588\textwidth][c]{MTSNMF}}&\multirow{1}*{\makebox[0.0588\textwidth][c]{LLRT}}&\multirow{1}*{\makebox[0.0588\textwidth][c]{NGMeet}}&\multirow{1}*{\makebox[0.0588\textwidth][c]{LRMR}}&\multirow{1}*{\makebox[0.0588\textwidth][c]{E-3DTV}}&\multirow{1}*{\makebox[0.0588\textwidth][c]{3DlogTNN}}&\multirow{1}*{\makebox[0.0588\textwidth][c]{SST}}&\multirow{1}*{\makebox[0.0588\textwidth][c]{TRQ3D}}&\multirow{1}*{\makebox[0.0588\textwidth][c]{SERT}}&\multirow{1}*{\makebox[0.0588\textwidth][c]{T3SC}}&\multirow{1}*{\makebox[0.0588\textwidth][c]{MTSNN++}}&\makebox[0.063\textwidth][c]{\textbf{DECSC}}&\makebox[0.063\textwidth][c]{\textbf{DECSC}}\\
    &\multirow{1}*{\makebox[0.0588\textwidth][c]{\cite{Maggioni2013BM4D}}}&\multirow{1}*{\makebox[0.0588\textwidth][c]{\cite{Ye2015MTSNMF}}}&\multirow{1}*{\makebox[0.0588\textwidth][c]{\cite{Chang2017LLRT}}}&\multirow{1}*{\makebox[0.0588\textwidth][c]{\cite{He2022NGMeet}}}&\multirow{1}*{\makebox[0.0588\textwidth][c]{\cite{Zhang2014LRMR}}}&\multirow{1}*{\makebox[0.0588\textwidth][c]{\cite{Peng2020E-3DTV}}}&\multirow{1}*{\makebox[0.0588\textwidth][c]{\cite{Zheng20203DlogTNN}}}&\multirow{1}*{\makebox[0.0588\textwidth][c]{\cite{li2022spatialspectral}}}&\multirow{1}*{\makebox[0.0588\textwidth][c]{\cite{Pang2022TRQ3DNet}}}&\multirow{1}*{\makebox[0.0588\textwidth][c]{\cite{li2023spectral}}}&\multirow{1}*{\makebox[0.0588\textwidth][c]{\cite{bodrito2021T3SC}}}&\multirow{1}*{\makebox[0.0588\textwidth][c]{\cite{xiong2023multitask}}}&\multirow{1}*{\makebox[0.07\textwidth][c]{\textbf{(Swin)}}}&\multirow{1}*{\makebox[0.07\textwidth][c]{\textbf{(Mamba)}}}\\
    \Xhline{1.2pt}
    PSNR      & 34.71 & 34.81 & 31.89 & 27.62 & 27.00 & 37.80 & 24.53  & 44.83 & 43.54 & 44.47 & 43.10 & 42.15 & 45.64 & 45.70 \\
    Param     & - & - & - & - & - & - & - & 4.14 & 0.67 & 1.91 & 0.83 & 1.95 & 4.29 & 5.05 \\
    Time (s)  & 198.05 & 40 & 1673.8 & 512.13 & 274.92 & 55.55 & 188.51 & 1.83 & 1.44 & 0.36 & 0.57 & 56.96 & 36.95 & 29.71 \\
    FLOPS     & - & - & - & - & - & - & - & 67.61G & 66.74G & 29.92G & 365.24M & 34.71G & 56.39T & 65.93T \\
    \Xhline{1.2pt}	
    \end{tabular}}}
            %  \vspace{-0.1cm}
\end{table*}

\begin{table}
    \caption{Ablation Study on the Contribution of Each Module.}\label{tab:breakdown}
    % \vspace{-0.3cm}
    \centering
    \begin{tabular}{c|c|c|ccc}
            \Xhline{1.2pt}
    % \multirow{2}*{Low-Rank Module}&\multirow{2}*{Index}
    \multirow{2}*{Swin}&\multicolumn{2}{c|}{\vphantom{\Big|}Detail Enhancement Module}&\multirow{2}*{PSNR$\uparrow$}&\multirow{2}*{SSIM$\uparrow$}&\multirow{2}*{SAM$\downarrow$}\\
    \cline{2-3}
    &\vphantom{\Big|}Difference Conv&Attention&&&\\
    \Xhline{1.2pt}  		
    \multicolumn{1}{c}{\Checkmark} & \multicolumn{1}{c}{\Checkmark} & \multicolumn{1}{c|}{\Checkmark}   & \multicolumn{1}{c}{45.64} & \multicolumn{1}{c}{.9848} & \multicolumn{1}{c}{.0387} \\
    \hline
    \multicolumn{1}{c}{\XSolidBrush} & \multicolumn{1}{c}{\Checkmark} & \multicolumn{1}{c|}{\Checkmark}   & \multicolumn{1}{c}{44.87} & \multicolumn{1}{c}{.9823} & \multicolumn{1}{c}{.0431} \\
    \hline
    \multicolumn{1}{c}{\Checkmark} & \multicolumn{1}{c}{\XSolidBrush} & \multicolumn{1}{c|}{\Checkmark}   & \multicolumn{1}{c}{45.32} & \multicolumn{1}{c}{.9839} & \multicolumn{1}{c}{.0410} \\
    \hline
    \multicolumn{1}{c}{\Checkmark} & \multicolumn{1}{c}{\Checkmark} & \multicolumn{1}{c|}{\XSolidBrush}   & \multicolumn{1}{c}{44.99} & \multicolumn{1}{c}{.9829} & \multicolumn{1}{c}{.0434} \\
    \Xhline{1.2pt}
    \end{tabular}
            %  \vspace{-0.1cm}
\end{table}

\subsubsection{The Impact of Dictionary Size}
\begin{figure}[!t]
    \centering
    \includegraphics[width=0.32\linewidth]{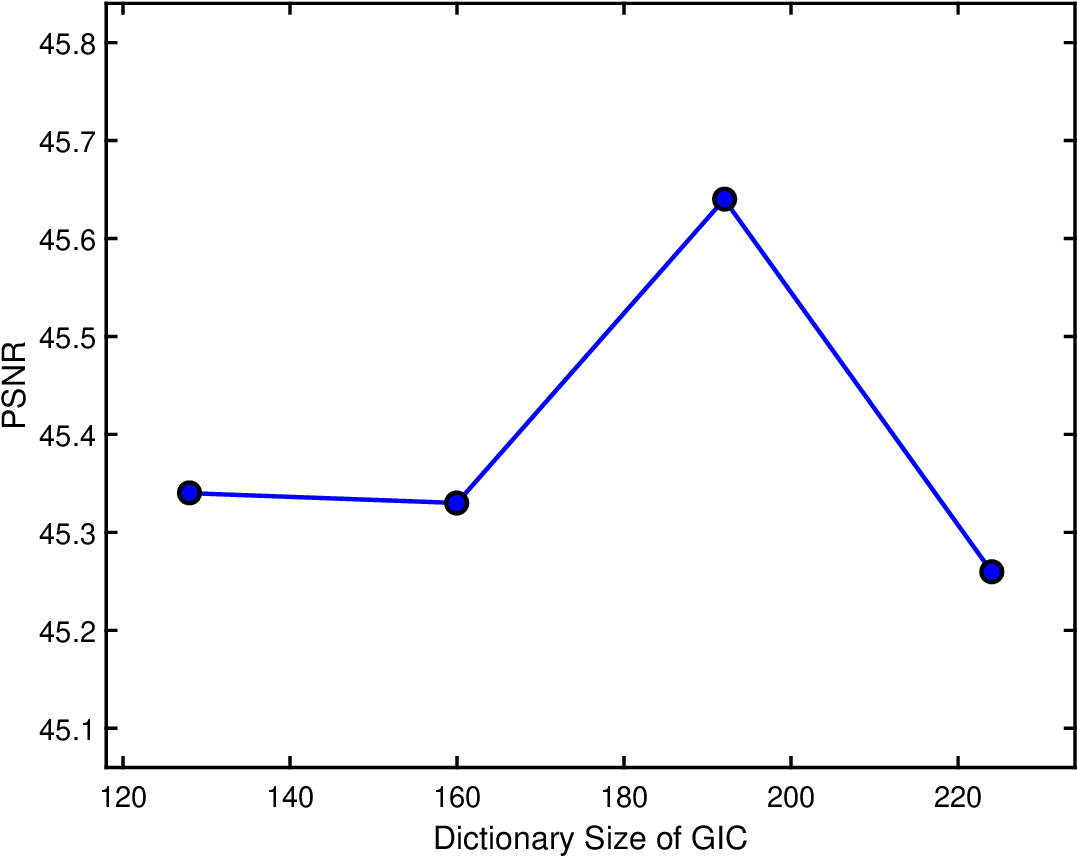}
    \includegraphics[width=0.32\linewidth]{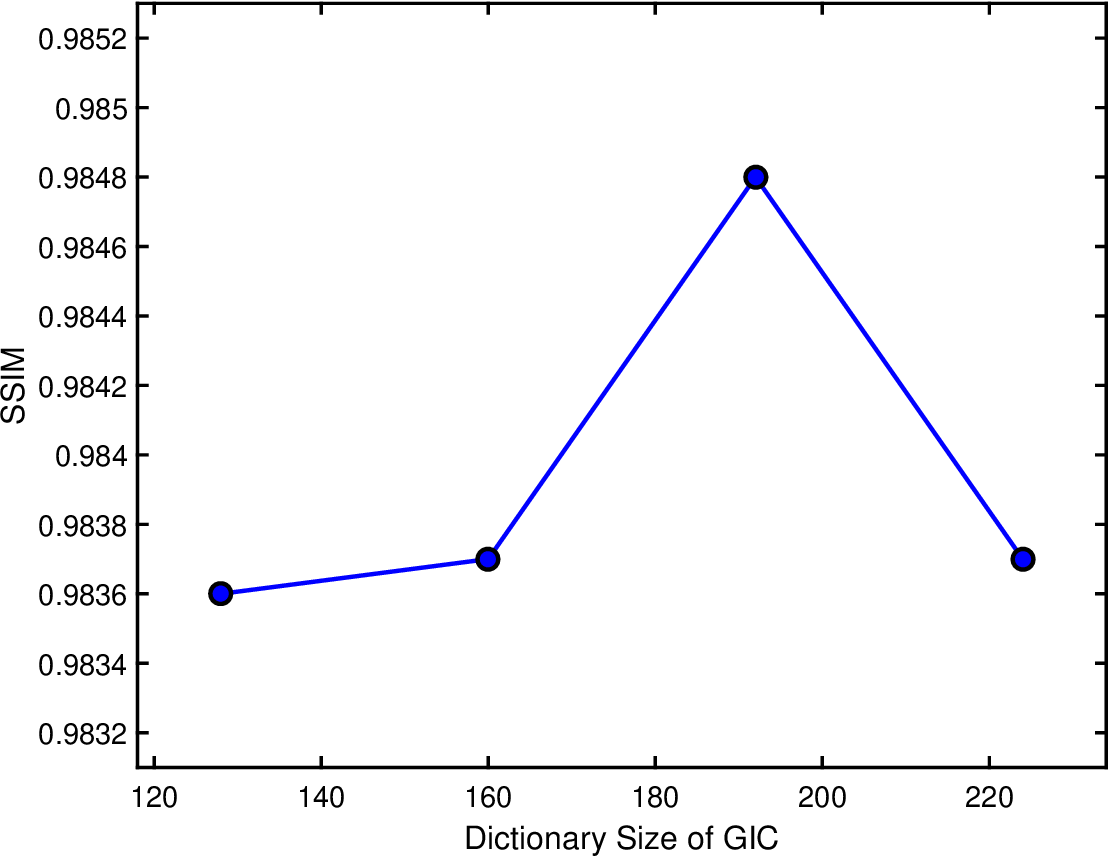}
    \includegraphics[width=0.32\linewidth]{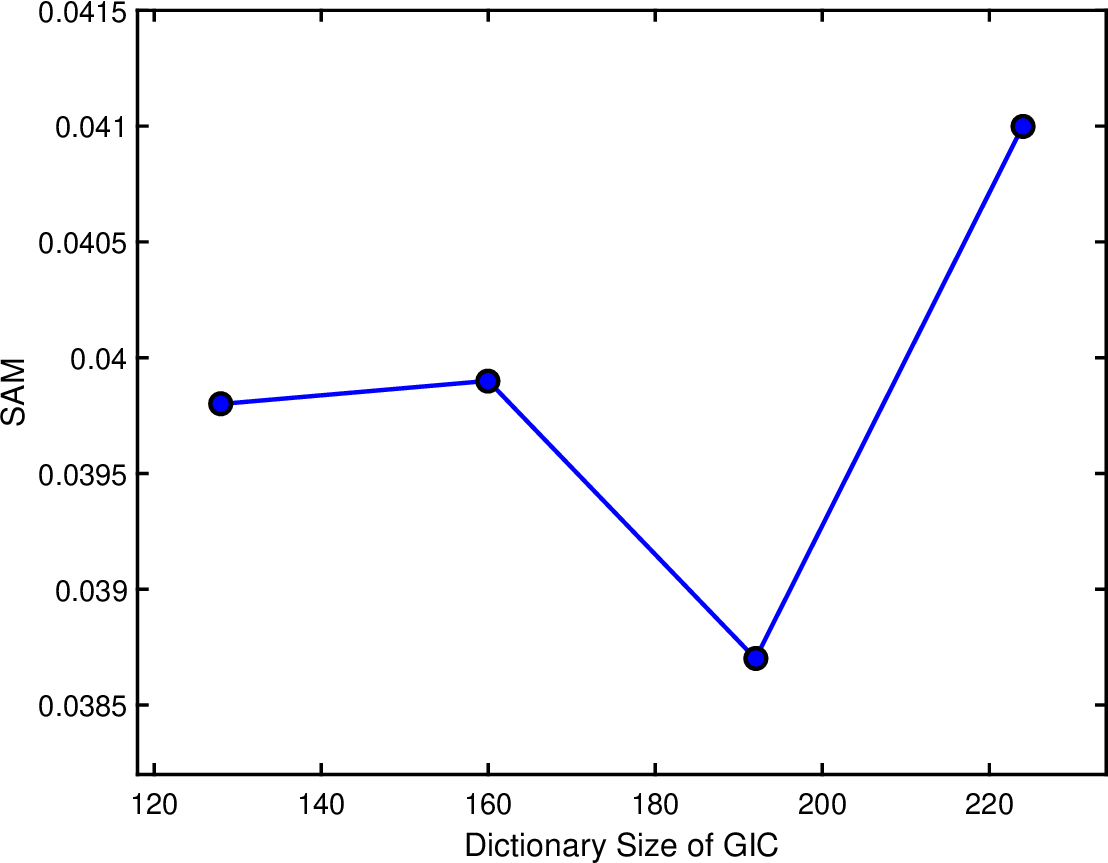}
    \caption{Impact of the number of atoms in the GIC component on the denoising performance.} \label{fig:gic}
 %    \vspace{-0.5cm}
 %   \vspace{-0.3cm}
\end{figure}

\begin{figure}[!t]
    \centering
    \includegraphics[width=0.32\linewidth]{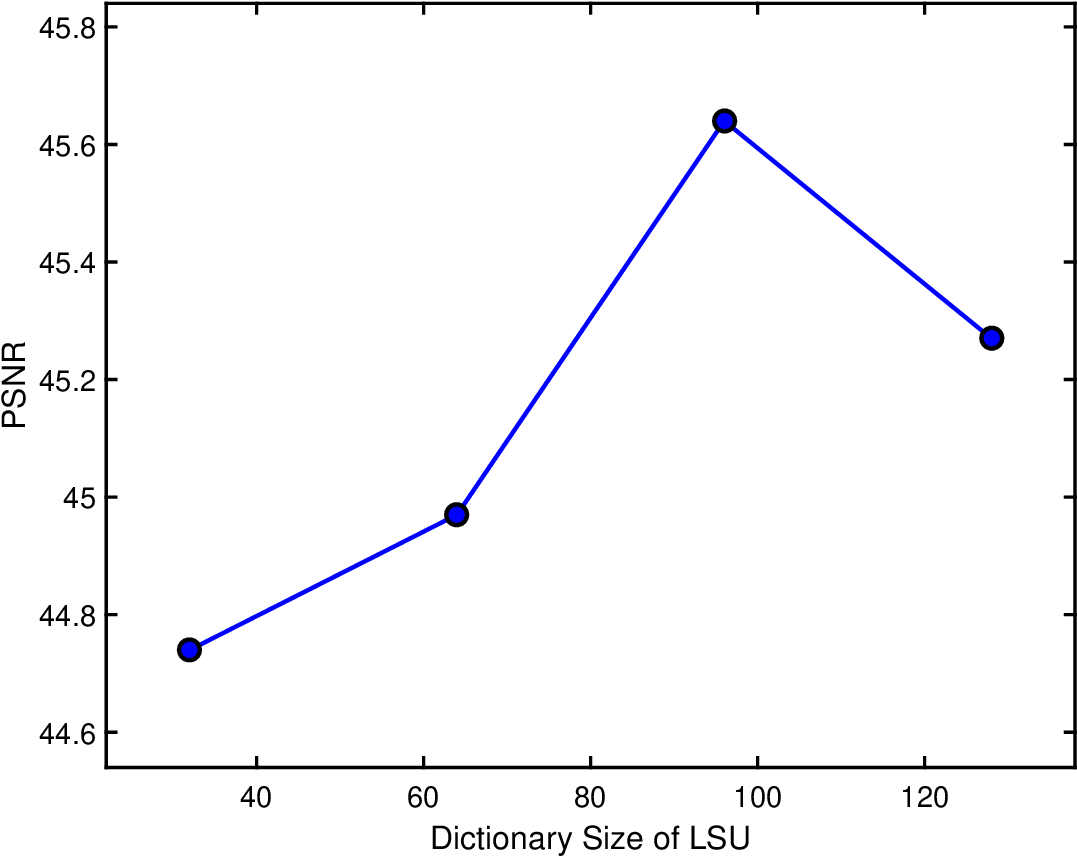}
    \includegraphics[width=0.32\linewidth]{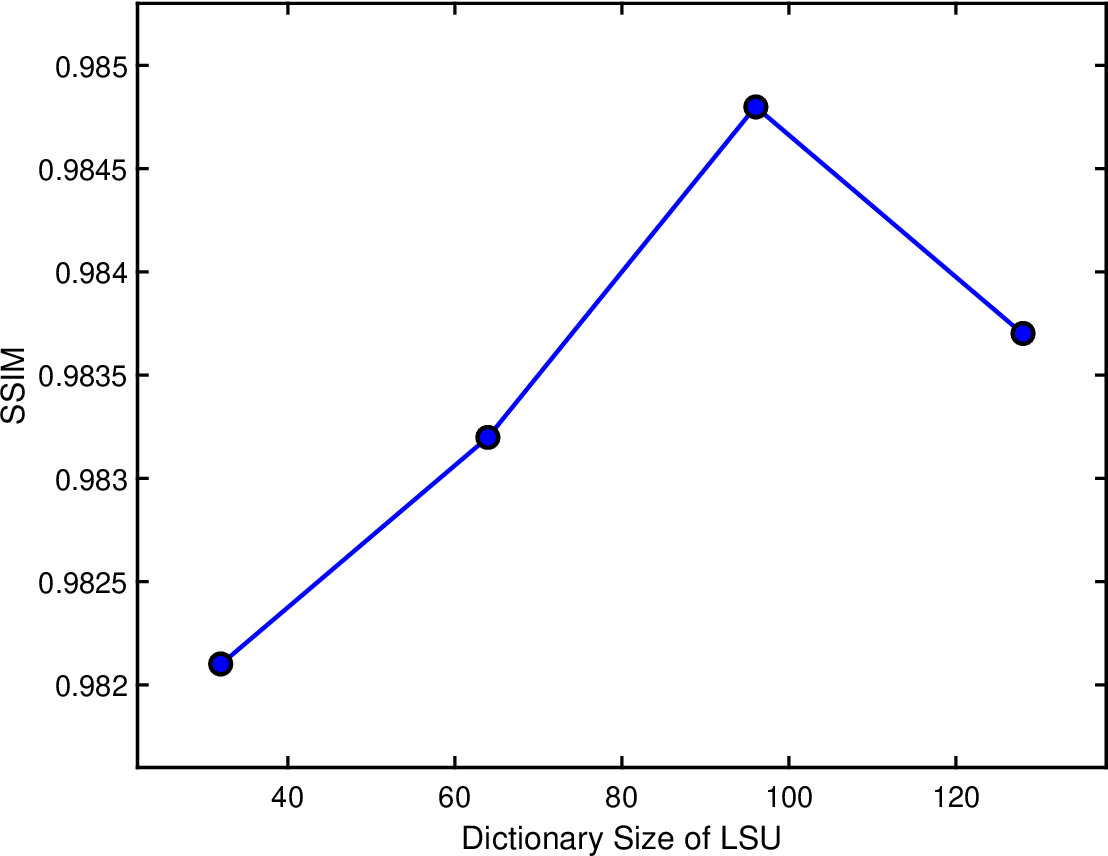}
    \includegraphics[width=0.32\linewidth]{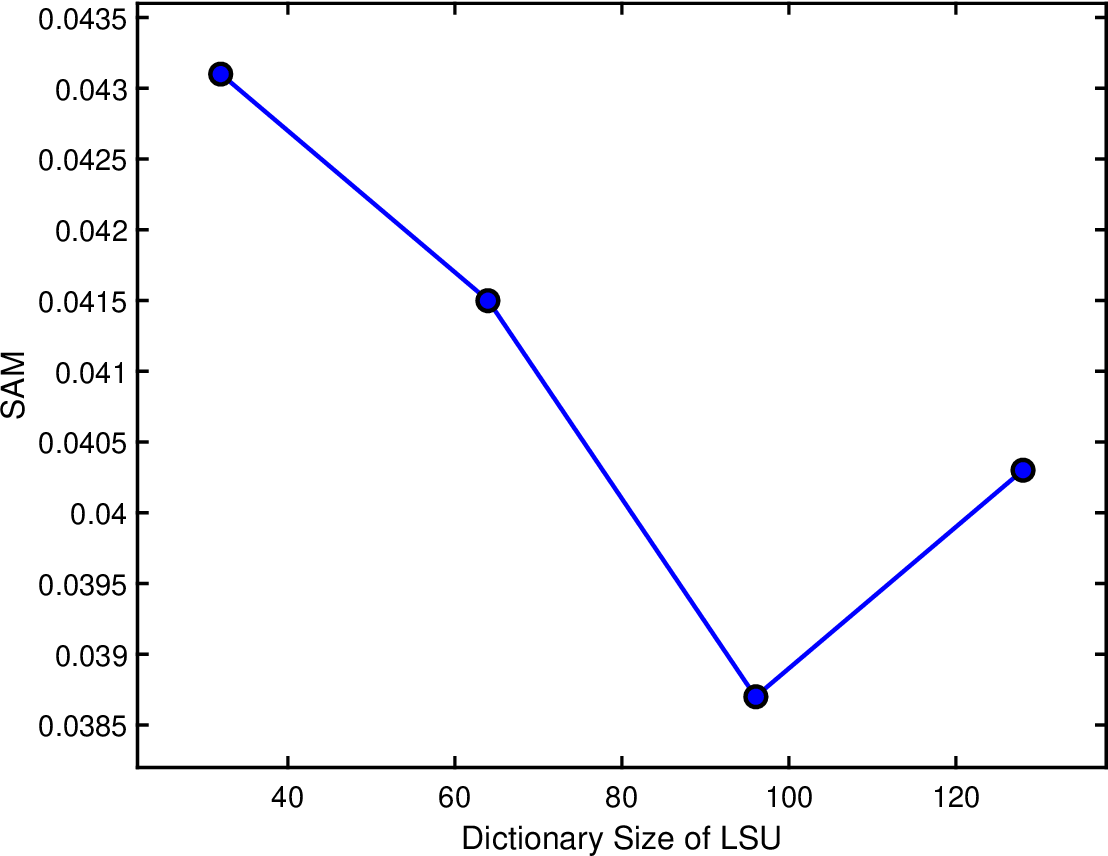}
    \caption{Impact of the number of atoms in the LSU component on the denoising performance.} \label{fig:lsu}
 %    \vspace{-0.5cm}
 %   \vspace{-0.3cm}
\end{figure}

To evaluate the impact of the number of atoms on denoising performance, we conducted ablation experiments on both the GIC and LSU components. Specifically, we first fixed the number of atoms in the LSU component to 96 and varied the number in the GIC component from 128 to 224 in increments of 32. As shown in Fig.\ref{fig:gic}, the denoising performance generally improved with an increasing dictionary size, with the best performance achieved at 192 atoms. Secondly, with the number of atoms in the GIC component fixed at 192, we varied the number of atoms in the LSU component from 32 to 128 in increments of 32. As shown in Fig.\ref{fig:lsu}, the optimal performance was obtained when the number of atoms was set to 96.
\subsubsection{The Impact of Neumann Series}
\begin{figure}[!t]
    \centering
    \includegraphics[width=0.32\linewidth]{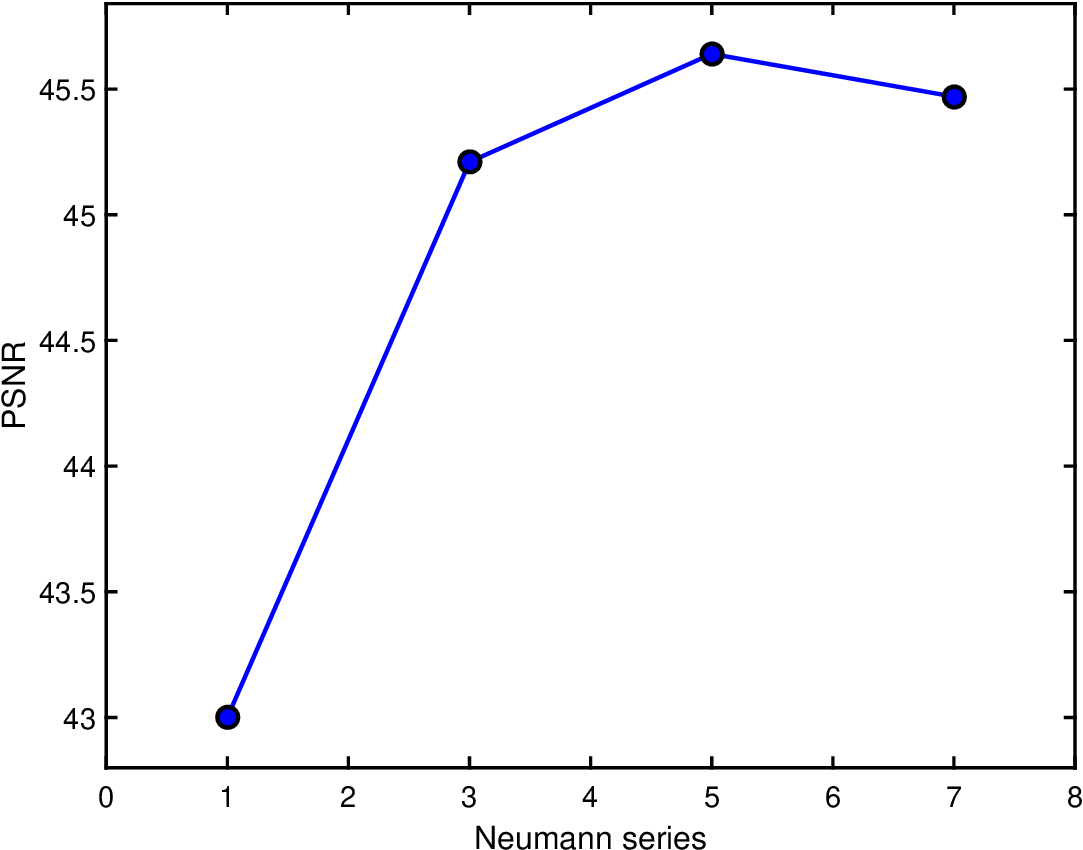}
    \includegraphics[width=0.32\linewidth]{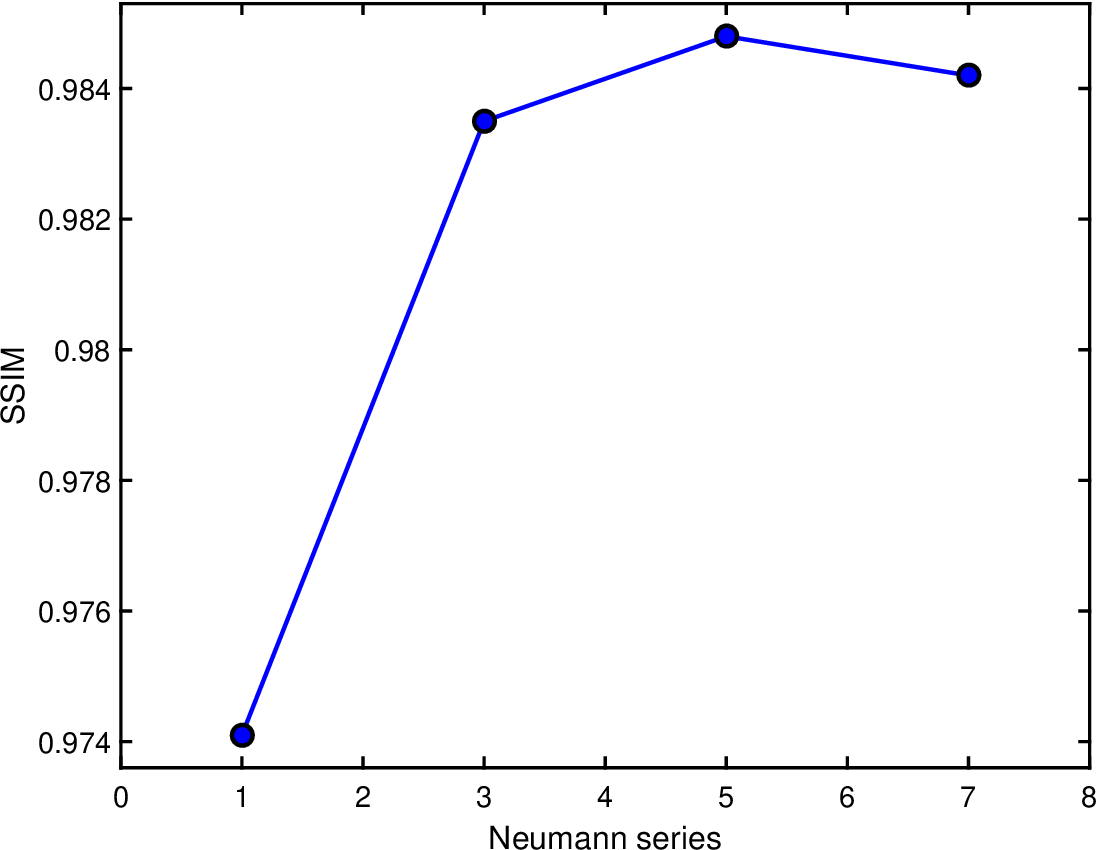}
    \includegraphics[width=0.32\linewidth]{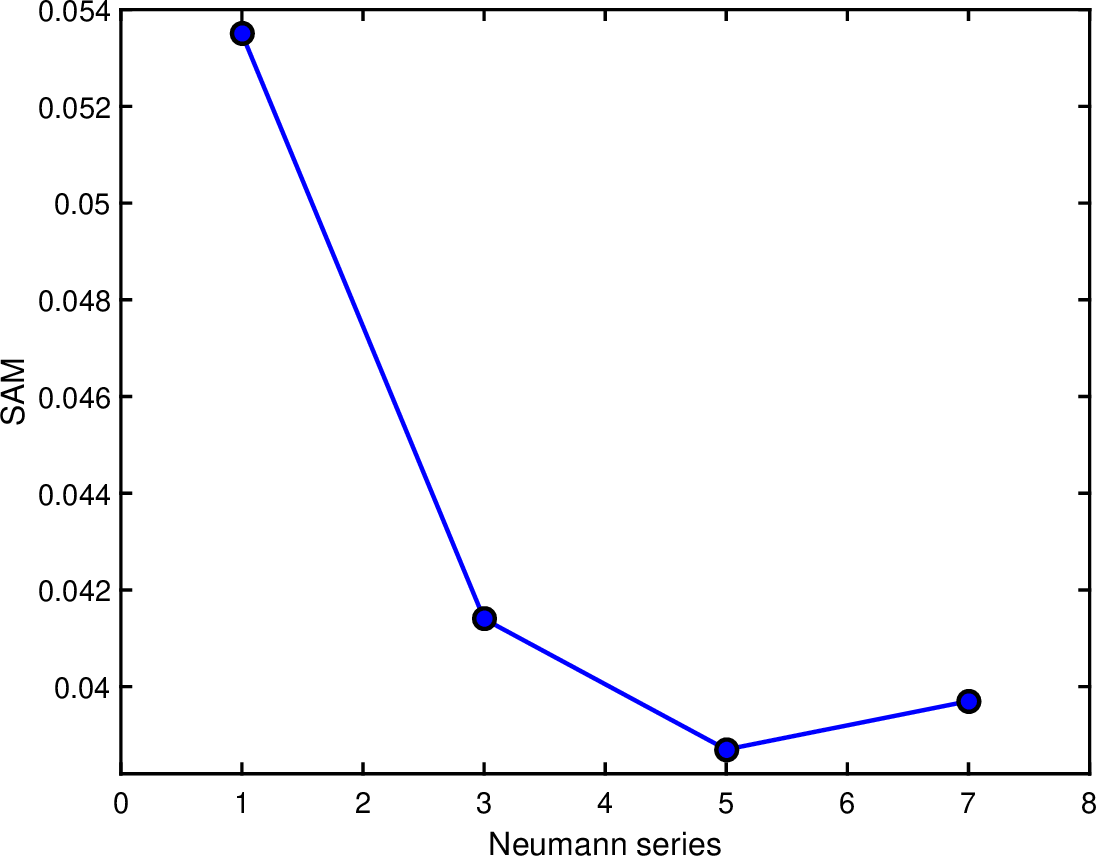}
    \caption{Impact of the Neumann series parameter L on the denoising performance.} \label{fig:L}
  %    \vspace{-0.5cm}
  %   \vspace{-0.3cm}
\end{figure}
We conducted an ablation study on the Neumann series to investigate its impact on denoising performance. As shown in Fig.~\ref{fig:L}, the overall denoising performance improves with increasing approximation accuracy, and the best performance is achieved when L=5.
\subsubsection{Comparison of Model Complexity and Efficiency}
As shown in the table, our DECSC model maintains a comparable parameter count to the SST model while achieving significantly better performance. However, we acknowledge that our method lags behind in running time due to its iterative forward pass, which stops only upon reaching an equilibrium point. This process is inherently more time-consuming and requires more FLOPs than conventional neural networks with a fixed number of layers. Given the substantial computational overhead introduced by the quadratic complexity of Transformers when processing long sequences, this work introduces Mamba to explore its capability in modeling global dependencies. Specifically, we adopt the VMamba architecture, which performs directional scanning along four spatial orientations to effectively capture global interactions within the GIC. To ensure a fair comparison, both the Transformer and VMamba models are implemented using an identical four-layer stacking configuration. The Mamba-based variant achieves performance comparable to that of the Transformer while benefiting from lower computational complexity, resulting in reduced inference time compared to the Transformer-based variant.

\section{Conclusion}\label{conc}

This paper introduces a novel DECSC framework for robust HSI denoising. By modeling the global shared spatial structure and local spatial-spectral structure, the denoising performance is significantly improved. In addition, the integration of transformer blocks and a detail enhancement module further boosts denoising by capturing nonlocal spatial self-similarities and local details. Thanks to the DEQ approach, the iterative optimization of the CSC model is effectively transformed into a learnable network that maintains physical interpretability and convergence guarantees. In future work, we plan to explore additional priors to better capture the unique structural properties of HSIs.

\appendices
\bibliographystyle{IEEEtran}
\bibliography{CSCNet}
\end{document}